\newcommand{\gsim}{\;\lower.6ex\hbox{$\sim$}\kern-7.75pt\raise.65ex\hbox{$>$}\;}
\newcommand{\lsim}{\;\lower.6ex\hbox{$\sim$}\kern-7.75pt\raise.65ex\hbox{$<$}\;}

\documentclass{aa}  

\usepackage{graphicx}
\usepackage{txfonts}
\usepackage[english]{babel}  
\usepackage[utf8]{inputenc}  
\usepackage{microtype}  
\usepackage{booktabs}  
\usepackage{rotating}  
\usepackage{lscape}
\usepackage[squaren]{SIunits}
\usepackage{natbib}
\usepackage{color}
\usepackage{amsmath}
\usepackage{amssymb}
\usepackage{placeins}
\begin{document}

   \title{The SOUL view of IRAS\,20126+4104.}

   \subtitle{Kinematics and variability of the H$_2$ jet from a massive protostar}

   \author{F. Massi
         \inst{1}
          \and
          A. Caratti o Garatti\inst{2,3}
          \and
          R. Cesaroni\inst{1}
          \and
          T. K. Sridharan\inst{4,5}
          \and
          E. Ghose\inst{1}
          \and 
          E. Pinna\inst{1}
          \and
          M. T. Beltr\'an\inst{1}
          \and
          S. Leurini
          \inst{6}
          \and
          L. Moscadelli\inst{1}
          \and
          A. Sanna\inst{6}
          \and 
          G. Agapito\inst{1}
          \and
          R. Briguglio\inst{1}
          \and
          J. Christou\inst{7}
          \and
          S. Esposito\inst{1}
          \and
          T. Mazzoni\inst{1}
          \and
          D. Miller\inst{7}
          \and
          C. Plantet\inst{1}
          \and
          J. Power\inst{7}
          \and
          A. Puglisi\inst{1}
          \and
          F. Rossi\inst{1}
          \and
          B. Rothberg\inst{7,8}
          \and
          G. Taylor\inst{7}
          \and
          C. Veillet\inst{7}
          }

   \institute{INAF-Osservatorio Astrofisico di Arcetri, 
              Largo E. Fermi 5, I-50125 Firenze, Italy\\
              \email{fabrizio.massi@inaf.it}
         \and
             INAF-Osservatorio Astronomico di Capodimonte, 
             via Moiariello 16, I-80131 Napoli, Italy\\
             \email{alessio.caratti@inaf.it}
        \and
            Dublin Institute for Advanced Studies, School of Cosmic Physics, Astronomy and Astrophysics Section, 31 Fitzwilliam Place, Dublin 2, Ireland
        \and
            National Radio Astronomy Observatory,
            520 Edgemont Road, Charlottesville, VA 22903, USA
        \and 
            Harvard-Smithsonian Center for Astrophysics, 60 Garden Street, Cambridge, MA 02138, USA 
        \and
            INAF-Osservatorio Astronomico di Cagliari, 
            Via della Scienza 5, I-09047 Selargius (CA), Italy
        \and
            Large Binocular Telescope Observatory, 933 N.\ Cherry Ave \#552, Tucson, AZ 85721, USA
        \and
            George Mason University, Department of Physics \& Astronomy, MS 3F3, 4400 University Dr., Fairfax, VA 22030, USA
             }

   \date{Received ; accepted }

 
  \abstract
   {We exploit the increased sensitivity of the recently installed adaptive optics SOUL at the LBT to obtain new high-spatial-resolution 
near-infrared images of the massive young stellar object IRAS20126+4104 and its outflow.
   }
   {We aim to derive the jet proper motions and kinematics, as well as to study its photometric variability by combining the novel 
performances of SOUL together with previous near-infrared images.
   }
   {We used both broad-band ($K_{s}$, $K'$) and narrow-band (Br$\gamma$, H2) observations from a number of near-infrared cameras 
(UKIRT/UFTI, SUBARU/CIAO, TNG/NICS, LBT/PISCES, and LBT/LUCI1) to derive maps of the continuum and the H$_2$ emission in the $2.12$ 
$\mu$m line. Three sets of images, obtained with adaptive optics (AO) systems (CIAO, in 2003; FLAO, in 2012; SOUL, in 2020), 
allowed us to derive the proper motions of a large number of H$_2$ knots along the jet. Photometry from all images was used to study 
the jet variability.
   }
   {We derived knot proper motions in the range of $1.7$--$20.3$ mas yr$^{-1}$ (i.e. 13--158 km s$^{-1}$ at $1.64$ kpc), implying 
an average outflow tangential velocity of $\sim 80$ km s$^{-1}$. The derived knot dynamical age spans a $\sim 200$--$4000$ yr 
interval. A ring-like H$_2$ feature near the protostar location exhibits peculiar kinematics and may represent the outcome of 
a wide-angle wind impinging on the outflow cavity. 
   Both H$_2$ geometry and velocities agree with those inferred from proper motions of the H$_2$O masers, located at a smaller 
distance from the protostar. Although the total H$_2$ line emission from the knots does not exhibit time variations at a 
$\widetilde{>} 0.3$ mag level, we have found a clear continuum flux variation
   (radiation scattered by the dust in the cavity opened by the jet) which is anti-correlated between the blue-shifted and 
red-shifted lobes and may be periodic (with a period of $\sim 12-18$ yr). We suggest that the continuum variability might be 
related to inner-disc oscillations which have also caused the jet precession.
   }
 {}


\maketitle
%

\section{Introduction}

Collimated jets and outflows from low-mass young stars and protostars are ubiquitous and are now believed to be intimately linked 
with accretion discs. The most widely accepted paradigm is that they originate from magnetically collimated disc winds and help 
remove angular momentum from 
the disc, allowing the gas to accrete onto the star surface (see, e.\,g. \citealt{2021NewAR..9301615R} and references therein). 
Another well-established observational fact in low-mass star formation is the so-called Luminosity Problem, whereby protostars are 
usually much fainter than predicted for expected accretion rates of $\sim 10^{-4} - 10^{-5}$ M$_{\sun}$ yr$^{-1}$ (see, e.\,g.
\citealt{2011IAUS..270...73M}), pointing to a much lower accretion activity. A proposed
scenario to circumvent the problem is episodic accretion, namely protostars grow through short episodes of intense matter inflows. 
If accretion and outflows are linked together, this would in fact agree with the morphology of many optical and near-infrared (NIR) 
jets that appear composed of aligned knots (emitting in several optical and NIR spectral lines),
which, in this view, would be relics of past accretion bursts. In this
scenario, both protostars and very young stars should also exhibit photometric variability with augmented luminosity occurring 
during phases of intense accretion, which has actually been observed
in both low-mass (see, e.\,g. \citealt{2017ApJ...849...69Y}; \citealt{2019ASSP...55..111C}), intermediate-mass (\citealt{2010A&A...517L...3B}) and high-mass young stars (see, e.\,g. \citealt{2017NatPh..13..276C};
\citealt{2017ApJ...837L..29H}).

In high-mass star formation, the picture is observationally more problematic as high-mass protostars are rare and usually far away (at kilo-parsec distances), and embedded
in crowded regions where low-mass protostars are also present. 
This highlights the need for high-spatial-resolution observations. In addition, studying photometric variability of a few selected 
high-mass-star precursors and their associated outflows can be a valuable tool to understand how accretion occurs in these objects, 
circumventing the need to resolve the driving protostars. This is of course better achieved when multi-epoch high-spatial-resolution 
observations are available.

In this respect, \object{IRAS\,20126+4104} is a source of great interest. 
This massive protostar has a luminosity of $\sim 10^4~L_\sun$ (\citealt{2007A&A...465..197H}; \citealt{2011MNRAS.415.2953J}) 
and is surrounded by a disc; this was first detected by \cite{1997A&A...325..725C} and confirmed by subsequent higher-resolution 
studies (\citealt{1998ApJ...505L.151Z}; \citealt{1999A&A...345..949C}, \citeyear{2005A&A...434.1039C},
\citeyear{2014A&A...566A..73C}).
It appears to be powering a parsec-scale jet-outflow undergoing precession about the disc axis (\citealt{2000ApJ...535..833S}; \citealt{2005A&A...434.1039C}; \citealt{2008A&A...485..137C}).
 This jet-outflow has been imaged on scales from a few 100~au (see Fig.~\ref{3col:fig}; \citealt{2005A&A...438..889M};
 \citealt{2013A&A...549A.146C}) to 0.5~pc (\citealt{2000ApJ...535..833S}; \citealt{2006A&A...448.1037L}) and its 3D expansion 
velocity, close to the source, has been measured through maser proper motions (\citealt{2005A&A...438..889M}). Two radio sources 
(see Fig.~\ref{3col:fig}) have been detected at high-spatial resolution and are associated with the protostar (\citealt{1999A&A...345L..43H}, 
\citeyear{2007A&A...465..197H}). The brighter one, N1, has been detected at $3.6$, $1.3$, and $0.7$\,cm and exhibits a spectral 
index consistent with thermal emission. It is likely to be located $\sim 0\farcs3$ north-west of the protostar, whose position has 
been inferred by OH maser emission at 1665\,MHz, displaying an elongated structure with a north-eastern-south-western
velocity gradient reminiscent
of Keplerian rotation \citep{2005A&A...434..213E}. The fainter one, N2, has only been detected at $3.6$\,cm, is located $\sim 1 
\arcsec$ north-west of N1 and its morphology is reminiscent of a bow shock. Radio sources N1 and N2 are roughly aligned in the 
direction of the jet, are associated with the water maser system, and have been interpreted as a manifestation of gas that has been 
shocked by 
the jet from the protostar \citep{2007A&A...465..197H}. Another feature found is a faint radio source, S (see Fig.~\ref{3col:fig}, 
top panel), which has only been detected at $3.6$\,cm, and lies $\sim 1\arcsec$ south of N1 and is roughly elongated in a 
direction parallel to that of 
the jet. This has been interpreted as a distinct radio jet and its contribution to the outflow should be negligible 
(\citealt{2000ApJ...535..833S}; \citealt{2007A&A...465..197H}); \citeauthor{2007AAS...211.6213S} 
(\citeyear{2005ApJ...631L..73S}, \citeyear{2007AAS...211.6213S}) 
found a point-like source in $L$- and $M$-band high-resolution images coinciding with S, proving that the latter has a stellar 
counterpart. A clear X-ray detection was obtained with {\it Chandra} and the X-ray spectrum is consistent with an embedded early B-type 
star \citep{2011AJ....142..158A}. 
 
The distance to IRAS\,20126+4104 (1.64$\pm$0.05~kpc) has been accurately determined from parallax measurements of water masers 
\citep{2011A&A...526A..66M}, which prove that this is one of the closest discs around a B-type protostar and this allows one to 
observe its environment with excellent linear resolution. More recently, \cite{2015PASJ...67...66N} repeated the parallax 
measurement with VERA, obtaining a distance of 1.33$^{+0.19}_{-0.092}$~kpc. As the difference with the value of 
\cite{2011A&A...526A..66M} is not statistically significant, in the following, we continue to use a distance of 1.64$\pm$0.05~kpc.

Sub-arcsecond imaging \citep{2014A&A...566A..73C}  and model fitting \citep{2016ApJ...823..125C}
 have proven that the disc is undergoing Keplerian rotation around a $\sim 12 M_\sun$ protostar, as suggested by the
 butterfly-shaped position--velocity plot obtained in almost every hot molecular
 core tracer observed. Evidence of
 accretion onto the protostar has also been reported (\citealt{1997A&A...325..725C}; \citealt{2010MNRAS.406..102K}; \citealt{2011MNRAS.415.2953J}). The jet traced by the water masers over a few 100~au from the protostar appears to be co-rotating with the disc (\citealt{2005AJ....130.2206T}; \citealt{2014A&A...566A..73C}).
 
Imaging at infrared wavelengths \citep{2008ApJ...685.1005Q} suggested
 that IRAS\,20126+4104 was associated with an anomalously poor stellar cluster. However,
 subsequent X-ray observations by \cite{2015ApJS..219...41M} revealed an
 embedded stellar population that hints at the existence of a richer (and possibly very young) cluster.
 
Measurements of the magnetic field orientation both at the clump scale \citep{2012ApJ...750L..29S}
and at the disc scale \citep{2014A&A...563A..30S} show that the direction of the field lines is basically the same from $\sim 0.2$\,pc
 to $\sim 1000$\,au and roughly perpendicular to the outflow axis. This suggests
 that the accreting material could be flowing onto the star through the disc along the magnetic field lines.

With all the above in mind, we think that IRAS\,20126+4104 is a suitable target
for high-resolution imaging at NIR wavelengths, in order to measure the physical properties of its protostellar jet, as previously 
undertaken, in part, by \citet{2008A&A...485..137C} and \citet{2013A&A...549A.146C}. 
These works detected a well-collimated and structured precessing jet, with two lobes, a south-eastern one (red-shifted) and a 
north-western one (blue-shifted), composed of several knots and bow-shock structures (see top panel of Fig.~\ref{3col:fig}), 
with each lobe spanning an arc of $\sim 10\arcsec$ ($\sim 0.08$\,pc). Larger field images show
further distant knots which are not aligned with the lobes, but they are reminiscent of a wiggled configuration. Wiggling is also 
evident in the two lobes, suggesting that the outflowing direction is in fact precessing with a period whose estimates range from 
$\sim 1100$ yr \citep[for the closest knots;][]{2008A&A...485..137C} to $\sim 64000$ yr (for the outer flow; 
\citealt{2000ApJ...535..833S}; \citealt{2005A&A...434.1039C}).
A puzzling feature that stands out in the images is a small ($\sim 1 \arcsec$) ring-like patch (hereby, the {\it ring-like feature} 
or $X$, see Fig.~\ref{3col:fig}) that mainly emits in the H$_2$ lines. This is not the circumstellar disc, which appears to be 
heavily obscured and invisible in the NIR, but a structure located a few tens of arcsecs south-east of the disc, hence facing the 
red-shifted lobe of the jet (see Fig.~\ref{3col:fig}).

We therefore targeted IRAS\,20126+4104 for observations with the second generation adaptive optics (AO) system SOUL that is nearly 
commissioned at the Large Binocular Telescope (LBT). The complex jet structure and the presence of a suitable guide star are optimal 
features to test the performances of an AO system. In addition, we aimed to complement the new observations with other 
high-resolution images available from 2000 on to derive, for the first time, proper motions, 3D kinematics and photometric 
variability along the NIR jet of a massive protostar.

The paper layout is the following. Section~\ref{obso:dr} describes the observations and the data reduction, including all the 
ancillary data. The results of our data analysis are dealt with in Sect.~\ref{resu:sec}. The implications of our results are 
discussed in Sect.~\ref{discu:sec}.

\section{Observations and data reduction}
\label{obso:dr}

\subsection{SOUL AO at the LBT}

SOUL (\citealt{2016SPIE.9909E..3VP}, \citealt{2021arXiv210107091P}) is the upgrade of the FLAO systems \citep{2010SPIE.7736E..3HQ} 
at the LBT with an electro-multiplied camera for the wavefront sensor. As FLAO, SOUL is a single conjugated AO system, using the 
adaptive secondary mirror as a corrector and one natural guide star for a pyramid wavefront sensor. The upgrade allows us to fully exploit the light flux from the reference star, improving the correction. In this observation, we used the $R=13.49$ star (according to 
UCAC5; \citealt{2017AJ....153..166Z}) as the AO reference, which is 
located at the western edge of the field (Fig.~\ref{3col:fig}, bottom), correcting 91 modes at 850 Hz.

\subsection{LBT/SOUL NIR observations}

The new AO-assisted observations were carried out with the NIR imager LUCI1 at the beginning of November 2020, during the 
commissioning time of SOUL. On-source and off-source dithered images were obtained
through the narrow-band filters H2 and Br$\gamma$
(centred at $2.124$ $\mu$m and $2.170$ $\mu$m, respectively), and the broad-band filter $K_{s}$ (centred at $2.163$ $\mu$m)
with the N30 camera (plate scale $0\farcs015$ pix$^{-1}$,
field of view $30\arcsec \times 30\arcsec$). LUCI1 is equipped with a HAWAII-2RG HgCdTe detector with $2048 \times 2048$  pixels. 
The $K_{s}$ images were obtained using the standard double-correlated sample readout scheme (indicated as LIR in the instrument 
handbook), whereas the H2 and Br$\gamma$ images used a readout scheme called MER, consisting of a set of five readings at the 
beginning and at the end of each integration. 
In addition, each set of NDIT\footnote{A NIR image usually consists of a sequence of $n$ single integrations of time $t$ which are 
combined on-chip, where $t$ is indicated as DIT and $n$ as NDIT.} 
exposures through the H2 and Br$\gamma$ filters was saved as a 3D FITS file with NDIT planes. A log of the observations is given 
in Table~\ref{obs:log}.

%
\begin{table*}
\caption{Observations log.}             
\label{obs:log}      
\centering          
\begin{tabular}{l l l l l l l}     
\hline\hline       
Filter & Target & Date & DIT & NDIT\tablefootmark{a} & Number & Number \\
       & & of observations & & & of on-source & of off-source\\
       & & (UT) & (s) & & frames & frames \\
\hline                    
$K_{s}$ & IRAS\,20126+4104 & November 2 & 5 & 12 & 5 & 5 \\
$K_{s}$ & FS149\tablefootmark{b} & November 2 & 30 & 1 & 5 & 0 \\
$K_{s}$ & IRAS\,20126+4104 & November 4 & 5 & 12 & 5 & 5 \\
$K_{s}$ & FS149\tablefootmark{b}& November 4 & 30 & 1 & 5 & 0 \\
$K_{s}$ & IRAS\,20126+4104 & November 6 & 5 & 12 & 3 & 5 \\
$K_{s}$ & FS149\tablefootmark{b} & November 6 & 30 & 1 & 2 & 0 \\
H2 & IRAS\,20126+4104 & November 3 & 30 & 10 & 7 & 7 \\
Br$\gamma$ & IRAS\,20126+4104 & November 4 & 30 & 8 & 5 & 5 \\
\hline                  
\end{tabular}
\tablefoot{
\tablefoottext{a}{H2 and Br$\gamma$ were obtained using the detector reading mode MER and all single DITs in a set of NDIT 
exposures are saved as a 3D FITS file, so that each frame consists of a data-cube of NDIT planes}
\tablefoottext{b}{taken with an open loop}
}
\end{table*}

The target field was imaged through a sequence of
dithered exposures lasting DITxNDIT each, alternating an on-source integration and an off-source one. The centre of field
of each on-source frame was randomly shifted with offsets of $< 3\arcsec$ in the $x$ and $y$ directions of the detector
reference frame. A dark area $\sim 1\arcmin$ away was selected as an off-source position and a dithering scheme with random offsets 
$< 1 \arcmin$ was adopted. A standard star (FS149, \citealt{2001MNRAS.325..563H}) was imaged in $K_{s}$ 
with an open loop (i.e. with the AO offline) when observations were carried out in this band (see Table~\ref{obs:log}) using
a five-position dithering scheme with the star alternatively in the frame centre and in each of the four quadrants ($x$ and $y$ 
offsets of $5\arcsec$). On November 6, only two shifted images of FS149 were taken due to poor weather conditions.

\begin{figure}
            \includegraphics[width=9cm]{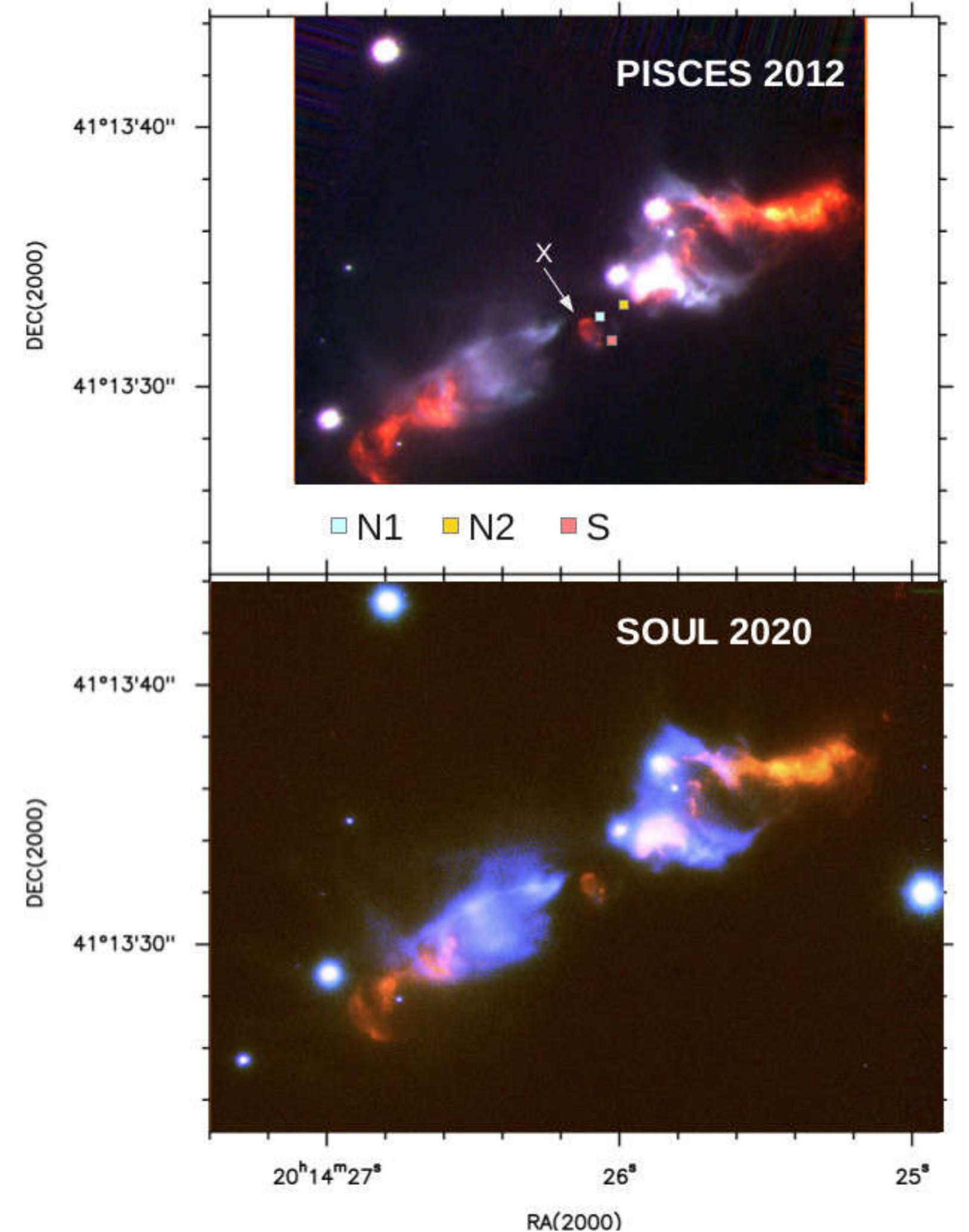}
      \caption{Three colour image (red H2 filter, green $K_{S}$,
      blue Br$\gamma$ filter) of IRAS\,20126+4104 obtained with PISCES and FLAO at the LBT in 2012 (top) and LUCI1 and the new 
      AO system SOUL at the LBT in November 2020 (bottom). The $K_{S}$ image used 
      for the bottom image is that taken on
      November 4, that is the one exhibiting the best spatial resolution. Approximate positions of radio sources
      N1 (blue square), N2 (yellow square), and S (orange square, see text) are marked in the top panel with small coloured 
      squares.
         \label{3col:fig}}
   \end{figure}

Data reduction was performed using standard IRAF routines. First, the frames were flat-field corrected using dome flat images. 
All images in each off-source 3D FITS file were averaged together in a single frame. Then, for each on-source 3D FITS file, a sky 
image was obtained as the median of the four nearest (in time) NDIT-averaged off-source images. For each on-source 3D FITS file, the 
single exposures were extracted, sky-subtracted, corrected for bad pixels, and checked for any change in point spread function (PSF) 
width and in flux. Deviant frames were discarded and the remaining ones were registered and averaged together. This was to compensate 
for any internal flexure from the LUCI instrument during the observations. The final averaged, sky-subtracted on-source images obtained 
from each 3D FITS file were registered and summed together. The same scheme was used for the $K_{s}$ frames, except that in this case 
the NDIT exposures were not saved as 3D FITS files. In the end, the final H2 image was the sum of 65 exposure (total integration time 
DIT $\times$ $65 = 1950$ s) with a PSF width $\sol 0.1\arcsec$ (mostly less than $0.083\arcsec$; measured with IRAF, for an accurate 
determination of the PSF widths see Appendix~\ref{soul:perf}), 
and the final Br$\gamma$ image is the sum of 40 exposures (total integration time DIT $\times$ $40 = 1200$ s) with a PSF width 
$\sol 0.09\arcsec$. The final $K_{s}$ images are the sum of five DIT$\times$NDIT exposures (total integration time 300 s, final PSF 
width $< 0.11\arcsec$; for an accurate assessment of the PSF widths
see Appendix~\ref{soul:perf}) for November 2, of five DIT$\times$NDIT exposures (total integration time 300 s, final PSF width 
$\sol 0.08\arcsec$) for November 4, of three DIT$\times$NDIT exposures (total integration time 180 s, final PSF width 
$\sog 0.12\arcsec$) for November 6 (used only for calibration purposes). A three-colour image (red H2 filter, green $K_{s}$, 
blue Br$\gamma$ filter) of IRAS\,20126+4104 is shown in Fig.~\ref{3col:fig}.

To perform photometry of the jet features, we first calibrated the stars in the small field of view (FoV). This was done using 
the $K_{s}$ images and the {\it daophot} package in IRAF. For each night with $K_{s}$ observations, we measured the brightness of 
the standard star FS149 in each position of the frame with different aperture radii. As the instrumental magnitude, we adopted
the value at which the measured brightness does not change significantly any more (aperture radius $\sim 1.5 \arcsec$). 
The zero point was obtained by averaging the instrumental magnitudes obtained at each position in the frame (and the standard 
deviation was taken as the error on the zero point), assuming - as $K_{s}$ standard magnitude - the value given in the 2Mass PSC 
($K_{s} = 10.052 \pm 0.017$).

In addition, we must take into account the fact that in the target images the Strehl ratio decreases with the distance to the 
AO guide star (the bright star west of the target in  Fig.~\ref{3col:fig}, i.e. star 8 in Fig.~\ref{nome:stelle}). To minimise 
this, we analysed the radial profile of the stellar PSF in the frames. This is composed of a narrow peak on top of a broader 
fainter feature, as expected (see Fig.~\ref{fit:psf}). The broader feature exhibits a width of $\sim 0\farcs3$ (November 2 
and 4) and $\sim 0\farcs45$ (November 6), which was taken as the aperture radius. We constructed the growth curve for the 
bright star lying to the north of the target
(star 4 in Fig.~\ref{nome:stelle}), which is isolated enough, and we derived a correction term for the aperture photometry. 
Using the zero point estimated from the standard star, we derived the $K_{s}$
magnitudes of the stars in the target image. As the airmass difference between the target field and standard was always $< 0.1$, 
no correction for airmass was needed.

\subsection{Ancillary data}

In order to derive the jet proper motion and its photometric variability, we retrieved all previous H2, Br$\gamma$ and $K$ 
images, both with low- and high-spatial resolution, from a number of archives. 
A $K$ image, presented in \cite{2005ApJ...631L..73S}, was obtained with UFTI at UKIRT from observations carried out from August 15--18,
2000 (seeing $\sim 0\farcs3$). Narrow-band H2, Br$\gamma$, and $K_{\rm cont}$ images, presented in \cite{2005A&A...434.1039C}, 
were obtained with NICS at the TNG in June 2001 (seeing $\sim 1\farcs2$). \citet{2008A&A...485..137C} studied 
a further series of H2, Br$\gamma$, and $K'$
images; namely, a dataset (first presented in \citealt{2007AAS...211.6213S})
acquired at the SUBARU telescope with the Coronagraphic Imager with Adaptive Optics (CIAO; \citealt{2003SPIE.4841..881M}), used as a      
high-resolution NIR imager, on July 10, 2003 (pixel scale of $\sim 0\farcs022$; seeing $\sim 0\farcs9$; PSF width $\sim 0\farcs15$),      
and a second dataset obtained with NICS at the TNG on August 6, 2006 (seeing $\sim 1.2 \arcsec$). Finally, high-spatial-resolution H2, 
Br$\gamma$, and $K_{s}$ images, analysed by \cite{2013A&A...549A.146C}, were obtained with PISCES and the AO system     
FLAO at the LBT on June 21, 2012 (PSF width $\sim 0\farcs09$).

Data reduction is described in the cited papers; in a few cases we redid it to improve the data quality and test photometric stability.
Notably, the CIAO, PISCES and SOUL maps roughly cover the same jet region around the source extending $\sim 30\arcsec$ (see Fig.~\ref{3col:fig}). 

Photometry on the large FoV images (UFTI, TNG) was performed with {\it daophot} using aperture radii $\sim 1$ full width at half maximum 
(FWHM)
of the PSF. Calibration was obtained matching the imaged stars and the 2Mass PSC and performing a linear fit. The fit formal error is 
typically $< 0.2$ mag ($\sim 0.08$ for UFTI), but one has to consider that this includes intrinsic variability in a number of stars and 
the zero point is then expected to be more accurate than $\sim 0.1$ mag.
As neither the CIAO images nor those obtained with PISCES are associated with observations of standard stars, in these cases we only 
performed relative photometry on the stars contained in the small FoV, with aperture radii of the order of 1 FWHM of the broad 
component of the PSF to minimise the effects of a varying Strehl ratio.

\subsection{Continuum subtraction}

To obtain H$_2$ and Br$\gamma$ pure line emission images, we first used the broad-band $K_{s}$ and $K'$, and the narrow-band
$K_{\rm cont}$ images, depending on the dataset. These were first rotated and shifted to the corresponding H2 and Br$\gamma$ 
images using the IRAF routines {\it geomap} and {\it geotran}. Then we performed aperture photometry on the common stars 
($> 600$ for the large FoV images, $4-15$ for the small FoV AO images) in the frames to scale the images by assuming that the 
intrinsic stellar fluxes do not vary with wavelength, a zeroth-order approximation. We note that this implies that images 
through different filters do not need to be taken during the same night, as telluric extinction differences are automatically 
taken into account. Short time stellar variability might affect the subtraction only for the small FoV images, where the 
number of stars is low. However, the r.m.s. of the average flux ratio is $\sol 10$ \% in these cases, which indicates no large 
stellar photometric variations between pairs of images. Finally, the scaled images were subtracted to yield continuum-corrected
images. We also note that when the continuum contribution is estimated from $K$, both images have to be scaled 
before subtraction to take into account that both include line emission as well as continuum emission.

We did not notice any significant Br$\gamma$ emission, so in the end we used the available Br$\gamma$ images to estimate the 
continuum contribution in the H2 filter. In fact, the broad-band $K$ filters encompass other H$_{2}$ emission lines and they
are therefore bound to overestimate the continuum contribution. In principle, this overestimate should be $\sim 10$ \%; in fact 
we found larger line fluxes when the continuum contribution was subtracted using a narrow-band filter rather than the $K$ band. 
However, we also found that the ratio of line flux (measurement with the continuum estimated using a narrow-band filter over 
the measurement with the continuum estimated using the $K$ band) increases when the ratio of the line to continuum ($K$) flux in a knot 
decreases. This indicates that the line flux may be significantly underestimated for faint line emission superimposed on a 
strong continuum ($K$) flux when the continuum contribution is evaluated using the $K$ band (see Appendix~\ref{poly:line:low}).

\subsection{Jet photometry}
\label{Jeph}

We performed photometry of the jet emission, both in the continuum and in the H$_2$ $2.12$ $\mu$m line, using the task 
{\it polyphot} in IRAF. We divided the jet into smaller components and defined a polygon for each
one, taking care of encompassing most of its emission (down to a $\sim 5\sigma$ level of the background as
determined far from the jet). 
Before doing the photometry, we registered  and projected all of the images onto the SOUL H2 frame grid as explained in 
Appendix~\ref{astro:map}, so we did not have to adapt the polygons to each image. The photometric zero points have been 
computed for each frame based on two bright stars which are common to all images. The variability of the two stars, the 
stability of the zero points, and the consistency of the calibration are discussed in Appendix~\ref{phot:stabi}, 
\ref{band:eff}, \ref{app:ana}.  

As for continuum emission, the selected polygons are described in Appendix~\ref{poly:des} and shown in 
Fig.~\ref{nome:poligoni:lowline}. Where possible, it was measured on the Br$\gamma$-filter images, which do not include 
the H$_2$ line emission and where Br$\gamma$ emission has not been detected. Only in two cases did we use $K$ images 
(UKIRT 2000 and TNG 2006) after subtracting the derived H$_2$ line emission (we used TNG 2001 to estimate the correction 
for UKIRT 2000). For the lower-resolution images (UKIRT and TNG), we also subtracted the stars in the frames by using 
{\it daophot} before performing the photometry. The sources of error involved in the various steps of the photometry are 
assessed in Appendix~\ref{poly:des}.

   \begin{figure}
   \centering
   \includegraphics[width=\hsize]{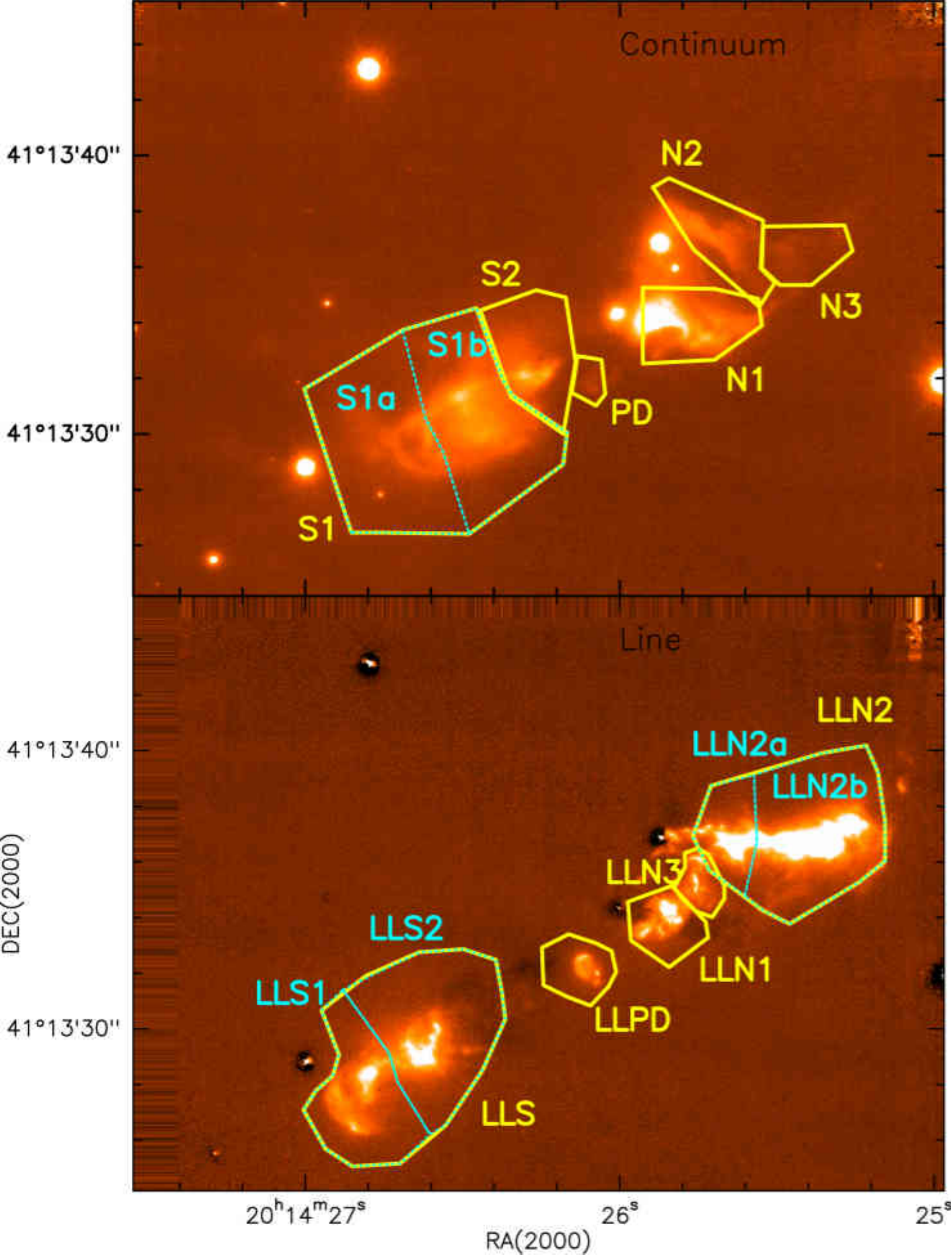}
      \caption{
Jet decomposition in polygons.
{\it Top panel}. Br$\gamma$-filter image of IRAS 20126+4104 obtained with LUCI1 and the AO system SOUL, overlaid with the 
polygons used for the photometry in the continuum. The adopted designation is labelled. We note that polygon S is composed of 
S1 and S2, and S1 is further subdivided into S1a and S1b.
{\it Bottom panel}. Pure H$_{2}$ line emission image of IRAS\,20126+4104 obtained with LUCI1 and the AO system SOUL, overlaid 
with the polygons used for the jet photometry. The adopted designation is labelled. We note that polygon LLS is composed of LLS1 
and LLS2, and LLN2 is subdivided into LLN2a and LLN2b.}
         \label{nome:poligoni:lowline}
   \end{figure}
%

Photometry of the jet emission in the H$_{2}$ $2.12$ $\mu$m line was performed in the same way as was done for continuum emission. 
However, in this case we used two different sets of polygons. First, we defined a set of larger polygons
(see Appendix~\ref{poly:line:low} and Fig.~\ref{nome:poligoni:lowline}.) to exploit both the higher and the lower spatial 
resolution images. Where possible we used the pure line emission images obtained by subtracting the continuum contribution 
as estimated from adjacent narrow-band filters.
In only one case was the correction estimated from the broad-band $K'$ filter. As shown in Appendix~\ref{poly:line:low}, 
while narrow-band filters (namely Br$\gamma$ and K$_{\rm cont}$) yield a consistent continuum correction, using broad-band 
$K$ images results in more discrepant continuum corrections compared to 
Br$\gamma$, leading to increasingly fainter magnitudes in the line emission with increasing continuum contamination.

\begin{figure*}
   \centering
   \includegraphics[width=16cm]{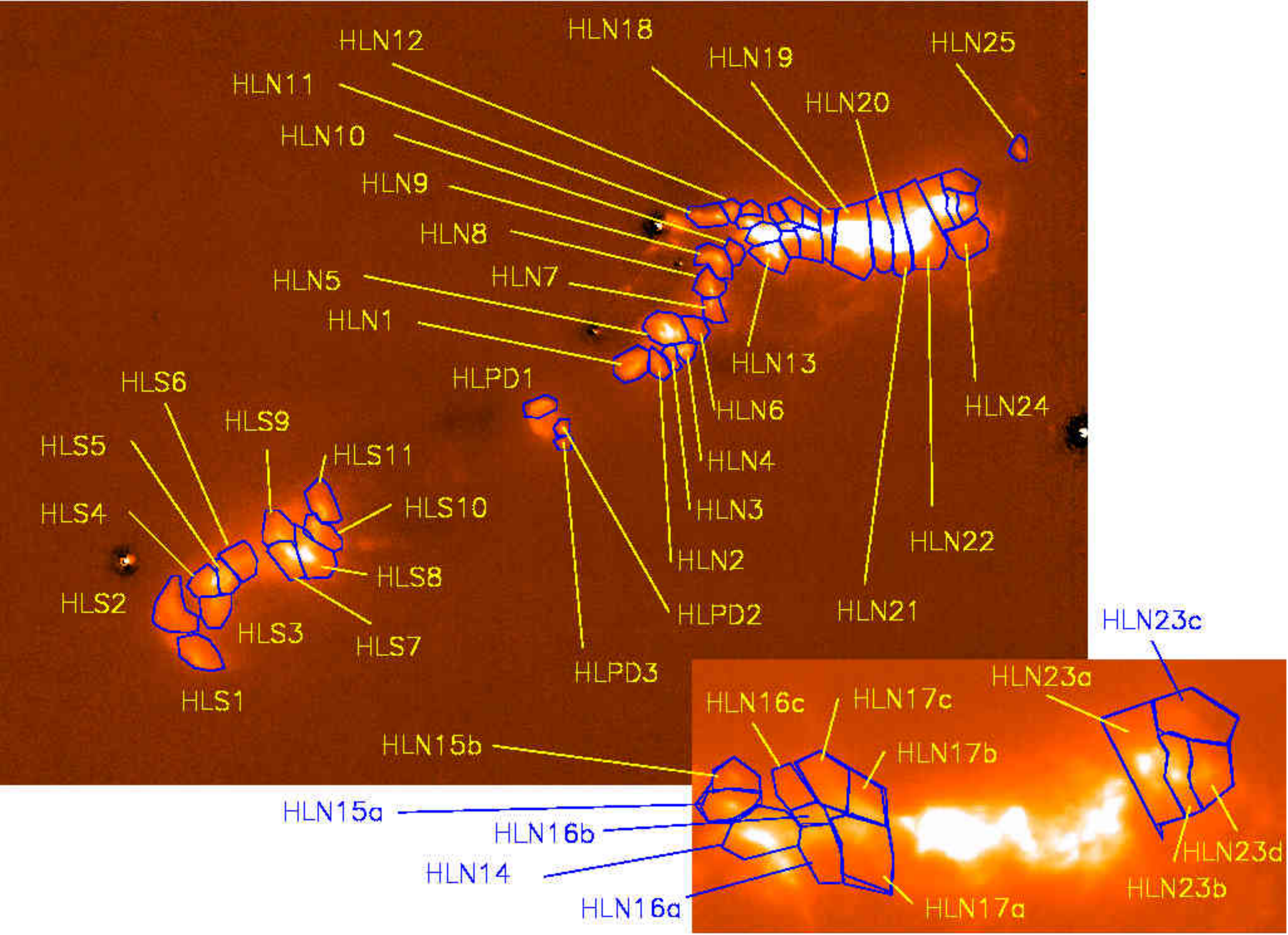}
      \caption{Pure H$_{2}$ line emission image of IRAS\,20126+4104 obtained with LUCI1 and the AO system SOUL, 
      overlaid with the polygons used
      for the high-spatial-resolution jet photometry. }
         \label{nome:poligoni:highline}
   \end{figure*}

Next, we analysed the high-spatial-resolution pure H$_{2}$ line emission images from the AO-assisted observations to identify 
single emission knots and defined a set of smaller polygons encompassing each knot (see Fig.~\ref{nome:poligoni:highline} for 
designation). In this case, we only used the high spatial resolution images (CIAO  2003, PISCES 2012, and SOUL 2020) with the 
continuum correction derived from the Br$\gamma$ images. To focus only on the smaller-scale structures, we adapted the 
multi-scale image decomposition described in \cite{2011A&A...527A.145B}, with seven levels of decomposition, to filter out 
large-scale emission structures.
This should have resulted in the efficient filtering out of $\sog 2\arcsec$ structures and a better background estimate.  
We note that the shape and location of a few polygons of this second set had to be slightly changed and adapted depending on 
the image because of the knots' proper motions.

%

\subsection{Proper motion analysis}
\label{PManalyis:sec}
We used the H2 continuum-subtracted high-angular resolution images from CIAO, PISCES and SOUL after flux calibration based on 
the photometric zero points (see Sect.~\ref{Jeph}), and registration to and projection onto the SOUL H2 frame grid 
(Appendix~\ref{astro:map}) to infer proper motions of knots and bow shocks along the jet. 
This dataset provides a time baseline of more than 17 years. We note that the proper motions are essentially derived in the 
reference frame of the protostar (see Appendix~\ref{astro:map}).
To compute the proper motions (PMs), we used the same strategy and our own developed software in python as described in 
\citet{2009A&A...502..579C}. 

Briefly, single shifts were computed between image pairs
(namely the SOUL image versus previous PISCES and CIAO images) keeping the latest epoch as a reference and using a 
cross-correlation method. After identifying each knot or structure in each image, a sub-image, enclosing the structure 
5\,$\sigma$ contour, was selected and cross-correlated with the corresponding sub-images obtained from the earlier epochs. 
In a given pair of sub-images, the earlier one was then shifted in steps of 0.1 pixel in $x$ (R.A.) and $y$ (Dec.), and for each 
shift ($\Delta x, \Delta y$) a product image was created.
The highest-likelihood shift for each epoch is that yielding the maximum correlation signal $f(x,y)$ integrated over the 
considered structure on the corresponding product image. 
To quantify the systematic errors for the shift measurements, the size and shape of each sub-image that encloses the 
considered structure was varied. The resulting range of shifts gives the systematic error, which depends on the structure 
signal-to-noise ratio (S/N), shape, and time variability. This uncertainty is typically much larger than the register error.
Finally, the PM value of each structure was derived from a weighted least square fit of the shifts derived for each epoch, 
fitting the motion in R.A. and Dec. simultaneously. This fit also provides the associated errors.
The results of the fits are detailed in Appendix~\ref{appendix:morphology&PMs}.

\section{Results}
\label{resu:sec}

\subsection{Morphology of the jet and outflow cavities}

The three-colour image (red H2 filter, green $K_{S}$,
blue Br$\gamma$ filter) obtained from the SOUL data, shown in the bottom panel of Fig.~\ref{3col:fig}, is quite similar to the 
PISCES one presented in Fig.~1 of \citet{2013A&A...549A.146C},
which has been adapted in the top panel of Fig.~\ref{3col:fig}. The continuum emission (Br$\gamma$ and $K_{s}$ filters) 
delineates the outflow cavities (in blue and cyan), which scatter the radiation from the central source, which
is undetected at 2\,$\mu$m. The H$_2$ shocked emission traces a precessing jet (in red; see also \citealt{2008A&A...485..137C}),
which is highly fragmented, especially towards the north-western blue-shifted side, displaying a large number of knots and 
bow shocks. It is worth pointing out that our high-resolution images show only the inner portion of the flow 
($\sim$30\,$\arcsec$ in size or $\sim$0.24\,pc), which extends further north and south (see, e.\,g. \citealt{2000ApJ...535..833S}, \citealt{2006A&A...448.1037L}).

To study the jet variability and kinematics, each sub-structure (knots and bow shocks) 
showing local emission peaks above 5\,$\sigma$ was identified and labelled in the SOUL H2 continuum-subtracted image. 
In our kinematical analysis we followed the general nomenclature used by \citet{2008A&A...485..137C}, who divided the flow 
regions into six different groups: A1, A2, X, B, C1, and C2 (see also Fig.~\ref{PMs:flow}d).

\subsection{Continuum variability}
\label{disk:cont:var}

   \begin{figure}
   \centering
   \includegraphics[width=\hsize]{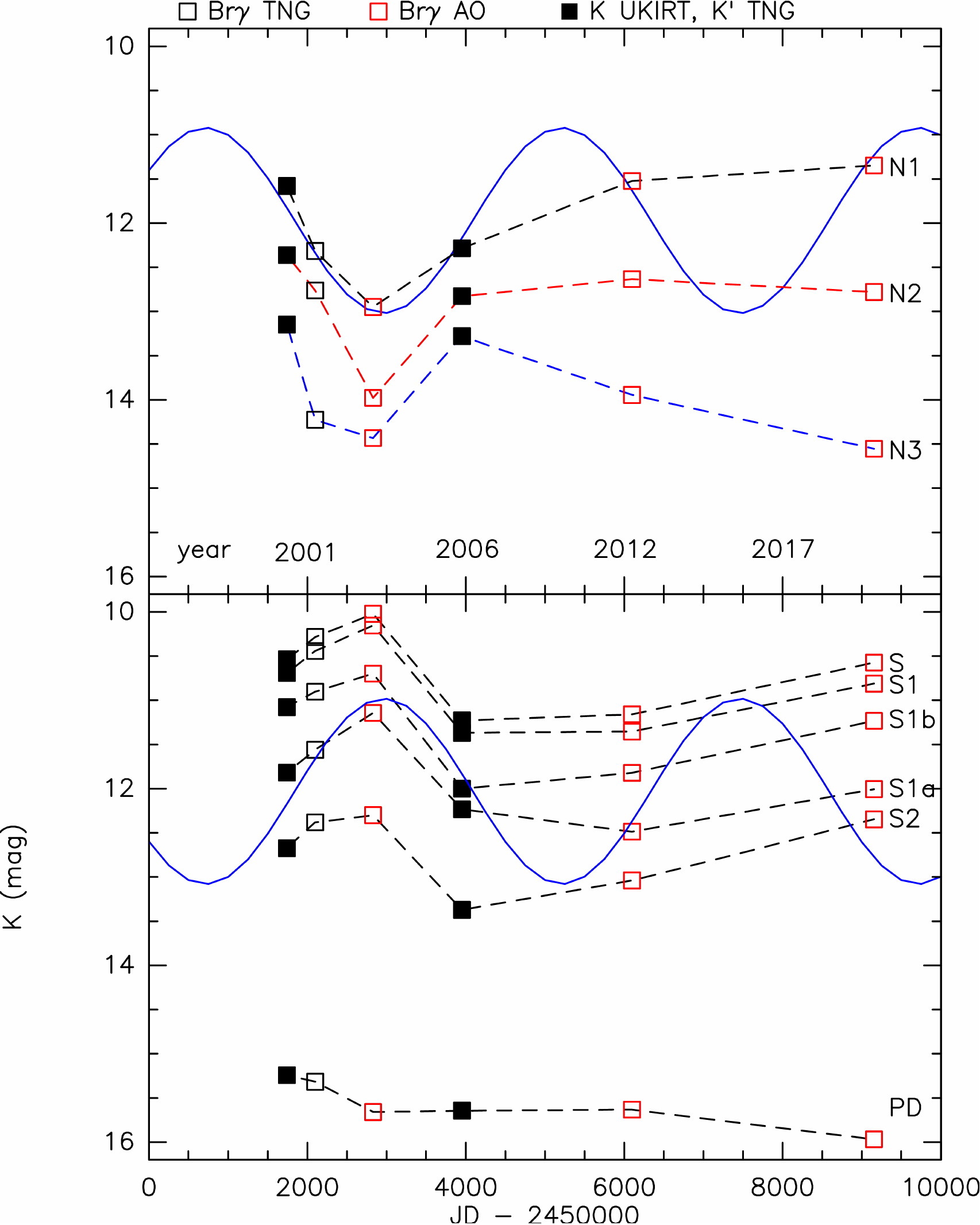}
      \caption{Photometric variability of the continuum features defined in Fig.~\ref{nome:poligoni:lowline}. We note that
N and S indicate the northern and southern lobe, respectively, and PD is the ring-like feature. The numbers increase from south-east
to north-west. We note that S is further subdivided into S1 and S2, and S1 is in turn subdivided into S1a and S1b. The symbol 
size is $\sim 0.2$ mag in the vertical direction, which is comparable with the photometric errors. A sinusoid of period $\sim 
12$ yr is overlaid on the N1 light curve. The same, but with a phase shift of $\pi$, is also drawn in the bottom box.
      }
         \label{fotopoli:continuo}
   \end{figure}
%
We found the most remarkable flux variations in the continuum emission.
To analyse such variability, we selected seven continuum emitting regions: three in the north-western and south-eastern cavity, respectively, and one enclosing the ring-like feature close to the source position
(see Appendix~\ref{poly:des} and Fig.~\ref{nome:poligoni:lowline}). The light curves of these regions are displayed in
Fig.~\ref{fotopoli:continuo}. The figure shows a clear decrease of the continuum emission in the north-western lobe between 
2000 and 2006 (see also Fig.~\ref{fotopoli:bright}, upper and bottom left panels,
and the short movie in Fig.~\ref{movie:cont} in Appendix~\ref{app:movies}),
with a subsequent increase. Conversely, the south-eastern lobe seems to have brightened up during the same time interval and 
then to have decreased in intensity. The ring-like feature (labelled PD in Fig.~\ref{fotopoli:continuo}) appears to exhibit a 
slow steady flux decrease. 

   \begin{figure*}
   \centering
   \includegraphics[width=\hsize]{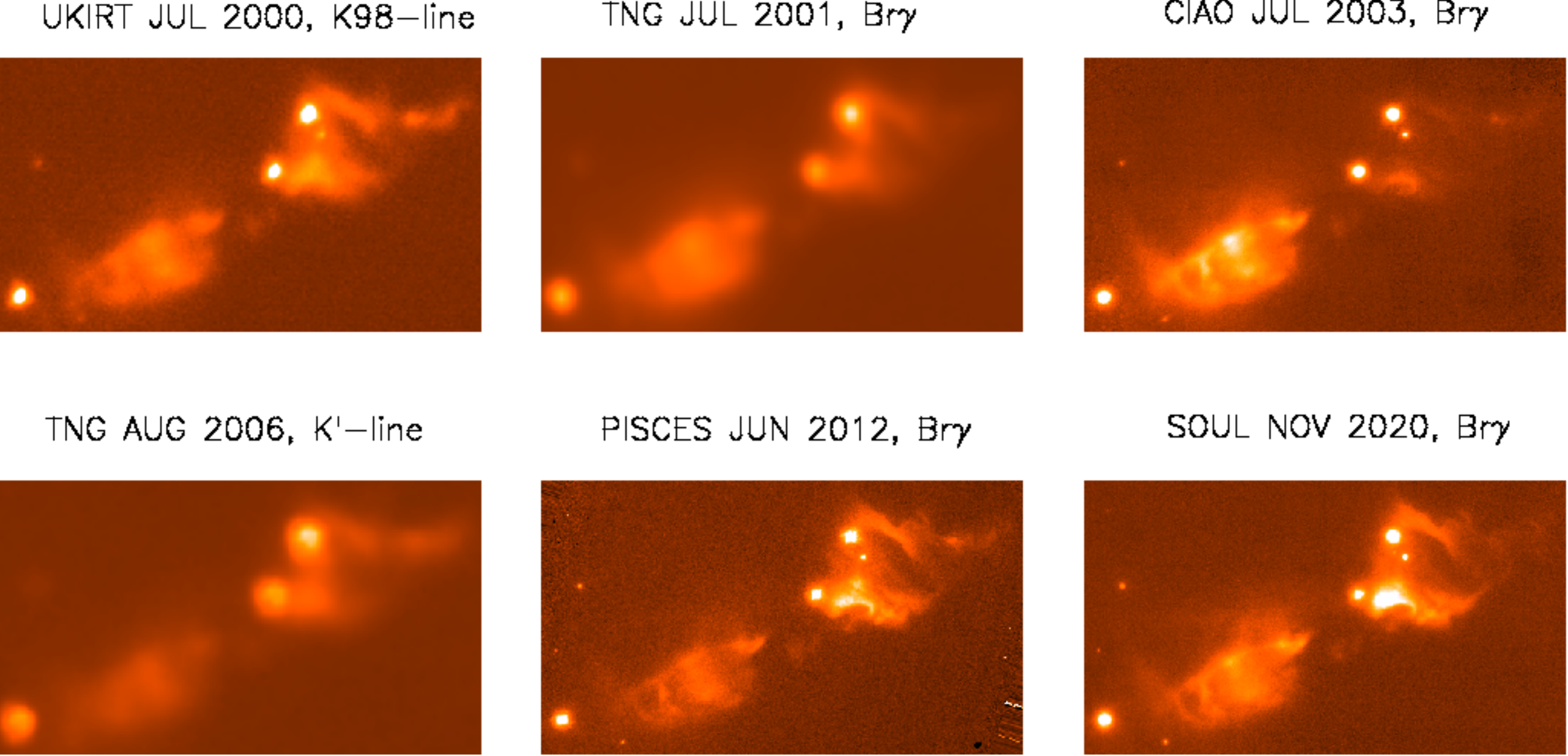}
      \caption{Comparison of the continuum emission throughout the various observed epochs.
      This has been approximated with the Br$\gamma$-filter images, except for the runs of July 2000 and August 2006 where broad-band $K$ images corrected for line emission have been used. The flux levels have been adjusted so that the
      mean of the counts of stars 2 and 4 is the same in all frames. The dimming of the northern lobe (especially region N1) and the simultaneous brightening of the southern lobe in July 2003 are evident.}
         \label{fotopoli:bright}
   \end{figure*}
%

\subsection{H$_2$ line variability}
\label{sec:line:var}

   \begin{figure}
   \centering
   \includegraphics[width=\hsize]{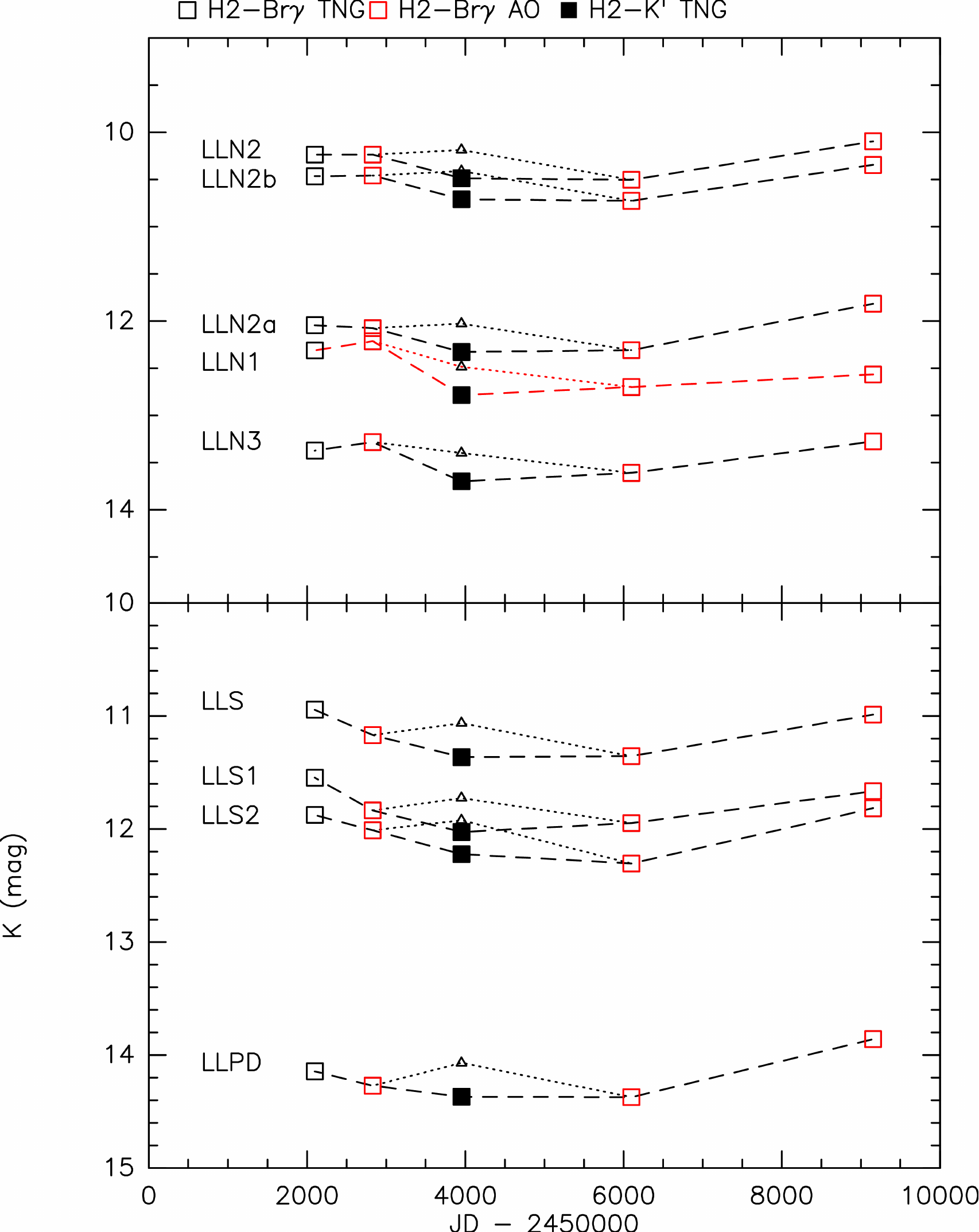}
      \caption{Photometric variability of jet areas in the H$_{2}$ $2.12$ $\mu$m line emission. We note that LLN and LLS indicate the 
northern and southern lobe, respectively, and PD is the ring-like feature (see Fig.~\ref{nome:poligoni:lowline}).In addition, 
LLS is further subdivided into LLS1 and LLS2, and LLN2 is in turn subdivided into LLN2a and LLN2b.
      The small open triangles (connected with dotted lines) indicate the fluxes with continuum subtraction from the $K'$ band 
when corrected following Appendix~\ref{poly:line:low}. The photometric errors are smaller than the symbol sizes. 
The magnitudes have been corrected for bandwidth differences.}
         \label{fotopoli:line:low}
   \end{figure}
%

The line photometry on the set of larger polygons (including the low spatial resolution images; see Appendix~\ref{poly:des} 
and Fig.~\ref{nome:poligoni:lowline}) is shown in Fig.~\ref{fotopoli:line:low}. We note that, as discussed in 
Appendix~\ref{poly:line:low}, the fluxes of the data points obtained by correcting the continuum contamination using the 
$K'$ image are expected to be underestimated, so the dip displayed in the plot on JD $2453953$ (black solid squares) is likely 
not real. 
A very rough correction of $0.3$ mag (open triangles in the figure), which should represent an upper limit, has been obtained 
following Appendix~\ref{poly:line:low}. Furthermore, the obtained magnitudes need to be corrected
for the different bandwidths of the H2 filters (see Appendix~\ref{poly:line:low}). We adopted the LUCI1 filter bandwidth as a standard
and corrected all other measurements using Eq.~(\ref{biro:biro}) and Eq.~(\ref{biro:biro2}). The line flux can be derived from 
the magnitude $m$ given in the figure through the following:
\begin{equation}
F = 1.06 \times 10^{-15} \times 10^{-0.4 m} \,\,\,\,{\rm W}\,{\rm cm}^{-2}.
\end{equation}
As we cannot rule out systematic errors in the zero points of $\sim 0.1 - 0.2$ mag in an unknown direction, the maximum variation 
intervals of $\sol 0.4$ mag per polygon indicate that
the total energy emitted in the $2.12$ $\mu$m line is probably almost constant in time within a few tenths of magnitude
in the sampled regions.

   \begin{figure*}
   \centering
   \includegraphics[width=16cm]{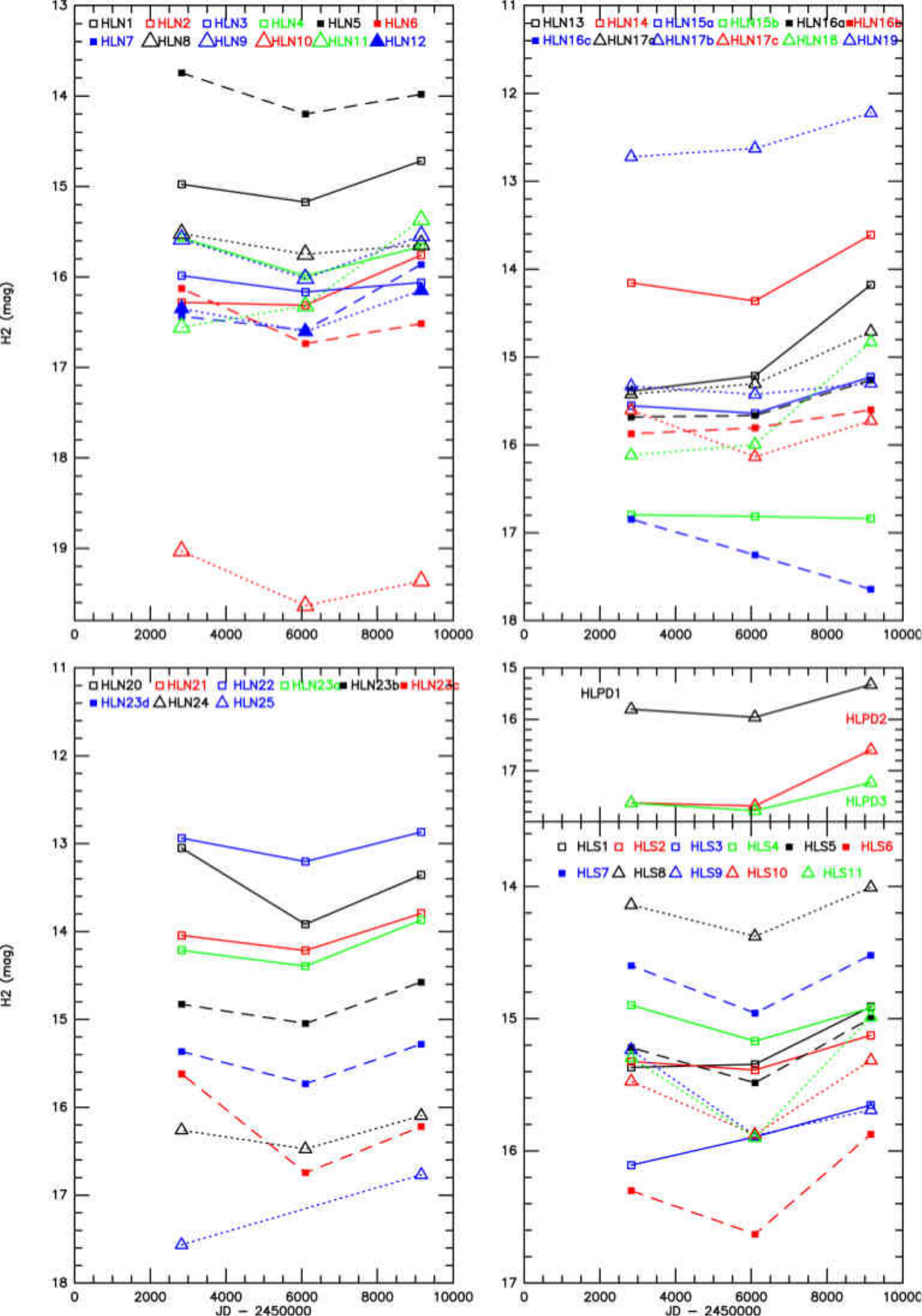}
      \caption{Photometric variability of jet knots in the H$_{2}$ $2.12$ $\mu$m line emission. We note that LHN and LHS indicate the 
       northern and southern lobe, respectively, and LHPD is the ring-like feature (see Fig.~\ref{nome:poligoni:highline}). The 
       magnitudes have been corrected to the LUCI1 H2 filter bandwidth.}
         \label{fotopoli:line:hres}
   \end{figure*}
%

The knot line variability on the smaller scale (excluding the lower spatial resolution images; see Fig.~\ref{nome:poligoni:highline} for 
polygon designations) is shown in Fig.~\ref{fotopoli:line:hres}.
Most knots exhibit a constant flux  within $\sim  0.2-0.4$ mag from 2003 to 2020 in a similar trend as found from the low spatial 
resolution photometry. Nevertheless, even assuming zero point systematic errors of $\sim 0.2$ mag in opposite directions, variations 
$> 0.4$ mag should be considered as real. In this respect, an emission increment $\geq$0.8\,mag  is evident for knot HLPD2 in the 
ring-like feature, the outermost north-western knot (HLN25), knot HLN11 adjacent to star 6, knot HLN13
connected with bow-shock C1h and C1k (see Sect.~\ref{b:s}), and knot HLN18 in the middle section of C1 (see Fig.~\ref{PMs:flow}). 
Another five knots exhibit an increase of $\sim 0.5$ mag, namely
HLN2, HLN14, and HLN19 and, in the south, HLS1 and HLS3 (see Figure~\ref{fotopoli:line:hres}). On the other hand, two knots exhibit a 
decrease of at last $\sim 0.5$ mag,
that is  HLN16c and HLN23c, and they are located in outer areas of C1.

An increase in line emission implies an increase in the column density of molecules in the upper level of the transition integrated 
over the knot area. In bow shocks, this column density increases with increasing shock speed in the range $\sim 5 - 15$ km s$^{-1}$ 
\citep{2018MNRAS.473.1472T}. A generalised trend of brightening therefore suggests that more and more gas is being entrained into 
the shocked regions. 

Due to the shape of the PSF in the AO-assisted images, the filtering is bound to cause some of the smaller-scale flux
to be missed
since  a residual small fraction of the PSF flux is spread on a larger size compared to the PSF diffraction-limited peak.
Thus, it is essential to ensure that this does not affect the time variations displayed by the single knots as well. By redoing 
the same photometry on the unfiltered images, we found the same trends as from the filtered images, with the main difference being 
that the knots are a few tenths of magnitude fainter in the filtered images (down to $\sim 1$ mag fainter in only a few cases). 
So we can confirm that the flux from most knots is stable within a few tenths of magnitude, with some exceptions.

\subsection{Proper motion, kinematics and dynamical age of the H$_2$ jet}
\label{prop:mo:sec}

\begin{figure*}
   \centering
    \includegraphics[width=\hsize]{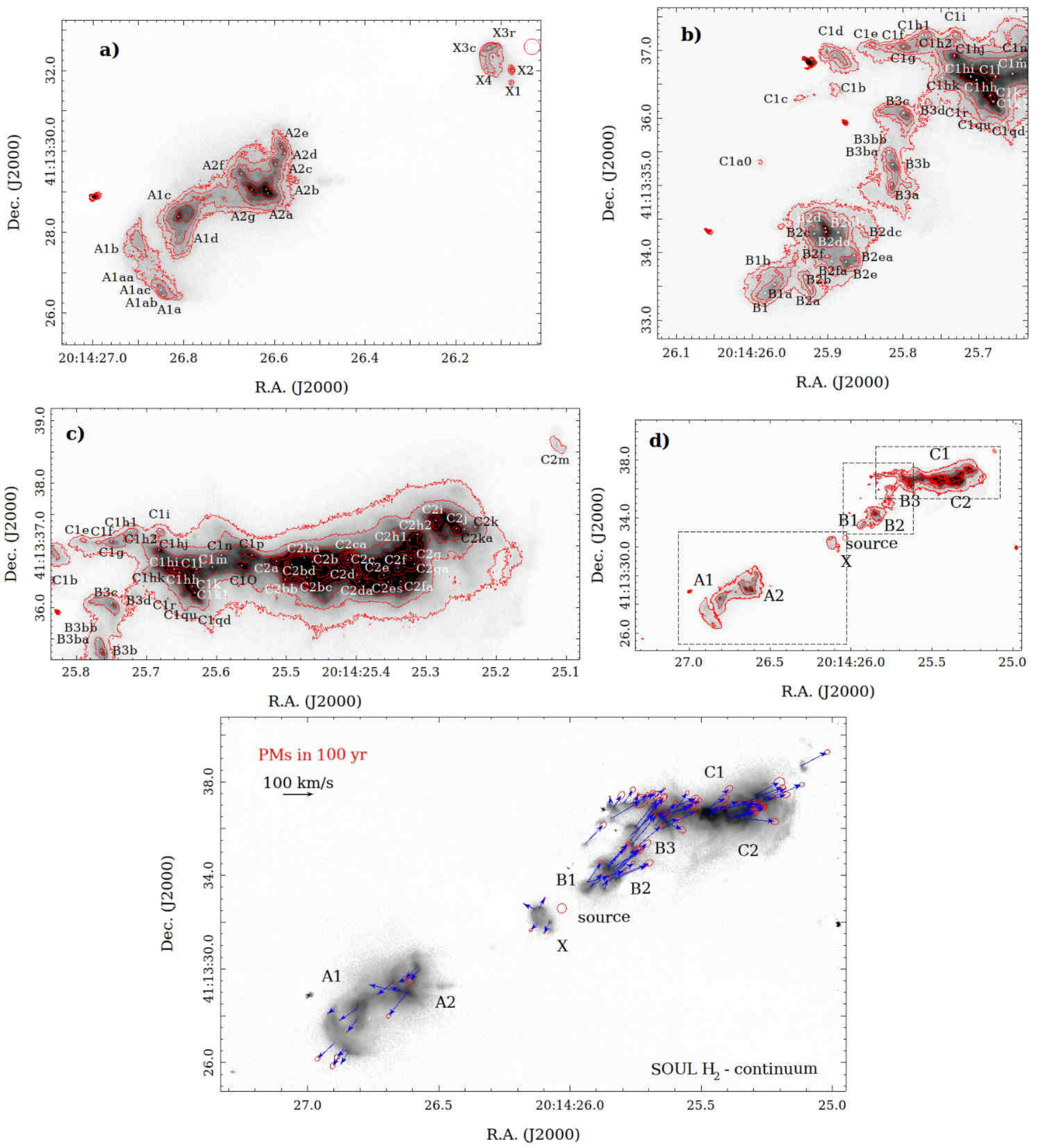}
      \caption{Knot identification in the SOUL H2 continuum-subtracted image of the IRAS\,20126-4104 flow and knot proper motions.
    {\it Panel a)}: Zoom-in on the A and X knot regions (red-shifted lobe). Contour levels at 5, 7, 10, 15, 20, and 25\,$\sigma$ are 
overlaid. We note that 1\,$\sigma$ corresponds to 8$\times$10$^{-23}$\,W\,cm$^{-2}$\,arcsec$^{-2}$. {\it Panel b)}: Zoom-in on the 
B and C1 knot regions (blue-shifted lobe). Contour levels at 5, 7, 10, 15, 30, and 40\,$\sigma$ are overlaid.  {\it Panel c)}: 
Zoom-in on the C1 and C2 knot regions (blue-shifted lobe). Contour levels at 5, 10, 20, 30, 40, 50, 60, 80, and 120\,$\sigma$ are 
overlaid. Knot peaks are indicated by white dots.
    {\it Panel d}: Overall view of the IRAS\,20126+4104 flow close to the source. The main structures as reported in 
\citet{2008A&A...485..137C} are labelled. {\it Panel e}: Proper motions (PMs) with their uncertainties (blue arrows and red 
ellipses) in 100\,yr of structures and sub-structures along the H$_2$ jet in IRAS20126+4104. The
actual observed shifts are approximately one fourth the length of the corresponding arrow. 
The red circle marks the position (along with its 
uncertainty) of the protostellar continuum emission at 1.4\,mm \citep{2014A&A...566A..73C}. The main structures as labelled in 
\citet{2008A&A...485..137C} are also indicated.
              }
         \label{PMs:flow}
   \end{figure*}

Figure~\ref{PMs:flow} (a-d panels) displays the knots identification and their measured proper motions (PMs; panel e) as derived from the analysed H2 continuum-subtracted images at high-spatial resolution (see Sect.~\ref{PManalyis:sec}). 
The shifts of some of the knots are evident in the short movie of Fig.~\ref{movie:h2} in Appendix~\ref{app:movies}. 
Thanks to the extremely high-spatial resolution of our images and their 17-year time baseline, we were able to measure knot PMs down 
to a few milliarcseconds (mas) per year. 
These values range from $\sim$2 to $\sim$20\,mas\,yr$^{-1}$. The results are listed in Table~\ref{PMs:tab} of 
Appendix~\ref{appendix:morphology&PMs}. In particular, the table provides knot IDs, coordinates (as derived from the SOUL map), 
proper motions (in mas yr$^{-1}$), tangential velocities (at a distance of 1.64\,kpc) and position angles (PAs, from north to east) 
of the corresponding vectors, and their uncertainties. The blue arrows in Fig.~\ref{PMs:flow} (panel e) show the average displacement 
of each knot in 100 years along with its uncertainty (red ellipses). 
Overall, both blue-shifted (B and C groups) and red-shifted (X and A groups) structures along the jet move away from the source position, roughly following the precession pattern described in \citet{2008A&A...485..137C} (see also their Figure\,11).
%

\begin{figure}
   \centering
   \includegraphics[width=9cm]{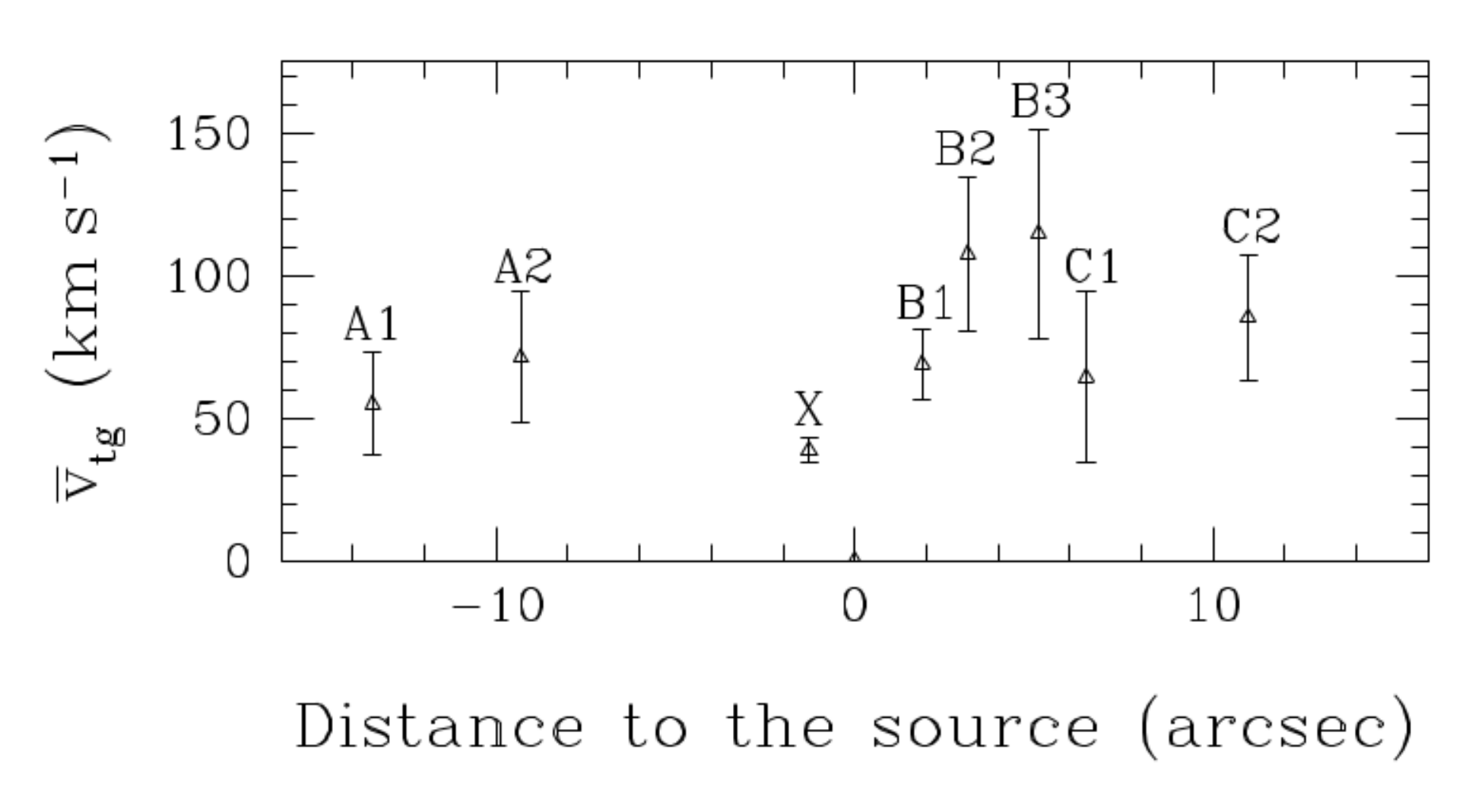}
    \caption{Average tangential velocity vs average projected distance to the source for each group of knots.}
         \label{v_vs_d:fig}
   \end{figure}

On average, velocities increase moving farther away from the source position (groups B1, B2 and B3 on the blue-shifted side with 
average $\varv_{\rm tg}\sim$ 69, 107, and 115\,km\,s$^{-1}$ and groups X and A2 on the red-shifted side with average 
$\varv_{\rm tg}\sim$ 40 and 70\,km\,s$^{-1}$) until they drop as the flow encounters a slower group moving ahead (i.\,e. group C1 and 
A1 with average $\varv_{\rm tg}\sim$ 64 and 55\,km\,s$^{-1}$, respectively) and collide against it (see Fig.~\ref{v_vs_d:fig}). 
Our analysis of the 3D velocities (see Sect.~\ref{3D:sec}) confirms that this is not just
a projection effect.

%

\begin{figure}
   \centering
   \includegraphics[width=9cm]{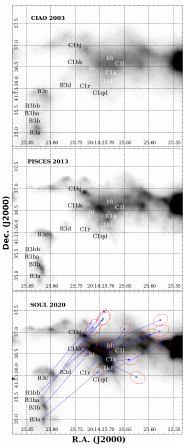}
    \caption{Evolution of the bright bow shock in the C1h and C1k regions. The top, middle and bottom panels show the same area of 
the sky at the three different epochs (2003, 2012, and 2020). The fast moving flow ($\varv_{\rm tg}\sim$100--120\,km\,s$^{-1}$) 
corresponding to B3 shocks against the C1h and C1k regions ($\varv_{\rm  tg}\sim$10--70\,km\,s$^{-1}$) producing a bright bow shock. 
The PMs of knots are shown as in Figure~\ref{PMs:flow}.             }
         \label{shocks:fig}
   \end{figure}

The clearest example is shown in Fig.~\ref{shocks:fig}. The fast flow in the B3 region is moving at $\varv_{\rm tg}  
\sim$100--120\,km\,s$^{-1}$ (see e.\,g. knots B3c and B3bb) and shocks against the C1hi, C1hj, and C1hk regions 
($\varv_{\rm tg}\sim$10$-$70\,km\,s$^{-1}$, corresponding to
polygons HLN13 and HLN14 in Fig.~\ref{nome:poligoni:highline}), producing a bow shock (knots C1hi, hk, hh, k, and k1), which 
increases in brightness from the 2003 to the 2020 epoch (middle and bottom panels of Fig.~\ref{shocks:fig},
see also the short movie of Fig.~\ref{movie:h2} in Appendix~\ref{app:movies}).
Indeed, one can see from Fig.~\ref{fotopoli:line:hres} that these two H$_2$ emitting regions are among those displaying the largest increase in luminosity (about 1.6\,mag).

%

\begin{figure}
   \centering
   \includegraphics[width=9cm]{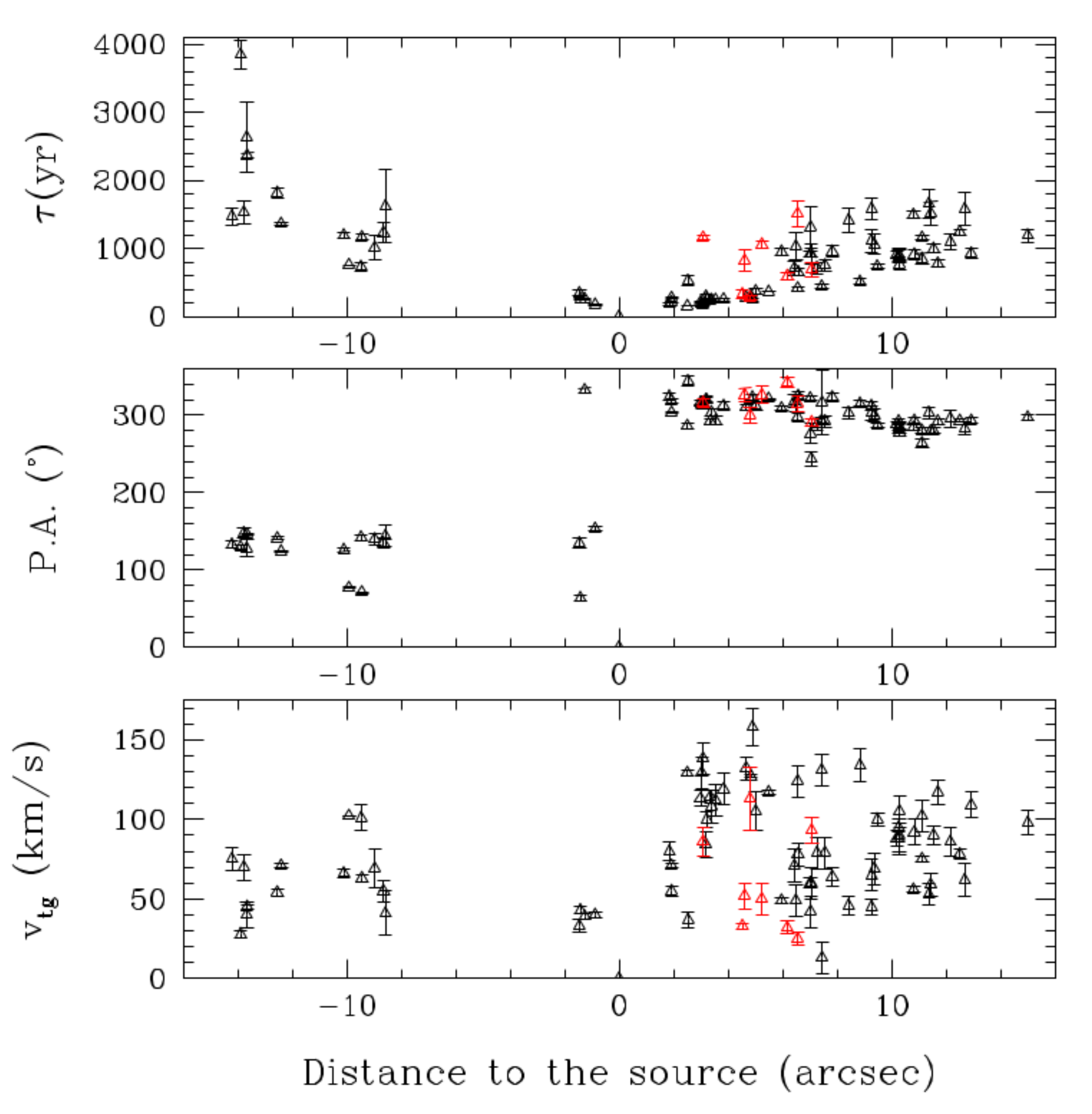}
    \caption{Tangential velocity (bottom panel), position angle (middle panel), and dynamical age (top panel) vs distance to the source for each knot. Red labels show knots in the C1 group that might belong to a different flow (see text).}
         \label{v_PA_Tdyn_vs_d:fig}
   \end{figure}

The average tangential velocity of the H$_2$ flow is 80$\pm$30\,km\,s$^{-1}$. Here the reported uncertainty is the standard deviation 
over the whole sample. In fact, by analysing the velocity vectors of all sub-structures in a group, it is clear that tangential 
velocities and position angles exhibit large spreads (namely tens of km\,s$^{-1}$ and degrees, i.\,e. much larger than the velocity 
uncertainty of each single knot) within the same group (see bottom and middle panels of Fig.~\ref{v_PA_Tdyn_vs_d:fig}). In addition, 
the scatter in velocity is larger in those regions where knots are more crowded and shocks are therefore expected to interact with 
each other. This reflects and possibly causes the observed fragmentation along the flow.

There are other interesting features arising from the proper motion analysis of the different regions along the jet. Group X, 
corresponding to the ring-like feature (PD) and as the H$_2$ region closest to the source position, looks like a structured expanding 
ring with the exception of knot X2 (polygon HLPD2 in Fig.~\ref{nome:poligoni:highline}), which first appeared in the 2012 image and 
increased in luminosity by 2020. The expansion is clearly visible from the vectors in Fig.~\ref{PMs:flow} and from the large scatter 
in their position angles (see middle panel of Fig.~\ref{v_PA_Tdyn_vs_d:fig}). 
On the other hand, knot X2, which first appeared in the 2012 image as well, is not co-moving with the expanding ring-like feature 
but it is rather moving straight away from the source position.

\begin{figure}
   \centering
   \includegraphics[width=9cm]{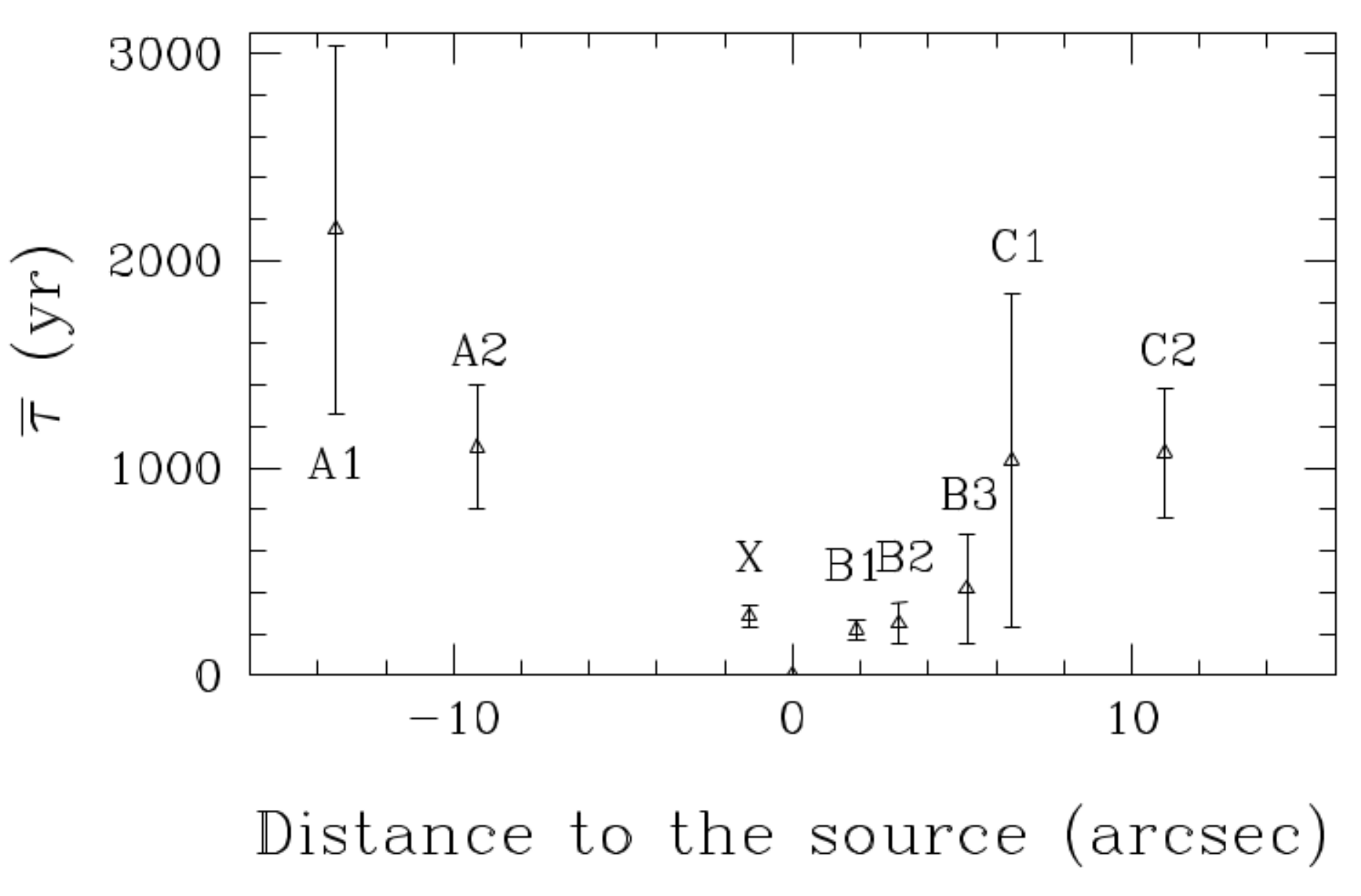}
    \caption{Average dynamical age vs average distance to the source for each group of knots.}
         \label{Tdyn_vs_d:fig}
   \end{figure}

By combining proper motions and projected distance ($d$) to the source, we also estimated the dynamical age ($\tau$=d/PM; in yr) of 
each knot. Values are reported in Column~8 of Table~\ref{PMs:tab} and plotted against their projected distance to IRAS\,20126+4104 
in the top panel of Fig.~\ref{v_PA_Tdyn_vs_d:fig}.  
The dynamical ages provide a raw estimate for the ejection time. Actually they provide a lower
limit, if matter actually accelerates soon after its ejection, as
Fig.~\ref{v_vs_d:fig} would suggest.
Figure~\ref{Tdyn_vs_d:fig} shows the average dynamical age of each group of knots versus their average projected distance to the 
protostar.
Notably, all the knots have been ejected recently. Their average $\tau$ ($\bar{\tau}$) ranges from 220$\pm$50\,yr for group B1 along 
the jet to 2200$\pm$900\,yr for the farthest group A1 towards the south-east. Group X (the ring-shaped region close to the source, 
labelled as PD in our photometry analysis) has a similar dynamical age as B1 ($\bar{\tau}_{X}=$280$\pm$50\,yr), whereas the 
$\bar{\tau}$ value of the farthest group towards the north-west (C2) is roughly consistent with that of A1, namely 
$\bar{\tau}_{C2}$=1070$\pm$300\,yr. This might indicate that the A1 group decelerated, as is also hinted at by the presence of a large 
bow shock at the front of the group. The large uncertainty on $\bar{\tau}$ of some groups reflects the large scatter in tangential 
velocities of the groups as seen in Fig.~\ref{v_PA_Tdyn_vs_d:fig}, in particular for groups A1, A2, B3, C1, and C2, namely where the 
dynamical interaction between knots may be more important.

%
%

\begin{figure}
   \centering
   \includegraphics[width=8cm]{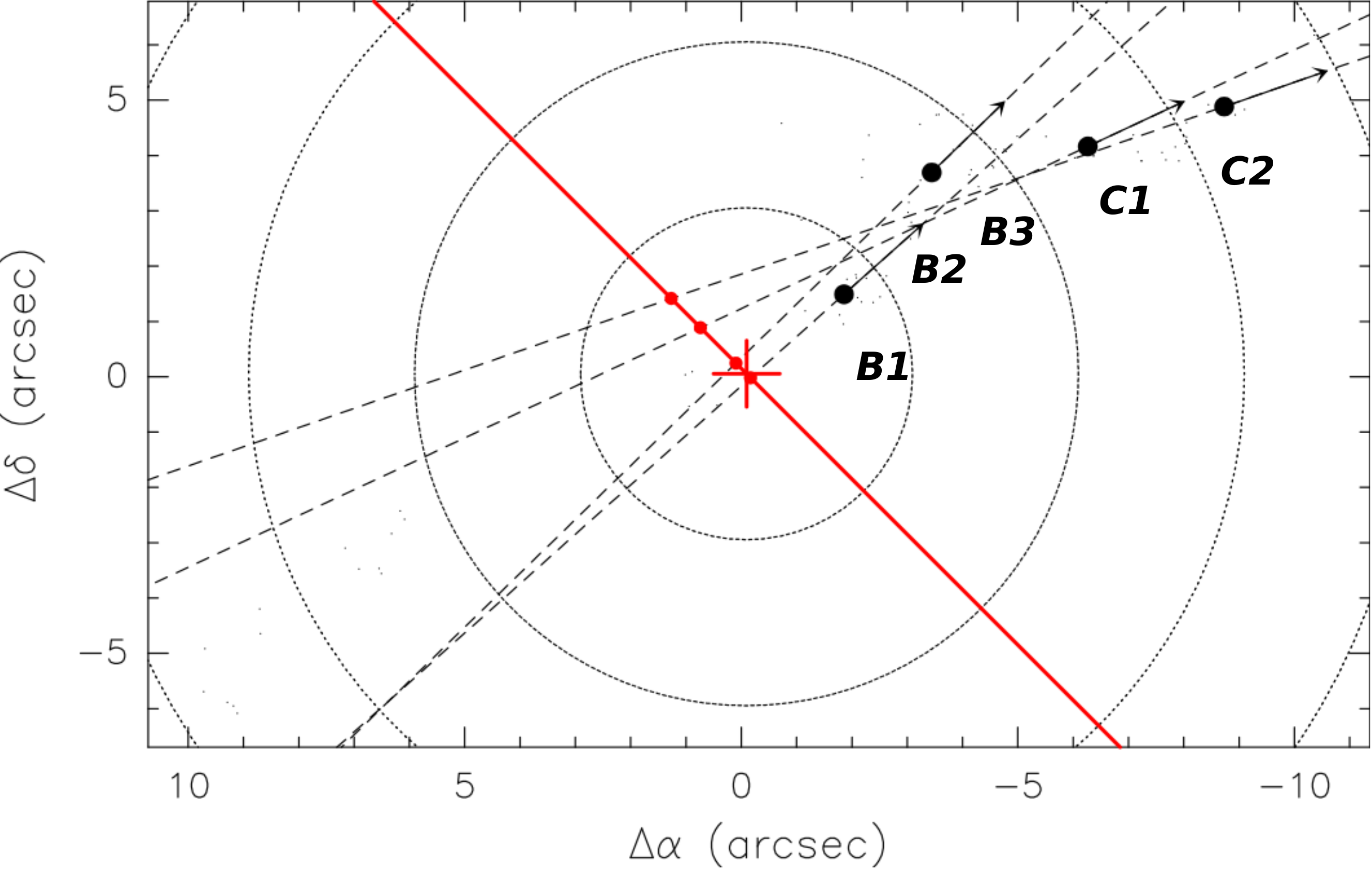}
   \includegraphics[width=8cm]{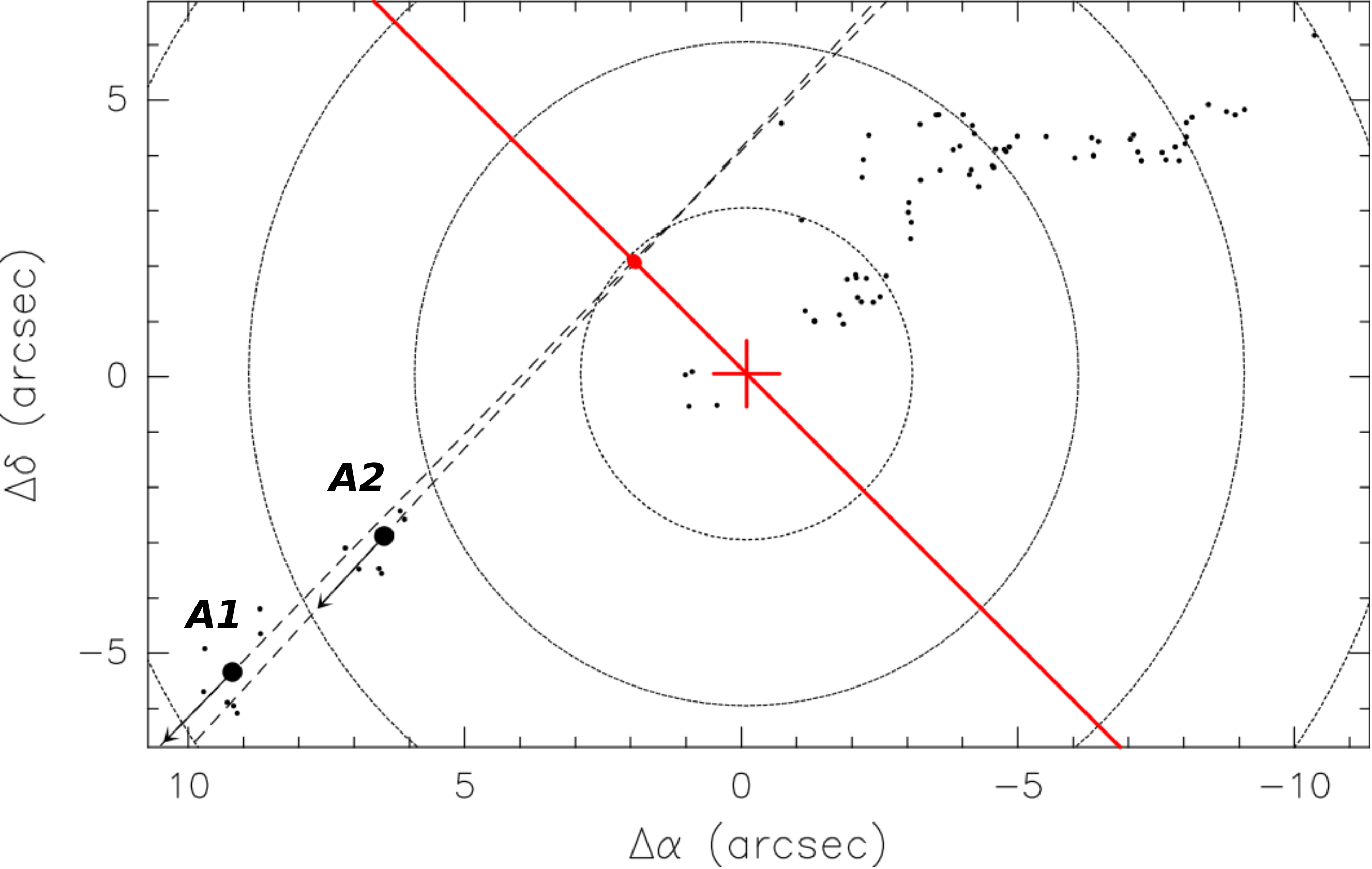}
    \caption{Mean directions of the proper motion vectors
    in the blue-shifted lobe (top panel) and in the
    red-shifted lobe (bottom panel). The proper motion
    track of the protostar is marked by the red line and its current location by the red cross. Clearly, the more 
    distant the knot group is from the protostar, the more
    distant (to the north-east) the intersection of its mean direction is with
    the track. The protostar is moving towards the south-west.}
         \label{prop_mo_sou:fig}
   \end{figure}

One puzzling feature displayed by the PM vectors is that the average direction of the farthest knots (A1, C1, and C2) does not intersect 
the current location of the protostar, but passes north-east of it. If a line coinciding with the proper motion direction of the 
protostar is drawn, the average directions of the knot proper motions cross it at the earliest locations (C1 and C2 $\sim 1\arcsec - 
2\arcsec$ north-east of the protostar and A1 and A2 $\sim 3\arcsec$ north-east of the protostar), in accordance with them being 
ejected at earlier epochs (see Fig.~\ref{prop_mo_sou:fig}). However, the knot PMs have been derived in a reference frame that 
appears to be similar to that of the protostar
(see Appendix~\ref{astro:map}), so such an effect should not be detected in our analysis. One possible explanation is that the 
protostar proper motion has decreased (in absolute value) with time.

%

\section{Discussion}
\label{discu:sec}

\subsection{H$_2$ 3D kinematics}
\label{3D:sec}

By combining the tangential velocities ($\varv_{\rm tg}$) derived in this work with the H$_2$ radial velocities ($\varv_{\rm r}$)  
obtained at high-spectral resolution ($\mathrm{R}$=18\,500) with UKIRT/CGS4 by \cite{2008A&A...485..137C}, we are able to infer both 
the total velocity ($\varv_{\rm tot}$) and the inclination of its vector with respect to the line of sight ($i$) for the knots 
encompassed by the CGS4 slit. It is worth noting, however, that the spectral images of CGS4  have a spatial resolution one order of 
magnitude worse (slit width $0\farcs5$, seeing $\sim 1\arcsec$) than the high-resolution images used here. As a result, the spectra 
extracted by \cite{2008A&A...485..137C} typically encompass more than one substructure and more than one velocity component is 
detected as well (see their Fig.~\,8 and Table~\,4). To associate radial and tangential velocities of each region, the spatial 
resolution of the CIAO image (closest in time to the spectral data) was first degraded to 1$\arcsec$ to match that of the 
spectral images of UKIRT/CGS4. Then, slit-like areas were cut out of the degraded image and the corresponding 1D profiles were
extracted and matched to those extracted from the CGS4 spectral images. We then assumed that the peak intensity of the extracted 
spectra originated from the brightest knot-substructure encompassed by the slit and combined its radial velocity with the tangential 
velocity of that knot or substructure (given in Table~\ref{PMs:tab}). 
The total velocity and inclination of each knot are then as follows:  
\begin{equation}
\mathrm{\varv_{\rm tot}} = \sqrt{\varv_{\rm tg}^2 + \varv_{\rm r}^2}   
\end{equation}
\begin{equation}
i=\arctan{\left(\frac{\varv_{\rm tg}}{\varv_{\rm r}}\right)}.
\end{equation}
The results are shown in Table~\ref{table:3Dkinematics} of Appendix~\ref{appendix:morphology&PMs}, where the identified substructures 
with total velocities and inclination $i$ with respect to the line of sight are listed.
The inclination $i$ of vectors in both lobes ranges from 76$\degr$ to 98$\degr$ 
with respect to the observer, yielding a weighted-mean inclination with respect to the plane of the sky of 
$i_{sky}$=8$\degr\pm$1$\degr$. This result is consistent with the value $i \sim 80\degr$ derived from the H$_2$O maser measurements 
obtained (towards the blue-shifted lobe) close to the source by \cite{2011A&A...526A..66M}. Our results therefore confirm that the 
outflow axis lies close to the plane of the sky at spatial scales ranging from a few hundred mas to $\sim 10\arcsec$. The 
$\sim$10$\degr$ spread in the $i$ values of the knots might arise from the jet precession or it might indicate that our original 
assumption in associating each peak $\varv_{\rm r}$ to the $\varv_{\rm tg}$ of the corresponding brightest substructure is not 
completely correct. In any case, such a spread does not really affect the fact that the reported tangential velocities of the knots 
can be assumed equal to the total velocities ($\varv_{\rm tg}/\sin{i} \simeq \varv_{\rm tg}\simeq \varv_{\rm tot}$) within a 
3\% uncertainty in the worst-case scenario (i.\,e. $i$=76$\degr$ towards group C2).

\subsection{Bow shocks}
\label{b:s}

We tested our data to check that the H$_2$ emission (morphology, fluxes, and projected velocities) of at least some knots is 
consistent with being originated in bow shocks.
We performed two tests based on a comparison with model predictions.
In the first test, the morphology, size, and brightness of the H$_2$ knots were compared with magneto-hydrodynamic
(MHD) models of bow shocks. As an example, we considered knots
A1, B3c, and C1h-C1k, as well as the 3D bow-shock models of \cite{2010A&A...513A...5G}. We converted the
knot photometry into brightness by using the 2MASS zero magnitude flux at $K_{s}$, a filter
width of $0.023$ $\mu$m (see Table~\ref{tab:bands}), and the solid angle of the relevant polygons
(correcting for extinction by adopting the $A_V$ values derived by \citealt{2008A&A...485..137C}).
The morphology of the selected knots is clearly reminiscent of that of bow shocks from a flow parallel to the plane of the sky. 
As for A1, both its width ($\sim 4000$ au) and its average brightness
($\sim 1-2 \times 10^{-6}$ W m$^{-2}$ sr$^{-1}$) roughly agree with a modelled bow shock impacting
a medium of density $\sim 10^{5}$ cm$^{-3}$ with a speed of $50-60$ km s$^{-1}$. Since the actual
speed that we have derived is $\sim 45$ km s$^{-1}$, A1 is likely to represent a terminal shock.
The width ($\sim 800$ au) and average brightness ($\sim 9 \times 10^{-7}$ W m$^{-2}$ sr$^{-1}$)
of B3c again points to a bow shock propagating in a medium of density $\sim 10^{5}$ cm$^{-3}$ with a speed of 
$50-60$\,km\,s$^{-1}$. Finally, C1h-C1k has a width of $\sim 1600$ au and an average
brightness of $\sim 0.4-1 \times 10^{-5}$\,W\,m$^{-2}$\,sr$^{-1}$, which is consistent with a bow shock
propagating in a medium of density $\sim 10^{5} - 10^{6}$ cm$^{-3}$ with a speed of $50-60$ km s$^{-1}$. We note that the models 
of \cite{2010A&A...513A...5G} are able to reproduce some of the
asymmetry exhibited by the knots on the basis of the magnetic field direction both on the
plane of the sky and along the line of sight. The magnetic field strength in the models of \cite{2010A&A...513A...5G} selected 
here range from between 500 and 5000 $\mu$G.
It is worth noting that
the most remarkable discrepancy between observations and models in the specific cases of
B3c and C1h-C1k is the measured knot velocity, which exceeds the hydrogen dissociation limit ($\sim 60$\,km\,s$^{-1}$). 
Indeed, most of the observed knot velocities exceed such a value (see Table~\ref{PMs:tab}). The most likely explanation, 
as already mentioned, is that
these features represent internal shocks, in other words that the flow here is impacting a parcel of gas already  in motion, 
resulting in a lower relative velocity. As a consequence, the measured tangential velocities (and thus total velocities) can be 
greater than the actual shock velocities.

\begin{figure*}
    \centering
    \includegraphics[width=16cm]{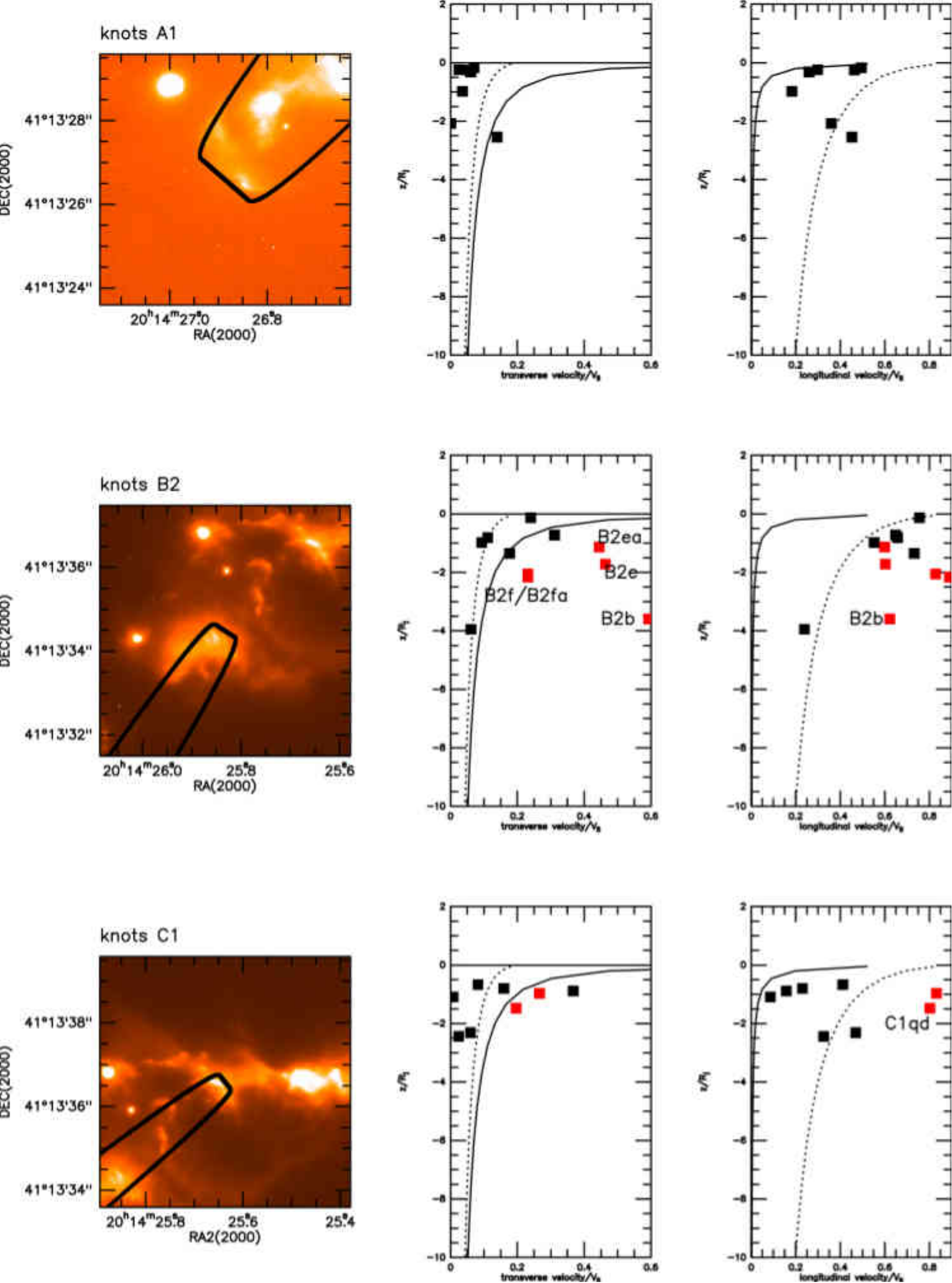}
    \caption{Ballistic bow-shock model of \citet{2001ApJ...557..443O}
    compared with knot chains A1 (top row), B2 (middle row), and C1
    (bottom row). In the column on the left $6\arcsec \times 6\arcsec$
    zoom-ins of the SOUL H2 image are shown with the bow-shock model shape overlaid. 
    In the middle column, the transverse velocities of the involved
    knots are compared with the mean shell (dotted line) and outer surface layer (solid
    line) values predicted by the model. The same comparison is made in the column on the right for the longitudinal
    velocities. 
    The knots whose speed departs from the model predictions the most are marked in red and labelled. For the sake of simplicity, 
    the outflow is assumed to lie on the plane of the sky.}
    \label{fig:bow}
\end{figure*}

In the second test, we compared knots A1, B2, and C1 with the ballistic bow-shock model of \citet{2001ApJ...557..443O}. 
In order to derive the proper coordinate system used by those authors (i.e., distance $z$ from the head of the shock along the 
jet axis and distance $R$ from the jet axis), we have developed a software routine that fits their Eq.~(22)
to the outermost shape of H$_2$ emission structures resembling bow shocks. For the sake of simplicity,
the fit assumes the jet to lie in the plane of the sky (a good assumption, as discussed in Sect.~\ref{3D:sec}), a bow-shock
speed $\varv_{\rm s} = 150$ km~s$^{-1}$, a sound speed $C_{\rm s} = 8$ km s$^{-1}$ (which is appropriate for a gas temperature of 
10000 K), and $\beta = 4.1$ \citep{2012ApJ...745..191S}. The fit yields the coordinates of the head of the shock (the centre of the 
working surface), the projected orientation of the shock axis, and $R_{j}$, the radius of the inner driving jet (and of the working 
surface). Using Table~\ref{PMs:tab}, one can now compute the coordinates $(R,z)$
of the associated knots in the bow-shock reference frame and the components of their projected
speed parallel (longitudinal) and perpendicular (transverse) to
the jet axis. In turn, these can be compared with the ones
predicted by the model (Eqs.~18, 19, 20, 21 of \citealt{2001ApJ...557..443O}). In particular, given a knot associated with a bow 
shock, its projected velocity component perpendicular to the jet axis should fall between zero and the predicted radial speed of 
the outer surface layer at the same distance $z$ from the shock head,
whereas the knot-projected velocity component parallel to the jet axis should fall between the predicted mean shell and outer 
surface layer longitudinal speeds at the same distance $z$ from the shock head.
The comparison and the spatial distribution of the shock envelope from the fit is displayed in Fig.~\ref{fig:bow} for knots 
A1, B2, and C1.  The results of the fit are also listed in Table~\ref{table:ostriker}. Figure~\ref{fig:bow} shows that
the morphology of knots A1, B2, and C1 is well fit by the bow-shock outer-shell model. The proper motions of the knots making up 
group A1 appear to be  consistent with the model prediction. As for C1, the velocity plots indicate that the
location of the working surface at the head of the shock has probably not been estimated correctly and that it should be shifted a 
little backwards along the jet. After this correction, the proper motions of the knots making up C1 are consistent with the model 
predictions as well (possibly
except for knot C1qd). Some of the proper motions of the knots making up B2 are consistent with the model predictions, while others 
are not. In this case, one has to
take into account that the region around B2 displays a complex pattern of shocks, thus the matter along the jet axis is probably 
soon entrained in other internal shocks due to subsequently ejected matter. In this respect, knot B2b, the most diverging one, is 
clearly a bright isolated patch of emission (see Fig.~\ref{PMs:flow}) located at the border of the fitted shock outer shell, marking 
a different shock episode with high probability.      

%
\begin{table}
\caption{Shape of three possible bow shocks (from fitting Eq.~(22) of \citealt{2001ApJ...557..443O}).}             
\label{table:ostriker}     
\centering                     
\begin{tabular}{l l l l l l l}       
\hline\hline                 
knot & $\varv_{\rm s}$ & $C_{\rm s}$ & $\beta$ & $\theta$ & $R_{j}$ & $\alpha$  \\    
ID  & (km\,s$^{-1}$) & (km\,s$^{-1}$) &  & ($\degr$) & (au) &  ($\degr$) \\
\hline                       
A1 & 150 & 8 & $4.1$ & $141.15$ & 1237 & 4 \\
B2 & 150 & 8 & $4.1$ & $-30.4$ & 469  &  $5.5$ \\
C1 & 150 & 8 & $4.1$ & $-49.25$ & 349  &  2\\
\hline                                   
\end{tabular}
\tablefoot{
Column 2: assumed bow-shock speed; column 3: assumed sound speed; column 4:
assumed $\beta$; column 5: jet axis position angle; column 6: inner jet radius; column 7; angle
to the driving source subtended by $R_{j}$.
}
\end{table}
%

Interestingly, the method also provides an estimate of the aperture angle of the jet from the radius of the inner jet driving the 
bow shock. The values obtained range from between $2-5.5\degr$ (half aperture, see Table~\ref{table:ostriker}), which is slightly 
less than
the value obtained from the water masers ($\sim 9\degr$; \citealt{2011A&A...526A..66M}). In addition, the bow-shock axes derived 
from A1 and C1 intersect the proper motion track of the protostar north-east of it, as found for the average proper motion 
directions.

\subsection{Jet precession}

A comparison between the continuum emission and the H$_2$ line emission in Fig.~\ref{3col:fig} clearly indicates that continuum 
emission (outlined by the blue-coloured emission) is detected on both the blue-shifted and the red-shifted sides of the outflow, 
meaning that scattered dust emission from the cavity dug by the protostellar jet is visible on both sides. This is consistent with 
a jet lying almost on the plane of the sky (as confirmed by our measured 3D velocities, see Table~\ref{table:3Dkinematics}), with 
a large enough opening angle (half opening $\sim 9\degr$ according to the water maser motion modelling by 
\citealt{2011A&A...526A..66M}). The relatively bright double-sided cavity, compared to the nearby fainter source S
suggests that the protostar may be located near to the front surface of its parental molecular clump, which could explain the low 
extinction towards both cavities.

\begin{figure}
    \centering
    \includegraphics[width=9cm]{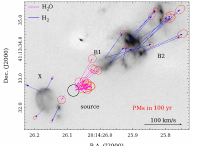}
    \caption{Zoom-in on Figure~\ref{PMs:flow} around the source showing H$_2$O maser proper motions (magenta arrows) as measured by \cite{2011A&A...526A..66M} and H$_2$ proper motions (blue arrows) from this paper. Red ellipses mark the uncertainties on PMs. The black circle shows the position of the protostellar continuum at 1.4\,mm from \cite{2014A&A...566A..73C}.}
    \label{fig:h2o}
\end{figure}

The proper motion analysis confirms that all the H$_2$ features in the north-western lobe
are moving away from areas close to the protostar location. In particular, the proper motions of group B knots agree well both in 
speed and direction with those of the cluster of water masers, which are closer to the protostar location \citep{2011A&A...526A..66M}, 
confirming that  the water masers and group B knots are tracing different episodes of mass ejection from the same source
(see Fig.~\ref{fig:h2o}). 

%
%

The knots of group A are roughly symmetrical to those of group C with respect to the protostar location and the velocities are 
red-shifted as expected.  In contrast, their morphology is different from that of knots in group C, possibly due to one or more of 
the following reasons: i) the mass ejection is asymmetric; ii) the south-eastern lobe is impinging on denser ambient gas; and iii) the 
south-eastern lobe is 
more extincted than the north-western one. The lower velocity of knots A (see Table~\ref{table:3Dkinematics}) would suggest 
deceleration, which is consistent with case ii. 

The knots of group B have no symmetrical counterparts in the south-eastern lobe, except for two tiny features south of the ring-like 
emission X (see Fig.~\ref{PMs:flow}).
However, a symmetrical counterpart has been detected at millimetre wavelengths.
In fact, B1 coincides with a patch of SO emission between $-26$ and $-12$ km s$^{-1}$ with respect to the systemic velocity 
\citep{2017MNRAS.467.2723P}, hence it is consistent with the radial velocities derived from the H$_2$ emission by 
\citet{2008A&A...485..137C}. A nearly symmetrical patch of SO emission between 11 and 32 km s$^{-1}$ south-east of the protostar was 
detected by \citet{2017MNRAS.467.2723P}. This of course discards the possibility of an asymmetric mass ejection for group B.

The HCO$^{+}$(1--0) outflow emission mapped with the IRAM Plateau de Bure (PdB) interferometer by \cite{1997A&A...325..725C} displays 
the same south-eastern-north-western morphology as seen in the $K$ band. In particular, the south-eastern lobe, when mapped at 
red-shifted velocities, exhibits the 'jaw-like' morphology found by \cite{2005MNRAS.357..579S} in their hydrodynamic simulations of 
slow-precessing protostellar jets. The knots of group A are located at the northern edge of the south-eastern molecular lobe, which 
corresponds to the most recent ejection episode when compared to the morphology of the hydrodynamical models. Interestingly, the knots 
of group A coincide with a blue-shifted patch of molecular emission (see Fig.~7 of \citealt{1997A&A...325..725C}), which confirms 
that the outflow is currently almost perpendicular to the line of sight. The reversal of blue-shifted and red-shifted lobes, 
when the HCO$^{+}$(1--0) radial velocities are closer to the systemic radial velocity, clearly indicates both the jet high 
inclination and its non-negligible opening angle. This is widely discussed in  \cite{2005A&A...434.1039C} and our proper motion 
analysis definitely confirms this scenario.    
The north-western outflow lobe does not exhibit a 'jaw-like' morphology as the south-eastern one does. This may be due to the fact 
that the outflow is still eroding the innermost parts of the parental cloud in the south-east, whereas it has partly emerged from the 
cloud 
in the north-west. This is confirmed by the 3D velocities of groups A and C, which are $< 100$ km s$^{-1}$ towards A, but still above 
$\sim 100$ km s$^{-1}$ towards C
(see Table~\ref{table:3Dkinematics}).

The global morphology of the H$_2$ emission resembles the models by \cite{2005MNRAS.357..579S} if one assumes that  in the case of 
IRAS\,20126+4104, the mass ejection is episodic rather than periodic. The interactions between different bow shocks and the sudden 
changes of direction predicted by the model are clearly visible. The models of \cite{2005MNRAS.357..579S} assume a precession angle 
of $10\degr$, a precession period of 400 yr, and a jet velocity of 100 km s$^{-1}$, which are close to the values of $\sim 7.6\degr$ 
and $1000$ yr derived by \cite{2008A&A...485..137C}, and the velocities obtained in this work. \cite{2005MNRAS.357..579S} also found 
that the total H$_2$ luminosity monotonically increases with time,
but apparently only less than $\sim 0.01$ mag yr$^{-1}$,
which is consistent with the results of Sect.~\ref{sec:line:var}. Thus, the similarity between models and observations further 
supports the idea of a pulsed precessing jet fed by episodic ejection in IRAS\,20126+4104.

\subsection{A possible wide-angle wind shocking the outflow cavity}

A clue as to the nature of the ring-like feature X can be found in the models of \cite{2005MNRAS.357..579S}. These authors find a 
'stagnation point' at the apex of the cone of matter which is not affected by the jet (inside the precession cone). Gas accumulates 
at the stagnation point originating a disc-like feature which then releases matter with large lateral speed. This might resemble the 
expansion of the ring-like feature shown in Fig.~\ref{PMs:flow}. 
Alternatively, an explanation for this expanding ring feature, here traced by H$_2$ shocked emission, is that we are actually seeing 
a wide-angle wind, which shocks the outflow-cavity walls and blows away the outer envelope, further opening the cavity. These 
wide-angle winds are predicted by MHD disc-wind models (see e.\,g. \citealt{1995ApJ...454..382K}, \citealt{2006ApJ...649..845S}) and 
have recently been detected at millimetre wavelengths with the ALMA interferometer in a few 
low-mass protostars (e.\,g. HH47 and DG Tau\,B; see 
\citealt{2019ApJ...883....1Z}, \citealt{2020A&A...634L..12D}). Indeed, such models predict that MHD disc winds have both a 
fast-moving collimated component (the jet itself) and a slow-moving less dense wide-angle component that blows into the ambient 
matter, forming an expanding outflow shell. 
In our case, both the ring-shaped geometry and the lower velocities observed in the ring-shaped structure of group X are consistent 
with this picture. Additionally, the fact that group X has lower H$_2$ column density and mass flux  (up to an order of magnitude) 
with respect to the main jet (see Table~5 of \citealt{2008A&A...485..137C}) further supports this idea. \cite{2017MNRAS.467.2723P} 
detected complex molecules at sub-arcsecond resolution using the PdB interferometer. Notably, one of the detections spatially matches 
the H$_2$ emitting region of group X.
By modelling the gas emission and from the X-shaped morphology of the SO low-velocity component and HNCO, \cite{2017MNRAS.467.2723P} 
conclude that such molecules are tracing shocks along the outflow cavity walls of IRAS\,20126+4104, which is consistent with our 
analysis. Furthermore, \cite{2019ApJ...883....1Z} suggest that these shells are the results of an intermittent wide-angle wind that 
is produced by episodic accretion and ejection. This would also be consistent with knot fragmentation seen in the IRAS\,20126+4104 jet.
Finally, by fitting an ellipse to the ring-shaped feature, we find that its major axis has a position angle of 29$\degr\pm$5$\degr$, 
namely orthogonal to the jet position angle (-60.9$\degr$) derived by \cite{2008A&A...485..137C}.
If one assumes circular symmetry for the ring-shaped feature,
it is also reasonable to expect its symmetry axis to be roughly parallel to the jet axis. In this case, the inclination of the 
ring-shaped feature to the line of sight  can be derived from the ratio of the ellipse axes, which is also equal to the symmetry axis inclination 
to the plane of the sky.
The feature inclination with respect to the observer ranges from between 68$\degr$ and 54$\degr$, meaning that under the hypothesis 
of circular symmetry the inclination of the feature symmetry axis with respect to the plane of the sky would be  quite different from 
$\sim$8$\degr$, as estimated for the jet in Sect.~\ref{3D:sec}. A possible explanation to this conundrum might be that we are not 
seeing the whole shocked region along the red-shifted outflow cavity, namely that part or most of the lobe surface towards the 
observer is obscured by high visual extinction.

\subsection{Disc oscillations and the anti-correlated continuum variability}

The continuum variability observed in the outflow cavity could in principle
have been produced by a short accretion increase that occurred between 2000--2003.
However, such an event would have either illuminated both outflow cavities or 
left one of them unaffected in the unlikely event that the burst had just occurred on one side of the disc.
This scenario is therefore difficult to reconcile with the clear
anti-correlation of the K-band continuum light curves between the north-western 
(dimming lobe, see Fig. 3) and the south-eastern lobe (brightening lobe, see Fig. 3).
Thus we have explored the alternative view of periodic flux oscillations
related to the disc and jet precession.
As the jet precession originates from a simultaneous disc precession, the inner circumstellar disc might periodically eclipse one 
hemisphere of the central protostar at different times, as seen from the north-west and south-east, producing such an out-of-phase 
variability. 
As shown in Fig.~\ref{fotopoli:continuo}, an anti-correlated  sinusoidal variation with a
period of $\sim 12-18$ yr roughly overlays the $K$-band continuum light curves.
Unfortunately, no clear-cut conclusions can be drawn concerning periodicity
due to the sparse photometric data coverage after 2003. Nevertheless,
in the following we assume that a periodic scenario holds to develop
a picture that can be easily checked by future observations.

By modelling the dynamics of a protostellar disc in a binary system, \cite{2000MNRAS.317..773B} found that a circumstellar disc 
misaligned with the orbital plane of the binary system exhibits a wobbling with a period about half the orbital period and precesses 
with a period nearly equal to 20 orbital periods. If the continuum variability is caused by the disc wobbling, this would indicate 
the presence of a companion with an orbital period of $\sim 24-36$\,yr, which would also produce a precession with a period of 
$\sim 480-720$\,yr, not too far from the inner-jet precession period estimated by \cite{2008A&A...485..137C}, that is $\sim$1000\,yr. 
It is worth noting, however, that these authors assumed a jet velocity of 80\,km\,s$^{-1}$ and a distance to the source of 1.7\,kpc. 
By using the revised, more accurate distance of 1.64\,kpc and a jet velocity of 110\,km\,s$^{-1}$, their precession period would be 
about 700\,yr, which nicely matches the period estimated by us using the recipe of \cite{2000MNRAS.317..773B}.

As the central protostellar mass derived by \cite{2016ApJ...823..125C} from the Keplerian velocity pattern of the disc roughly agrees 
with that derived from the system bolometric luminosity, the companion, if any, must be a low-mass star. Assuming a central 
protostellar mass of $\sim 12$ $M_\sun$, it should be located at a distance of $\sim 19-25$\,au (i.e. subtending an angle 
$<0\farcs02$). It is also worth noting that, according to \cite{2000MNRAS.317..773B}, the plane of the 
orbit of the supposed companion must 
be misaligned with respect to the  disc plane. We can use Eq.~(1) of \cite{1999ApJ...512L.131T} to obtain some more constraints. Of course, 
this equation cannot be applied to the circumstellar disc as a whole in our case, with a precession period $\sim 1000$ yr. In fact, 
it assumes that the disc precesses as a rigid body and this requires that the time needed to cross the disc at the speed of sound is 
much less than the precession period. However, a disc of $\sim 1000$\,au would be crossed in $\sim 5000$\,yr at a sound speed of 
1\,km\,s$^{-1}$.

By assuming a primary star mass of 12 $M_\sun$, an orbit inclination of $30\degr$ with respect to the disc, an orbital radius of 
19\,au, and  Eq.~(1) of \cite{1999ApJ...512L.131T} yields a disc radius of $\sim 15$\,au for a secondary star of 1 $M_\sun$ and a 
precession period of 1000 yr or a secondary star of 2\,$M_\sun$ and a precession period of 500\,yr. \cite{2009ApJ...702..567V} 
estimated that the dust sublimation radius is between 3--17\,au for a 10 $M_\sun$ protostar. However, \cite{2020A&A...635L..12G} 
and \cite{2021A&A...654A.109K} measured a smaller range of values (from 0.6 to 10\,au) for a sample of massive protostars 
($\sim$10 to 20\,$M_\sun$) using NIR interferometry. In the end, a jet precession with a period of $\sim 700$\,yr is consistent 
with the scenario of a nearby low-mass companion inducing oscillations in the innermost region of the circumstellar disc, which 
might also explain the anti-correlated $K$-band continuum brightness variations. An analysis of the NIR fluctuations could then 
become a powerful tool to understand the circumstellar structure of the protostar and would call for a more frequent monitoring.

\subsection{Origin of the parsec-scale outflow morphology}

The scenario pictured above does not entirely match the data on larger scales. On a scale of $\sim 1$ pc, both 
H$_2$ and CO emission display a more extended north-south outflow centred on the protostar position (see 
\citealt{2000ApJ...535..833S}). The southern lobe is red-shifted, has a jaw-like morphology,
and the south-eastern lobe of the HCO$^{+}$ and H$_2$ outflow lies at its south-eastern border, indicating, 
on a larger scale, a different precessing geometry with respect to the closest H$_2$ flow, which represents the 
most recent mass ejection episodes. 
The northern flow (blue-shifted) is smaller than the southern one, confirming that the outflow is still moving 
inside the cloud in the south, but has probably emerged outside in the north.
The jet axis position angles of the closest (-60.9$\degr$) and farthest (-37$\degr$) H$_2$ features differ by 
22.1$\degr$. 
This is difficult to reconcile with the precession period ($\sim$1000\,yr) and opening angle ($\sim$14$\degr$) 
of the inner jet derived by \cite{2008A&A...485..137C}, in that the CO morphology, along with the presence of 
optical and NIR line emitting knots further from the protostar, would imply a much larger precession period 
($\geq 64000$ yr) and opening angle ($37\degr$; \citealt{2005A&A...434.1039C}). Indeed, 
\cite{2008A&A...485..137C} were unable to find a superposition of two different precessions by simultaneously 
fitting the positions of the H$_2$ knots along the whole flow. Nevertheless, the flow agrees with a jet whose 
projected axis was initially in a roughly north-south direction with an inclination angle of $\sim 45\degr$ 
with respect 
to the observer, which has then moved to the current north-western-south-eastern direction (roughly on the 
plane of the sky). 

A possible explanation is that there are two (or more) 
different outflows, a younger one and an older one.
In fact, two other signatures of outflowing matter have been evidenced near IRAS20126+4104. However, one of 
them, source S, is unlikely to be the driving source of an older outflow, as jet signatures could not have 
remained obscured in the NIR after some 10000 yr of mass ejection activity out of the plane of the sky. 

An alternative explanation is that a close encounter with
a passing-by massive body has caused the disc axis to twist by
$\sim$22$\degr$ on the plane of the sky and by $\sim$35$\degr$ away from the observer line of sight.
According to \cite{2000MNRAS.317..773B}, a close encounter, particularly if occurring during a phase of intense 
accretion, can cause the outflow axis to tilt, mimicking a precession. 
The likelihood of such an encounter is of course higher if the
protostar is embedded in a young star cluster. In this case,
\citet{2009ApJS..185..486P} found that the number of close encounters
per cluster member is given by
\begin{equation}
    \Gamma = \Gamma_{0} [b/(1000 {\rm au}]^{\gamma}
\end{equation}
where b is the smallest encounter distance. These authors computed
$\Gamma_{0}$ and $\gamma$ for a wide parameter space populated by young stellar clusters,
including the number of members. As for IRAS20126+4104, \citet{2015ApJS..219...41M} estimate
that the cluster it is embedded in is made up of at least 43 young stars
within a radius of $1.2$ kpc. Given the low X-ray sensitivity this is bound to be
a lower limit, so that we can assume that the cluster has at least 100 members.
As they provide $\Gamma$ per million years, this needs to be multiplied by $0.1$
since a time span of 100000 years seems more appropriate for this case. Under these assumptions,
$\Gamma$ ranges from between $\sim 1-2$ \%  for $b = 1000$ au, which, although small, indicates that this scenario is not implausible.

It is unlikely that this body is the hypothesised nearby low-mass companion (see previous section), because a 
similar precession pattern is also detected in the southern and farthest part of jet (see left panel of 
Figure~7 in \citealt{2008A&A...485..137C}). This suggests that if the jet precession is caused by a companion, 
the latter has to be orbiting around the central source ever since. A likely candidate for such a close 
encounter could then be the intermediate-mass 
medium infrared- and radio-source S, located 1$\arcsec$ away. The observed tilt of the disc axis is compatible 
with the system geometry and with source S moving from the north-east to south-west and passing by 
IRAS\,20126+4104. The close encounter might also have caused the similar orientation of the jet-like radio 
structure associated with S. 
Indeed the magnitude of the disc axis tilt depends on the angle between the disc plane and the orbital plane, 
the mass of the star and the distance between the disc and the passing-by object (see e.\,g. \citealt{1995MNRAS.274..987P}).

If a close encounter with source S has caused the jet axis to tilt, then source S must have passed by the 
protostar near enough to affect the disc plane but at least at the escape velocity. Another obvious constraint 
is that the close encounter must have occurred earlier than the ejection of the outer knots of the H$_2$ inner
jet (the average dynamical ages of groups A1 and C2 are 2200$\pm$900\,yr and 1070$\pm$300, respectively, see 
Sect.~\ref{prop:mo:sec} and Fig.~\ref{Tdyn_vs_d:fig}), but later than a CO outflow dynamical time ago ($\sim 
64000$ yr, \citealt{2000ApJ...535..833S}). Assuming a closest distance of 1000 au, the escape velocity would be 
$\sim 6$ km s$^{-1}$ even for a total central mass of 20 $M_\sun$ (star plus disc). Typical stellar 1D velocity 
dispersions in young clusters are $\sim 1-3$ km s$^{-1}$ (e. g. \citealt{2019ApJ...870...32K}),
thus such a value would not be implausible for two young stellar objects that originated in the same region, 
whereas much higher values are unlikely.
The protostar and S are now at a projected distance of $\sim 1640$ au, which is a lower limit for the actual 
distance, and would then be crossed at the escape speed in $> 1300$ yr. This timing is consistent with the two 
age constraints.

A less likely alternative is that S has been trapped in an elliptic orbit, namely IRAS20126+4104 would then be 
a triple system, which, typically, is not stable.
In this respect, we note that \citet{2000ApJ...535..833S} found that a star roughly the same mass as the 
protostar in a circular orbit with $R \sim 1400$ au (i.e. similar to the projected distance between the 
protostar and S) would be able to cause the slower precession. In fact, \citet{2005ApJ...631L..73S} proposed 
that source S is the stellar companion responsible for the slower precession. In this case, the orbital motion would 
imprint the outflow as well. The effects of orbital motions on outflows were discussed by 
\citet{2002ApJ...568..733M}. By assuming that source S has a mass equal or less than the protostar (12 
$M_{\sun}$) and a circular orbit
with a radius greater than the projected distance of source S to the protostar, we can use Eq.~(1) of 
\citet{2002ApJ...568..733M} to estimate a period $> 38000$ yr, which is of course roughly the same as that of 
the induced precession for any outflow oscillations due to the source orbital motion. However, from Eq.~(13) of 
\citet{2002ApJ...568..733M}, one can see that the oscillations would be confined to a cone of aperture 
$\la 1\degr$, which is negligible compared to the observed wiggling cone. Thus, the existence of an orbital 
motion cannot be inferred from the outflow morphology.
Another possibility is that accretion of matter with different angular momenta has caused the disc inclination 
to change in time \citep{2017ApJ...839...69M}.

We can conclude that the jet from IRAS20126+4104 displays one faster precession motion and exhibits the 
signature of either a close stellar encounter that occurred less than $\sim 64000$ years ago, or of a slower 
precession motion. These scenarios need to be investigated through more detailed theoretical models of the 
circumstellar and stellar environments, in particular, by simulating the effects of a multiple system composed 
of a nearby low-mass star affecting the innermost region of the disc (this work) and a more massive, further 
companion or cluster member affecting the whole disc \citep{2000ApJ...535..833S}, and of asymmetric 
accretion   \citep{2000ApJ...535..833S} as well.

\section{Conclusions}
By means of new high-spatial-resolution NIR images ($K_s$, H2, and Br$\gamma$ filters) obtained at the LBT with 
the LUCI1 instrument and the new AO system SOUL, combined with published imaging and spectroscopic data, we 
have presented a multi-epoch kinematic and photometric study of the IRAS\,20126+4104 H$_2$ jet and outflow. 
Two other observing runs with AO (2003 and 2012) providing high-spatial-resolution data have been used for 
our analysis.

The main results of this paper are the following:
\begin{itemize}
     \item Thanks to the extremely high-spatial resolution of our images and their 17-year time baseline, 
we have been able to measure knot proper motions down to a few mas per year, with values ranging from 1.7 to 
20.3\,mas\,yr$^{-1}$, which translate into $\varv_{\rm tg}$=13--158\,km\,s$^{-1}$ at a distance of 1.64\,kpc. 
The average $\varv_{\rm tg}$ of the flow is 80$\pm$30\,km\,s$^{-1}$. The large scatter in velocity is likely 
due to the strong interaction between knots along the flow. 
     \item The dynamical age of the knots increases moving away from the source, ranging from $\sim$200 to 
$\sim$4000\,yr, indicating that the inner jet is relatively young.
     \item By combining tangential and radial velocities (the latter inferred from \citealt{2008A&A...485..137C}) 
of the H$_2$ knots, we have reconstructed the 3D kinematics of the jet, which roughly expands along an axis 
with an average inclination angle of 8$\degr\pm$1$\degr$ with respect to the plane of the sky. 
      Overall the inferred 3D kinematics is consistent with the jet precession model presented by 
\cite{2008A&A...485..137C} with a $\sim$700\,yr period and agrees with the motion of the water masers studied 
by \citet{2011A&A...526A..66M}.
      \item Some of the larger knot groups are consistent with the prediction of bow-shock models. A fit to a 
ballistic bow-shock model provides estimates for the half-aperture of the inner driving jet in the $2-5.5\degr$ 
range.
      \item Both the average proper motion directions and the bow-shock axes of the
      outer knot groups do not intersect the protostar position, but cross its
      proper motion projected track $\sim 1\arcsec - 3 \arcsec$ to the north-east. As the knot proper motions 
are roughly computed in the protostar framework and the
      protostar moves towards the south-west, this may suggest that the protostar proper motion has decreased in time.  
    \item Our multi-epoch photometry of the north-eastern and south-western outflow cavities shows an 
anti-correlated variability of the NIR continuum with a possible periodicity of 12--18\,yr. We argue that such 
a variability could be caused by the wobbling of the inner disc, which would alternately eclipse the two 
hemispheres of the central protostar. Such wobbling could be caused by a nearby low-mass star companion with 
an orbital period in the range of 24--36\,yr, which would also produce the precession of the jet.
    \item The total H$_2$ $2.12$ $\mu$m luminosity does not exhibit time variations at a $\ga 0.4$ mag level, 
which is consistent with hydrodynamical models of outflows. However, some knots exhibit clear line flux 
increases (of $\sim 1$ mag), notably one associated with the bow-shock C1, and two knots exhibit a flux 
decrease $> 0.5$ mag.
    \item The longer precession time of 64000 yr found on the basis of CO emission imaged on the parsec scale by 
other authors can be reconciled with our estimate by assuming either a past close encounter with a relatively 
massive object or the presence of a second precession caused by a companion orbiting outside the disc. We have 
shown that in both cases the nearby intermediate-mass source S is a suitable candidate. 
   \end{itemize}

Our analysis demonstrates that multi-epoch high-spatial-resolution observations are a powerful tool to unveil 
the physical properties of highly embedded massive protostars. 
In this respect, a photometric monitoring
of IRAS\,20126+4104 could provide valuable insights into its circumstellar environment.

\begin{acknowledgements}
A.C.G. has been supported by PRIN-INAF MAIN-STREAM 2017 “Protoplanetary disks seen through the eyes of new-
generation instruments”. F. M. and A.C.G. also acknowledge support from PRIN-INAF 2019 “Spectroscopically tracing
the disk dispersal evolution (STRADE)”. 
The LBTO AO group would like to acknowledge the assistance of R.T. Gatto with night observations.
This  work  is  based  on observations made with the Large Binocular Telescope. The LBT is an international collaboration among institutions in the United States, Italy, and Germany. LBT Corporation partners are: The University of  Arizona  on  behalf  of  the  Arizona  Board  of  Regents;  Istituto Nazionale di Astrofisica, Italy; LBT Beteiligungsgesellschaft, Germany, representing the Max-Planck Society, The Leibniz Institute for Astrophysics Potsdam, and Heidelberg University; The Ohio State University, representing OSU, University of Notre Dame, University of  Minnesota,  and  University  of  Virginia. 
This publication makes use of data products from the Two Micron All Sky Survey, which is a joint project of the University of Massachusetts and the Infrared Processing and Analysis Center/California Institute of Technology, funded by the National Aeronautics and Space Administration and the National Science Foundation. This work has also made use of data from the European Space Agency (ESA) mission
{\it Gaia} (\url{https://www.cosmos.esa.int/gaia}), processed by the {\it Gaia}
Data Processing and Analysis Consortium (DPAC,
\url{https://www.cosmos.esa.int/web/gaia/dpac/consortium}). Funding for the DPAC
has been provided by national institutions, in particular the institutions
participating in the {\it Gaia} Multilateral Agreement.
\end{acknowledgements}

\bibliographystyle{aa}
\bibliography{master_biblio.bib}

\begin{thebibliography}{66}
\expandafter\ifx\csname natexlab\endcsname\relax\def\natexlab#1{#1}\fi

\bibitem[{{Anderson} {et~al.}(2011){Anderson}, {Hofner}, {Shepherd}, \&
  {Creech-Eakman}}]{2011AJ....142..158A}
{Anderson}, C.~N., {Hofner}, P., {Shepherd}, D., \& {Creech-Eakman}, M. 2011,
  \aj, 142, 158

\bibitem[{{Bate} {et~al.}(2000){Bate}, {Bonnell}, {Clarke}, {Lubow}, {Ogilvie},
  {Pringle}, \& {Tout}}]{2000MNRAS.317..773B}
{Bate}, M.~R., {Bonnell}, I.~A., {Clarke}, C.~J., {et~al.} 2000, \mnras, 317,
  773

\bibitem[{{Belloche} {et~al.}(2011){Belloche}, {Schuller}, {Parise},
  {Andr{\'e}}, {Hatchell}, {J{\o}rgensen}, {Bontemps}, {Wei{\ss}}, {Menten}, \&
  {Muders}}]{2011A&A...527A.145B}
{Belloche}, A., {Schuller}, F., {Parise}, B., {et~al.} 2011, \aap, 527, A145

\bibitem[{{Benisty} {et~al.}(2010){Benisty}, {Malbet}, {Dougados}, {Natta}, {Le
  Bouquin}, {Massi}, {Bonnefoy}, {Bouvier}, {Chauvin}, {Chesneau}, {Garcia},
  {Grankin}, {Isella}, {Ratzka}, {Tatulli}, {Testi}, {Weigelt}, \&
  {Whelan}}]{2010A&A...517L...3B}
{Benisty}, M., {Malbet}, F., {Dougados}, C., {et~al.} 2010, \aap, 517, L3

\bibitem[{{Caratti o Garatti} \& {Eisl{\"o}ffel}(2019)}]{2019ASSP...55..111C}
{Caratti o Garatti}, A. \& {Eisl{\"o}ffel}, J. 2019, in Astrophysics and Space
  Science Proceedings, Vol.~55, JET Simulations, Experiments, and Theory: Ten
  Years After JETSET. What Is Next?, ed. C.~{Sauty}, 111

\bibitem[{{Caratti o Garatti} {et~al.}(2009){Caratti o Garatti},
  {Eisl{\"o}ffel}, {Froebrich}, {Nisini}, {Giannini}, \&
  {Calzoletti}}]{2009A&A...502..579C}
{Caratti o Garatti}, A., {Eisl{\"o}ffel}, J., {Froebrich}, D., {et~al.} 2009,
  \aap, 502, 579

\bibitem[{{Caratti o Garatti} {et~al.}(2008){Caratti o Garatti}, {Froebrich},
  {Eisl{\"o}ffel}, {Giannini}, \& {Nisini}}]{2008A&A...485..137C}
{Caratti o Garatti}, A., {Froebrich}, D., {Eisl{\"o}ffel}, J., {Giannini}, T.,
  \& {Nisini}, B. 2008, \aap, 485, 137

\bibitem[{{Caratti o Garatti} {et~al.}(2017){Caratti o Garatti}, {Stecklum},
  {Garcia Lopez}, {Eisl{\"o}ffel}, {Ray}, {Sanna}, {Cesaroni}, {Walmsley},
  {Oudmaijer}, {de Wit}, {Moscadelli}, {Greiner}, {Krabbe}, {Fischer}, {Klein},
  \& {Iba{\~n}ez}}]{2017NatPh..13..276C}
{Caratti o Garatti}, A., {Stecklum}, B., {Garcia Lopez}, R., {et~al.} 2017,
  Nature Physics, 13, 276

\bibitem[{{Cesaroni} {et~al.}(1999){Cesaroni}, {Felli}, {Jenness}, {Neri},
  {Olmi}, {Robberto}, {Testi}, \& {Walmsley}}]{1999A&A...345..949C}
{Cesaroni}, R., {Felli}, M., {Jenness}, T., {et~al.} 1999, \aap, 345, 949

\bibitem[{{Cesaroni} {et~al.}(1997){Cesaroni}, {Felli}, {Testi}, {Walmsley}, \&
  {Olmi}}]{1997A&A...325..725C}
{Cesaroni}, R., {Felli}, M., {Testi}, L., {Walmsley}, C.~M., \& {Olmi}, L.
  1997, \aap, 325, 725

\bibitem[{{Cesaroni} {et~al.}(2014){Cesaroni}, {Galli}, {Neri}, \&
  {Walmsley}}]{2014A&A...566A..73C}
{Cesaroni}, R., {Galli}, D., {Neri}, R., \& {Walmsley}, C.~M. 2014, \aap, 566,
  A73

\bibitem[{{Cesaroni} {et~al.}(2013){Cesaroni}, {Massi}, {Arcidiacono},
  {Beltr{\'a}n}, {McCarthy}, {Kulesa}, {Boutsia}, {Paris},
  {Quir{\'o}s-Pacheco}, \& {Xompero}}]{2013A&A...549A.146C}
{Cesaroni}, R., {Massi}, F., {Arcidiacono}, C., {et~al.} 2013, \aap, 549, A146

\bibitem[{{Cesaroni} {et~al.}(2005){Cesaroni}, {Neri}, {Olmi}, {Testi},
  {Walmsley}, \& {Hofner}}]{2005A&A...434.1039C}
{Cesaroni}, R., {Neri}, R., {Olmi}, L., {et~al.} 2005, \aap, 434, 1039

\bibitem[{{Chen} {et~al.}(2016){Chen}, {Keto}, {Zhang}, {Sridharan}, {Liu}, \&
  {Su}}]{2016ApJ...823..125C}
{Chen}, H.-R.~V., {Keto}, E., {Zhang}, Q., {et~al.} 2016, \apj, 823, 125

\bibitem[{{de Valon} {et~al.}(2020){de Valon}, {Dougados}, {Cabrit}, {Louvet},
  {Zapata}, \& {Mardones}}]{2020A&A...634L..12D}
{de Valon}, A., {Dougados}, C., {Cabrit}, S., {et~al.} 2020, \aap, 634, L12

\bibitem[{{Edris} {et~al.}(2005){Edris}, {Fuller}, {Cohen}, \&
  {Etoka}}]{2005A&A...434..213E}
{Edris}, K.~A., {Fuller}, G.~A., {Cohen}, R.~J., \& {Etoka}, S. 2005, \aap,
  434, 213

\bibitem[{{Gaia Collaboration} {et~al.}(2021){Gaia Collaboration}, {Brown},
  {Vallenari}, {Prusti}, {de Bruijne}, {Babusiaux}, {Biermann}, {Creevey},
  {Evans}, {Eyer}, {Hutton}, {Jansen}, {Jordi}, {Klioner}, {Lammers},
  {Lindegren}, {Luri}, {Mignard}, {Panem}, {Pourbaix}, {Randich}, {Sartoretti},
  {Soubiran}, {Walton}, {Arenou}, {Bailer-Jones}, {Bastian}, {Cropper},
  {Drimmel}, {Katz}, {Lattanzi}, {van Leeuwen}, {Bakker}, {Cacciari},
  {Casta{\~n}eda}, {De Angeli}, {Ducourant}, {Fabricius}, {Fouesneau},
  {Fr{\'e}mat}, {Guerra}, {Guerrier}, {Guiraud}, {Jean-Antoine Piccolo},
  {Masana}, {Messineo}, {Mowlavi}, {Nicolas}, {Nienartowicz}, {Pailler},
  {Panuzzo}, {Riclet}, {Roux}, {Seabroke}, {Sordo}, {Tanga}, {Th{\'e}venin},
  {Gracia-Abril}, {Portell}, {Teyssier}, {Altmann}, {Andrae}, {Bellas-Velidis},
  {Benson}, {Berthier}, {Blomme}, {Brugaletta}, {Burgess}, {Busso}, {Carry},
  {Cellino}, {Cheek}, {Clementini}, {Damerdji}, {Davidson}, {Delchambre},
  {Dell'Oro}, {Fern{\'a}ndez-Hern{\'a}ndez}, {Galluccio}, {Garc{\'\i}a-Lario},
  {Garcia-Reinaldos}, {Gonz{\'a}lez-N{\'u}{\~n}ez}, {Gosset}, {Haigron},
  {Halbwachs}, {Hambly}, {Harrison}, {Hatzidimitriou}, {Heiter},
  {Hern{\'a}ndez}, {Hestroffer}, {Hodgkin}, {Holl}, {Jan{\ss}en}, {Jevardat de
  Fombelle}, {Jordan}, {Krone-Martins}, {Lanzafame}, {L{\"o}ffler}, {Lorca},
  {Manteiga}, {Marchal}, {Marrese}, {Moitinho}, {Mora}, {Muinonen}, {Osborne},
  {Pancino}, {Pauwels}, {Petit}, {Recio-Blanco}, {Richards}, {Riello},
  {Rimoldini}, {Robin}, {Roegiers}, {Rybizki}, {Sarro}, {Siopis}, {Smith},
  {Sozzetti}, {Ulla}, {Utrilla}, {van Leeuwen}, {van Reeven}, {Abbas}, {Abreu
  Aramburu}, {Accart}, {Aerts}, {Aguado}, {Ajaj}, {Altavilla}, {{\'A}lvarez},
  {{\'A}lvarez Cid-Fuentes}, {Alves}, {Anderson}, {Anglada Varela}, {Antoja},
  {Audard}, {Baines}, {Baker}, {Balaguer-N{\'u}{\~n}ez}, {Balbinot}, {Balog},
  {Barache}, {Barbato}, {Barros}, {Barstow}, {Bartolom{\'e}}, {Bassilana},
  {Bauchet}, {Baudesson-Stella}, {Becciani}, {Bellazzini}, {Bernet}, {Bertone},
  {Bianchi}, {Blanco-Cuaresma}, {Boch}, {Bombrun}, {Bossini}, {Bouquillon},
  {Bragaglia}, {Bramante}, {Breedt}, {Bressan}, {Brouillet}, {Bucciarelli},
  {Burlacu}, {Busonero}, {Butkevich}, {Buzzi}, {Caffau}, {Cancelliere},
  {C{\'a}novas}, {Cantat-Gaudin}, {Carballo}, {Carlucci}, {Carnerero},
  {Carrasco}, {Casamiquela}, {Castellani}, {Castro-Ginard}, {Castro Sampol},
  {Chaoul}, {Charlot}, {Chemin}, {Chiavassa}, {Cioni}, {Comoretto}, {Cooper},
  {Cornez}, {Cowell}, {Crifo}, {Crosta}, {Crowley}, {Dafonte}, {Dapergolas},
  {David}, {David}, {de Laverny}, {De Luise}, {De March}, {De Ridder}, {de
  Souza}, {de Teodoro}, {de Torres}, {del Peloso}, {del Pozo}, {Delbo},
  {Delgado}, {Delgado}, {Delisle}, {Di Matteo}, {Diakite}, {Diener},
  {Distefano}, {Dolding}, {Eappachen}, {Edvardsson}, {Enke}, {Esquej}, {Fabre},
  {Fabrizio}, {Faigler}, {Fedorets}, {Fernique}, {Fienga}, {Figueras},
  {Fouron}, {Fragkoudi}, {Fraile}, {Franke}, {Gai}, {Garabato},
  {Garcia-Gutierrez}, {Garc{\'\i}a-Torres}, {Garofalo}, {Gavras}, {Gerlach},
  {Geyer}, {Giacobbe}, {Gilmore}, {Girona}, {Giuffrida}, {Gomel}, {Gomez},
  {Gonzalez-Santamaria}, {Gonz{\'a}lez-Vidal}, {Granvik},
  {Guti{\'e}rrez-S{\'a}nchez}, {Guy}, {Hauser}, {Haywood}, {Helmi}, {Hidalgo},
  {Hilger}, {H{\l}adczuk}, {Hobbs}, {Holland}, {Huckle}, {Jasniewicz},
  {Jonker}, {Juaristi Campillo}, {Julbe}, {Karbevska}, {Kervella}, {Khanna},
  {Kochoska}, {Kontizas}, {Kordopatis}, {Korn}, {Kostrzewa-Rutkowska},
  {Kruszy{\'n}ska}, {Lambert}, {Lanza}, {Lasne}, {Le Campion}, {Le Fustec},
  {Lebreton}, {Lebzelter}, {Leccia}, {Leclerc}, {Lecoeur-Taibi}, {Liao},
  {Licata}, {Lindstr{\o}m}, {Lister}, {Livanou}, {Lobel}, {Madrero Pardo},
  {Managau}, {Mann}, {Marchant}, {Marconi}, {Marcos Santos}, {Marinoni},
  {Marocco}, {Marshall}, {Martin Polo}, {Mart{\'\i}n-Fleitas}, {Masip},
  {Massari}, {Mastrobuono-Battisti}, {Mazeh}, {McMillan}, {Messina},
  {Michalik}, {Millar}, {Mints}, {Molina}, {Molinaro}, {Moln{\'a}r},
  {Montegriffo}, {Mor}, {Morbidelli}, {Morel}, {Morris}, {Mulone}, {Munoz},
  {Muraveva}, {Murphy}, {Musella}, {Noval}, {Ord{\'e}novic}, {Orr{\`u}},
  {Osinde}, {Pagani}, {Pagano}, {Palaversa}, {Palicio}, {Panahi}, {Pawlak},
  {Pe{\~n}alosa Esteller}, {Penttil{\"a}}, {Piersimoni}, {Pineau}, {Plachy},
  {Plum}, {Poggio}, {Poretti}, {Poujoulet}, {Pr{\v{s}}a}, {Pulone}, {Racero},
  {Ragaini}, {Rainer}, {Raiteri}, {Rambaux}, {Ramos}, {Ramos-Lerate}, {Re
  Fiorentin}, {Regibo}, {Reyl{\'e}}, {Ripepi}, {Riva}, {Rixon}, {Robichon},
  {Robin}, {Roelens}, {Rohrbasser}, {Romero-G{\'o}mez}, {Rowell}, {Royer},
  {Rybicki}, {Sadowski}, {Sagrist{\`a} Sell{\'e}s}, {Sahlmann}, {Salgado},
  {Salguero}, {Samaras}, {Sanchez Gimenez}, {Sanna}, {Santove{\~n}a},
  {Sarasso}, {Schultheis}, {Sciacca}, {Segol}, {Segovia}, {S{\'e}gransan},
  {Semeux}, {Shahaf}, {Siddiqui}, {Siebert}, {Siltala}, {Slezak}, {Smart},
  {Solano}, {Solitro}, {Souami}, {Souchay}, {Spagna}, {Spoto}, {Steele},
  {Steidelm{\"u}ller}, {Stephenson}, {S{\"u}veges}, {Szabados}, {Szegedi-Elek},
  {Taris}, {Tauran}, {Taylor}, {Teixeira}, {Thuillot}, {Tonello}, {Torra},
  {Torra}, {Turon}, {Unger}, {Vaillant}, {van Dillen}, {Vanel}, {Vecchiato},
  {Viala}, {Vicente}, {Voutsinas}, {Weiler}, {Wevers}, {Wyrzykowski}, {Yoldas},
  {Yvard}, {Zhao}, {Zorec}, {Zucker}, {Zurbach}, \&
  {Zwitter}}]{2021A&A...649A...1G}
{Gaia Collaboration}, {Brown}, A.~G.~A., {Vallenari}, A., {et~al.} 2021, \aap,
  649, A1

\bibitem[{{Gaia Collaboration} {et~al.}(2016){Gaia Collaboration}, {Brown},
  {Vallenari}, {Prusti}, {de Bruijne}, {Mignard}, {Drimmel}, {Babusiaux},
  {Bailer-Jones}, {Bastian}, {Biermann}, {Evans}, {Eyer}, {Jansen}, {Jordi},
  {Katz}, {Klioner}, {Lammers}, {Lindegren}, {Luri}, {O'Mullane}, {Panem},
  {Pourbaix}, {Randich}, {Sartoretti}, {Siddiqui}, {Soubiran}, {Valette}, {van
  Leeuwen}, {Walton}, {Aerts}, {Arenou}, {Cropper}, {H{\o}g}, {Lattanzi},
  {Grebel}, {Holland}, {Huc}, {Passot}, {Perryman}, {Bramante}, {Cacciari},
  {Casta{\~n}eda}, {Chaoul}, {Cheek}, {De Angeli}, {Fabricius}, {Guerra},
  {Hern{\'a}ndez}, {Jean-Antoine-Piccolo}, {Masana}, {Messineo}, {Mowlavi},
  {Nienartowicz}, {Ord{\'o}{\~n}ez-Blanco}, {Panuzzo}, {Portell}, {Richards},
  {Riello}, {Seabroke}, {Tanga}, {Th{\'e}venin}, {Torra}, {Els},
  {Gracia-Abril}, {Comoretto}, {Garcia-Reinaldos}, {Lock}, {Mercier},
  {Altmann}, {Andrae}, {Astraatmadja}, {Bellas-Velidis}, {Benson}, {Berthier},
  {Blomme}, {Busso}, {Carry}, {Cellino}, {Clementini}, {Cowell}, {Creevey},
  {Cuypers}, {Davidson}, {De Ridder}, {de Torres}, {Delchambre}, {Dell'Oro},
  {Ducourant}, {Fr{\'e}mat}, {Garc{\'\i}a-Torres}, {Gosset}, {Halbwachs},
  {Hambly}, {Harrison}, {Hauser}, {Hestroffer}, {Hodgkin}, {Huckle}, {Hutton},
  {Jasniewicz}, {Jordan}, {Kontizas}, {Korn}, {Lanzafame}, {Manteiga},
  {Moitinho}, {Muinonen}, {Osinde}, {Pancino}, {Pauwels}, {Petit},
  {Recio-Blanco}, {Robin}, {Sarro}, {Siopis}, {Smith}, {Smith}, {Sozzetti},
  {Thuillot}, {van Reeven}, {Viala}, {Abbas}, {Abreu Aramburu}, {Accart},
  {Aguado}, {Allan}, {Allasia}, {Altavilla}, {{\'A}lvarez}, {Alves},
  {Anderson}, {Andrei}, {Anglada Varela}, {Antiche}, {Antoja}, {Ant{\'o}n},
  {Arcay}, {Bach}, {Baker}, {Balaguer-N{\'u}{\~n}ez}, {Barache}, {Barata},
  {Barbier}, {Barblan}, {Barrado y Navascu{\'e}s}, {Barros}, {Barstow},
  {Becciani}, {Bellazzini}, {Bello Garc{\'\i}a}, {Belokurov}, {Bendjoya},
  {Berihuete}, {Bianchi}, {Bienaym{\'e}}, {Billebaud}, {Blagorodnova},
  {Blanco-Cuaresma}, {Boch}, {Bombrun}, {Borrachero}, {Bouquillon}, {Bourda},
  {Bouy}, {Bragaglia}, {Breddels}, {Brouillet}, {Br{\"u}semeister},
  {Bucciarelli}, {Burgess}, {Burgon}, {Burlacu}, {Busonero}, {Buzzi}, {Caffau},
  {Cambras}, {Campbell}, {Cancelliere}, {Cantat-Gaudin}, {Carlucci},
  {Carrasco}, {Castellani}, {Charlot}, {Charnas}, {Chiavassa}, {Clotet},
  {Cocozza}, {Collins}, {Costigan}, {Crifo}, {Cross}, {Crosta}, {Crowley},
  {Dafonte}, {Damerdji}, {Dapergolas}, {David}, {David}, {De Cat}, {de Felice},
  {de Laverny}, {De Luise}, {De March}, {de Martino}, {de Souza}, {Debosscher},
  {del Pozo}, {Delbo}, {Delgado}, {Delgado}, {Di Matteo}, {Diakite},
  {Distefano}, {Dolding}, {Dos Anjos}, {Drazinos}, {Duran}, {Dzigan},
  {Edvardsson}, {Enke}, {Evans}, {Eynard Bontemps}, {Fabre}, {Fabrizio},
  {Faigler}, {Falc{\~a}o}, {Farr{\`a}s Casas}, {Federici}, {Fedorets},
  {Fern{\'a}ndez-Hern{\'a}ndez}, {Fernique}, {Fienga}, {Figueras}, {Filippi},
  {Findeisen}, {Fonti}, {Fouesneau}, {Fraile}, {Fraser}, {Fuchs}, {Gai},
  {Galleti}, {Galluccio}, {Garabato}, {Garc{\'\i}a-Sedano}, {Garofalo},
  {Garralda}, {Gavras}, {Gerssen}, {Geyer}, {Gilmore}, {Girona}, {Giuffrida},
  {Gomes}, {Gonz{\'a}lez-Marcos}, {Gonz{\'a}lez-N{\'u}{\~n}ez},
  {Gonz{\'a}lez-Vidal}, {Granvik}, {Guerrier}, {Guillout}, {Guiraud},
  {G{\'u}rpide}, {Guti{\'e}rrez-S{\'a}nchez}, {Guy}, {Haigron},
  {Hatzidimitriou}, {Haywood}, {Heiter}, {Helmi}, {Hobbs}, {Hofmann}, {Holl},
  {Holland}, {Hunt}, {Hypki}, {Icardi}, {Irwin}, {Jevardat de Fombelle},
  {Jofr{\'e}}, {Jonker}, {Jorissen}, {Julbe}, {Karampelas}, {Kochoska},
  {Kohley}, {Kolenberg}, {Kontizas}, {Koposov}, {Kordopatis}, {Koubsky},
  {Krone-Martins}, {Kudryashova}, {Kull}, {Bachchan}, {Lacoste-Seris}, {Lanza},
  {Lavigne}, {Le Poncin-Lafitte}, {Lebreton}, {Lebzelter}, {Leccia}, {Leclerc},
  {Lecoeur-Taibi}, {Lemaitre}, {Lenhardt}, {Leroux}, {Liao}, {Licata},
  {Lindstr{\o}m}, {Lister}, {Livanou}, {Lobel}, {L{\"o}ffler}, {L{\'o}pez},
  {Lorenz}, {MacDonald}, {Magalh{\~a}es Fernandes}, {Managau}, {Mann},
  {Mantelet}, {Marchal}, {Marchant}, {Marconi}, {Marinoni}, {Marrese},
  {Marschalk{\'o}}, {Marshall}, {Mart{\'\i}n-Fleitas}, {Martino}, {Mary},
  {Matijevi{\v{c}}}, {Mazeh}, {McMillan}, {Messina}, {Michalik}, {Millar},
  {Miranda}, {Molina}, {Molinaro}, {Molinaro}, {Moln{\'a}r}, {Moniez},
  {Montegriffo}, {Mor}, {Mora}, {Morbidelli}, {Morel}, {Morgenthaler},
  {Morris}, {Mulone}, {Muraveva}, {Musella}, {Narbonne}, {Nelemans},
  {Nicastro}, {Noval}, {Ord{\'e}novic}, {Ordieres-Mer{\'e}}, {Osborne},
  {Pagani}, {Pagano}, {Pailler}, {Palacin}, {Palaversa}, {Parsons}, {Pecoraro},
  {Pedrosa}, {Pentik{\"a}inen}, {Pichon}, {Piersimoni}, {Pineau}, {Plachy},
  {Plum}, {Poujoulet}, {Pr{\v{s}}a}, {Pulone}, {Ragaini}, {Rago}, {Rambaux},
  {Ramos-Lerate}, {Ranalli}, {Rauw}, {Read}, {Regibo}, {Reyl{\'e}}, {Ribeiro},
  {Rimoldini}, {Ripepi}, {Riva}, {Rixon}, {Roelens}, {Romero-G{\'o}mez},
  {Rowell}, {Royer}, {Ruiz-Dern}, {Sadowski}, {Sagrist{\`a} Sell{\'e}s},
  {Sahlmann}, {Salgado}, {Salguero}, {Sarasso}, {Savietto}, {Schultheis},
  {Sciacca}, {Segol}, {Segovia}, {Segransan}, {Shih}, {Smareglia}, {Smart},
  {Solano}, {Solitro}, {Sordo}, {Soria Nieto}, {Souchay}, {Spagna}, {Spoto},
  {Stampa}, {Steele}, {Steidelm{\"u}ller}, {Stephenson}, {Stoev}, {Suess},
  {S{\"u}veges}, {Surdej}, {Szabados}, {Szegedi-Elek}, {Tapiador}, {Taris},
  {Tauran}, {Taylor}, {Teixeira}, {Terrett}, {Tingley}, {Trager}, {Turon},
  {Ulla}, {Utrilla}, {Valentini}, {van Elteren}, {Van Hemelryck}, {van
  Leeuwen}, {Varadi}, {Vecchiato}, {Veljanoski}, {Via}, {Vicente}, {Vogt},
  {Voss}, {Votruba}, {Voutsinas}, {Walmsley}, {Weiler}, {Weingrill}, {Wevers},
  {Wyrzykowski}, {Yoldas}, {{\v{Z}}erjal}, {Zucker}, {Zurbach}, {Zwitter},
  {Alecu}, {Allen}, {Allende Prieto}, {Amorim}, {Anglada-Escud{\'e}},
  {Arsenijevic}, {Azaz}, {Balm}, {Beck}, {Bernstein}, {Bigot}, {Bijaoui},
  {Blasco}, {Bonfigli}, {Bono}, {Boudreault}, {Bressan}, {Brown}, {Brunet},
  {Bunclark}, {Buonanno}, {Butkevich}, {Carret}, {Carrion}, {Chemin},
  {Ch{\'e}reau}, {Corcione}, {Darmigny}, {de Boer}, {de Teodoro}, {de Zeeuw},
  {Delle Luche}, {Domingues}, {Dubath}, {Fodor}, {Fr{\'e}zouls}, {Fries},
  {Fustes}, {Fyfe}, {Gallardo}, {Gallegos}, {Gardiol}, {Gebran}, {Gomboc},
  {G{\'o}mez}, {Grux}, {Gueguen}, {Heyrovsky}, {Hoar}, {Iannicola}, {Isasi
  Parache}, {Janotto}, {Joliet}, {Jonckheere}, {Keil}, {Kim}, {Klagyivik},
  {Klar}, {Knude}, {Kochukhov}, {Kolka}, {Kos}, {Kutka}, {Lainey}, {LeBouquin},
  {Liu}, {Loreggia}, {Makarov}, {Marseille}, {Martayan}, {Martinez-Rubi},
  {Massart}, {Meynadier}, {Mignot}, {Munari}, {Nguyen}, {Nordlander}, {Ocvirk},
  {O'Flaherty}, {Olias Sanz}, {Ortiz}, {Osorio}, {Oszkiewicz}, {Ouzounis},
  {Palmer}, {Park}, {Pasquato}, {Peltzer}, {Peralta}, {P{\'e}turaud},
  {Pieniluoma}, {Pigozzi}, {Poels}, {Prat}, {Prod'homme}, {Raison}, {Rebordao},
  {Risquez}, {Rocca-Volmerange}, {Rosen}, {Ruiz-Fuertes}, {Russo}, {Sembay},
  {Serraller Vizcaino}, {Short}, {Siebert}, {Silva}, {Sinachopoulos}, {Slezak},
  {Soffel}, {Sosnowska}, {Strai{\v{z}}ys}, {ter Linden}, {Terrell}, {Theil},
  {Tiede}, {Troisi}, {Tsalmantza}, {Tur}, {Vaccari}, {Vachier}, {Valles}, {Van
  Hamme}, {Veltz}, {Virtanen}, {Wallut}, {Wichmann}, {Wilkinson}, {Ziaeepour},
  \& {Zschocke}}]{2016A&A...595A...2G}
{Gaia Collaboration}, {Brown}, A.~G.~A., {Vallenari}, A., {et~al.} 2016, \aap,
  595, A2

\bibitem[{{Grankin} {et~al.}(2008){Grankin}, {Bouvier}, {Herbst}, \&
  {Melnikov}}]{2008A&A...479..827G}
{Grankin}, K.~N., {Bouvier}, J., {Herbst}, W., \& {Melnikov}, S.~Y. 2008, \aap,
  479, 827

\bibitem[{{Grankin} {et~al.}(2007){Grankin}, {Melnikov}, {Bouvier}, {Herbst},
  \& {Shevchenko}}]{2007A&A...461..183G}
{Grankin}, K.~N., {Melnikov}, S.~Y., {Bouvier}, J., {Herbst}, W., \&
  {Shevchenko}, V.~S. 2007, \aap, 461, 183

\bibitem[{{Gravity Collaboration} {et~al.}(2020){Gravity Collaboration},
  {Caratti o Garatti}, {Fedriani}, {Garcia Lopez}, {Koutoulaki}, {Perraut},
  {Linz}, {Brandner}, {Garcia}, {Klarmann}, {Henning}, {Labadie},
  {Sanchez-Bermudez}, {Lazareff}, {van Dishoeck}, {Caselli}, {de Zeeuw}, {Bik},
  {Benisty}, {Dougados}, {Ray}, {Amorim}, {Berger}, {Cl{\'e}net}, {Coud{\'e} Du
  Foresto}, {Duvert}, {Eckart}, {Eisenhauer}, {Gao}, {Gendron}, {Genzel},
  {Gillessen}, {Gordo}, {Jocou}, {Horrobin}, {Kervella}, {Lacour}, {Le
  Bouquin}, {L{\'e}na}, {Grellmann}, {Ott}, {Paumard}, {Perrin}, {Rousset},
  {Scheithauer}, {Shangguan}, {Stadler}, {Straub}, {Straubmeier}, {Sturm},
  {Thi}, {Vincent}, \& {Widmann}}]{2020A&A...635L..12G}
{Gravity Collaboration}, {Caratti o Garatti}, A., {Fedriani}, R., {et~al.}
  2020, \aap, 635, L12

\bibitem[{{Gustafsson} {et~al.}(2010){Gustafsson}, {Ravkilde}, {Kristensen},
  {Cabrit}, {Field}, \& {Pineau Des For{\^e}ts}}]{2010A&A...513A...5G}
{Gustafsson}, M., {Ravkilde}, T., {Kristensen}, L.~E., {et~al.} 2010, \aap,
  513, A5

\bibitem[{{Hawarden} {et~al.}(2001){Hawarden}, {Leggett}, {Letawsky},
  {Ballantyne}, \& {Casali}}]{2001MNRAS.325..563H}
{Hawarden}, T.~G., {Leggett}, S.~K., {Letawsky}, M.~B., {Ballantyne}, D.~R., \&
  {Casali}, M.~M. 2001, \mnras, 325, 563

\bibitem[{{Hofner} {et~al.}(2007){Hofner}, {Cesaroni}, {Olmi},
  {Rodr{\'\i}guez}, {Mart{\'\i}}, \& {Araya}}]{2007A&A...465..197H}
{Hofner}, P., {Cesaroni}, R., {Olmi}, L., {et~al.} 2007, \aap, 465, 197

\bibitem[{{Hofner} {et~al.}(1999){Hofner}, {Cesaroni}, {Rodr{\'\i}guez}, \&
  {Mart{\'\i}}}]{1999A&A...345L..43H}
{Hofner}, P., {Cesaroni}, R., {Rodr{\'\i}guez}, L.~F., \& {Mart{\'\i}}, J.
  1999, \aap, 345, L43

\bibitem[{{Hunter} {et~al.}(2017){Hunter}, {Brogan}, {MacLeod}, {Cyganowski},
  {Chandler}, {Chibueze}, {Friesen}, {Indebetouw}, {Thesner}, \&
  {Young}}]{2017ApJ...837L..29H}
{Hunter}, T.~R., {Brogan}, C.~L., {MacLeod}, G., {et~al.} 2017, \apjl, 837, L29

\bibitem[{{Johnston} {et~al.}(2011){Johnston}, {Keto}, {Robitaille}, \&
  {Wood}}]{2011MNRAS.415.2953J}
{Johnston}, K.~G., {Keto}, E., {Robitaille}, T.~P., \& {Wood}, K. 2011, \mnras,
  415, 2953

\bibitem[{{Keto} \& {Zhang}(2010)}]{2010MNRAS.406..102K}
{Keto}, E. \& {Zhang}, Q. 2010, \mnras, 406, 102

\bibitem[{{Koumpia} {et~al.}(2021){Koumpia}, {de Wit}, {Oudmaijer}, {Frost},
  {Lumsden}, {Caratti o Garatti}, {Goodwin}, {Stecklum}, {Mendigut{\'\i}a},
  {Ilee}, \& {Vioque}}]{2021A&A...654A.109K}
{Koumpia}, E., {de Wit}, W.~J., {Oudmaijer}, R.~D., {et~al.} 2021, \aap, 654,
  A109

\bibitem[{{Kuhn} {et~al.}(2019){Kuhn}, {Hillenbrand}, {Sills}, {Feigelson}, \&
  {Getman}}]{2019ApJ...870...32K}
{Kuhn}, M.~A., {Hillenbrand}, L.~A., {Sills}, A., {Feigelson}, E.~D., \&
  {Getman}, K.~V. 2019, \apj, 870, 32

\bibitem[{{Kwan} \& {Tademaru}(1995)}]{1995ApJ...454..382K}
{Kwan}, J. \& {Tademaru}, E. 1995, \apj, 454, 382

\bibitem[{{Lebr{\'o}n} {et~al.}(2006){Lebr{\'o}n}, {Beuther}, {Schilke}, \&
  {Stanke}}]{2006A&A...448.1037L}
{Lebr{\'o}n}, M., {Beuther}, H., {Schilke}, P., \& {Stanke}, T. 2006, \aap,
  448, 1037

\bibitem[{{Masciadri} \& {Raga}(2002)}]{2002ApJ...568..733M}
{Masciadri}, E. \& {Raga}, A.~C. 2002, \apj, 568, 733

\bibitem[{{Matsumoto} {et~al.}(2017){Matsumoto}, {Machida}, \&
  {Inutsuka}}]{2017ApJ...839...69M}
{Matsumoto}, T., {Machida}, M.~N., \& {Inutsuka}, S.-i. 2017, \apj, 839, 69

\bibitem[{{McKee} \& {Offner}(2011)}]{2011IAUS..270...73M}
{McKee}, C.~F. \& {Offner}, S. R.~R. 2011, in Computational Star Formation, ed.
  J.~{Alves}, B.~G. {Elmegreen}, J.~M. {Girart}, \& V.~{Trimble}, Vol. 270,
  73--80

\bibitem[{{Montes} {et~al.}(2015){Montes}, {Hofner}, {Anderson}, \&
  {Rosero}}]{2015ApJS..219...41M}
{Montes}, V.~A., {Hofner}, P., {Anderson}, C., \& {Rosero}, V. 2015, \apjs,
  219, 41

\bibitem[{{Moscadelli} {et~al.}(2005){Moscadelli}, {Cesaroni}, \&
  {Rioja}}]{2005A&A...438..889M}
{Moscadelli}, L., {Cesaroni}, R., \& {Rioja}, M.~J. 2005, \aap, 438, 889

\bibitem[{{Moscadelli} {et~al.}(2011){Moscadelli}, {Cesaroni}, {Rioja},
  {Dodson}, \& {Reid}}]{2011A&A...526A..66M}
{Moscadelli}, L., {Cesaroni}, R., {Rioja}, M.~J., {Dodson}, R., \& {Reid},
  M.~J. 2011, \aap, 526, A66

\bibitem[{{Murakawa} {et~al.}(2003){Murakawa}, {Suto}, {Tamura}, {Takami},
  {Takato}, {Hayashi}, {Doi}, {Kaifu}, {Hayano}, {Gaessler}, \&
  {Kamata}}]{2003SPIE.4841..881M}
{Murakawa}, K., {Suto}, H., {Tamura}, M., {et~al.} 2003, in Society of
  Photo-Optical Instrumentation Engineers (SPIE) Conference Series, Vol. 4841,
  Instrument Design and Performance for Optical/Infrared Ground-based
  Telescopes, ed. M.~{Iye} \& A.~F.~M. {Moorwood}, 881--888

\bibitem[{{Nagayama} {et~al.}(2015){Nagayama}, {Omodaka}, {Handa}, {Burns},
  {Chibueze}, {Kobayashi}, {Sato}, {Ueno}, \&
  {Shizugami}}]{2015PASJ...67...66N}
{Nagayama}, T., {Omodaka}, T., {Handa}, T., {et~al.} 2015, \pasj, 67, 66

\bibitem[{{Ostriker} {et~al.}(2001){Ostriker}, {Lee}, {Stone}, \&
  {Mundy}}]{2001ApJ...557..443O}
{Ostriker}, E.~C., {Lee}, C.-F., {Stone}, J.~M., \& {Mundy}, L.~G. 2001, \apj,
  557, 443

\bibitem[{{Palau} {et~al.}(2017){Palau}, {Walsh}, {S{\'a}nchez-Monge},
  {Girart}, {Cesaroni}, {Jim{\'e}nez-Serra}, {Fuente}, {Zapata}, \&
  {Neri}}]{2017MNRAS.467.2723P}
{Palau}, A., {Walsh}, C., {S{\'a}nchez-Monge}, {\'A}., {et~al.} 2017, \mnras,
  467, 2723

\bibitem[{{Papaloizou} \& {Terquem}(1995)}]{1995MNRAS.274..987P}
{Papaloizou}, J. C.~B. \& {Terquem}, C. 1995, \mnras, 274, 987

\bibitem[{{Pinna} {et~al.}(2016){Pinna}, {Esposito}, {Hinz}, {Agapito},
  {Bonaglia}, {Puglisi}, {Xompero}, {Riccardi}, {Briguglio}, {Arcidiacono},
  {Carbonaro}, {Fini}, {Montoya}, \& {Durney}}]{2016SPIE.9909E..3VP}
{Pinna}, E., {Esposito}, S., {Hinz}, P., {et~al.} 2016, in Society of
  Photo-Optical Instrumentation Engineers (SPIE) Conference Series, Vol. 9909,
  Adaptive Optics Systems V, ed. E.~{Marchetti}, L.~M. {Close}, \& J.-P.
  {V{\'e}ran}, 99093V

\bibitem[{{Pinna} {et~al.}(2021){Pinna}, {Rossi}, {Puglisi}, {Agapito},
  {Bonaglia}, {Plantet}, {Mazzoni}, {Briguglio}, {Carbonaro}, {Xompero},
  {Grani}, {Riccardi}, {Esposito}, {Hinz}, {Vaz}, {Ertel}, {Montoya}, {Durney},
  {Christou}, {Miller}, {Taylor}, {Cavallaro}, \&
  {Lefebvre}}]{2021arXiv210107091P}
{Pinna}, E., {Rossi}, F., {Puglisi}, A., {et~al.} 2021, arXiv e-prints,
  arXiv:2101.07091

\bibitem[{{Proszkow} \& {Adams}(2009)}]{2009ApJS..185..486P}
{Proszkow}, E.-M. \& {Adams}, F.~C. 2009, \apjs, 185, 486

\bibitem[{{Qiu} {et~al.}(2008){Qiu}, {Zhang}, {Megeath}, {Gutermuth},
  {Beuther}, {Shepherd}, {Sridharan}, {Testi}, \& {De
  Pree}}]{2008ApJ...685.1005Q}
{Qiu}, K., {Zhang}, Q., {Megeath}, S.~T., {et~al.} 2008, \apj, 685, 1005

\bibitem[{{Quir{\'o}s-Pacheco} {et~al.}(2010){Quir{\'o}s-Pacheco}, {Busoni},
  {Agapito}, {Esposito}, {Pinna}, {Puglisi}, \&
  {Riccardi}}]{2010SPIE.7736E..3HQ}
{Quir{\'o}s-Pacheco}, F., {Busoni}, L., {Agapito}, G., {et~al.} 2010, in
  Society of Photo-Optical Instrumentation Engineers (SPIE) Conference Series,
  Vol. 7736, Adaptive Optics Systems II, ed. B.~L. {Ellerbroek}, M.~{Hart},
  N.~{Hubin}, \& P.~L. {Wizinowich}, 77363H

\bibitem[{{Ray} \& {Ferreira}(2021)}]{2021NewAR..9301615R}
{Ray}, T.~P. \& {Ferreira}, J. 2021, \nar, 93, 101615

\bibitem[{{Sanna} {et~al.}(2012){Sanna}, {Reid}, {Carrasco-Gonz{\'a}lez},
  {Menten}, {Brunthaler}, {Moscadelli}, \& {Rygl}}]{2012ApJ...745..191S}
{Sanna}, A., {Reid}, M.~J., {Carrasco-Gonz{\'a}lez}, C., {et~al.} 2012, \apj,
  745, 191

\bibitem[{{Shang} {et~al.}(2006){Shang}, {Allen}, {Li}, {Liu}, {Chou}, \&
  {Anderson}}]{2006ApJ...649..845S}
{Shang}, H., {Allen}, A., {Li}, Z.-Y., {et~al.} 2006, \apj, 649, 845

\bibitem[{{Shepherd} {et~al.}(2000){Shepherd}, {Yu}, {Bally}, \&
  {Testi}}]{2000ApJ...535..833S}
{Shepherd}, D.~S., {Yu}, K.~C., {Bally}, J., \& {Testi}, L. 2000, \apj, 535,
  833

\bibitem[{{Shinnaga} {et~al.}(2012){Shinnaga}, {Novak}, {Vaillancourt},
  {Machida}, {Kataoka}, {Tomisaka}, {Davidson}, {Phillips}, {Dowell}, {Leeuw},
  \& {Houde}}]{2012ApJ...750L..29S}
{Shinnaga}, H., {Novak}, G., {Vaillancourt}, J.~E., {et~al.} 2012, \apjl, 750,
  L29

\bibitem[{{Smith} \& {Rosen}(2005)}]{2005MNRAS.357..579S}
{Smith}, M.~D. \& {Rosen}, A. 2005, \mnras, 357, 579

\bibitem[{{Sridharan} {et~al.}(2007){Sridharan}, {Saito}, {Fuller}, \&
  {Kandori}}]{2007AAS...211.6213S}
{Sridharan}, T.~K., {Saito}, M., {Fuller}, G.~A., \& {Kandori}, R. 2007, in
  American Astronomical Society Meeting Abstracts, Vol. 211, American
  Astronomical Society Meeting Abstracts, 62.13

\bibitem[{{Sridharan} {et~al.}(2005){Sridharan}, {Williams}, \&
  {Fuller}}]{2005ApJ...631L..73S}
{Sridharan}, T.~K., {Williams}, S.~J., \& {Fuller}, G.~A. 2005, \apjl, 631, L73

\bibitem[{{Surcis} {et~al.}(2014){Surcis}, {Vlemmings}, {van Langevelde},
  {Moscadelli}, \& {Hutawarakorn Kramer}}]{2014A&A...563A..30S}
{Surcis}, G., {Vlemmings}, W.~H.~T., {van Langevelde}, H.~J., {Moscadelli}, L.,
  \& {Hutawarakorn Kramer}, B. 2014, \aap, 563, A30

\bibitem[{{Terquem} {et~al.}(1999){Terquem}, {Eisl{\"o}ffel}, {Papaloizou}, \&
  {Nelson}}]{1999ApJ...512L.131T}
{Terquem}, C., {Eisl{\"o}ffel}, J., {Papaloizou}, J.~C.~B., \& {Nelson}, R.~P.
  1999, \apjl, 512, L131

\bibitem[{{Townsley} {et~al.}(2018){Townsley}, {Broos}, {Garmire}, {Anderson},
  {Feigelson}, {Naylor}, \& {Povich}}]{2018ApJS..235...43T}
{Townsley}, L.~K., {Broos}, P.~S., {Garmire}, G.~P., {et~al.} 2018, \apjs, 235,
  43

\bibitem[{{Tram} {et~al.}(2018){Tram}, {Lesaffre}, {Cabrit}, {Gusdorf}, \&
  {Nhung}}]{2018MNRAS.473.1472T}
{Tram}, L.~N., {Lesaffre}, P., {Cabrit}, S., {Gusdorf}, A., \& {Nhung}, P.~T.
  2018, \mnras, 473, 1472

\bibitem[{{Trinidad} {et~al.}(2005){Trinidad}, {Curiel}, {Migenes}, {Patel},
  {Torrelles}, {G{\'o}mez}, {Rodr{\'\i}guez}, {Ho}, \&
  {Cant{\'o}}}]{2005AJ....130.2206T}
{Trinidad}, M.~A., {Curiel}, S., {Migenes}, V., {et~al.} 2005, \aj, 130, 2206

\bibitem[{{Vaidya} {et~al.}(2009){Vaidya}, {Fendt}, \&
  {Beuther}}]{2009ApJ...702..567V}
{Vaidya}, B., {Fendt}, C., \& {Beuther}, H. 2009, \apj, 702, 567

\bibitem[{{Yoo} {et~al.}(2017){Yoo}, {Lee}, {Mairs}, {Johnstone}, {Herczeg},
  {Kang}, {Kang}, {Cho}, \& {JCMT Transient Team}}]{2017ApJ...849...69Y}
{Yoo}, H., {Lee}, J.-E., {Mairs}, S., {et~al.} 2017, \apj, 849, 69

\bibitem[{{Zacharias} {et~al.}(2017){Zacharias}, {Finch}, \&
  {Frouard}}]{2017AJ....153..166Z}
{Zacharias}, N., {Finch}, C., \& {Frouard}, J. 2017, \aj, 153, 166

\bibitem[{{Zhang} {et~al.}(1998){Zhang}, {Hunter}, \&
  {Sridharan}}]{1998ApJ...505L.151Z}
{Zhang}, Q., {Hunter}, T.~R., \& {Sridharan}, T.~K. 1998, \apjl, 505, L151

\bibitem[{{Zhang} {et~al.}(2019){Zhang}, {Arce}, {Mardones}, {Cabrit},
  {Dunham}, {Garay}, {Noriega-Crespo}, {Offner}, {Raga}, \&
  {Corder}}]{2019ApJ...883....1Z}
{Zhang}, Y., {Arce}, H.~G., {Mardones}, D., {et~al.} 2019, \apj, 883, 1

\end{thebibliography}

\begin{appendix}

\section{SOUL performances}
\label{soul:perf}

\begin{figure}
   \centering
   \includegraphics[width=\hsize]{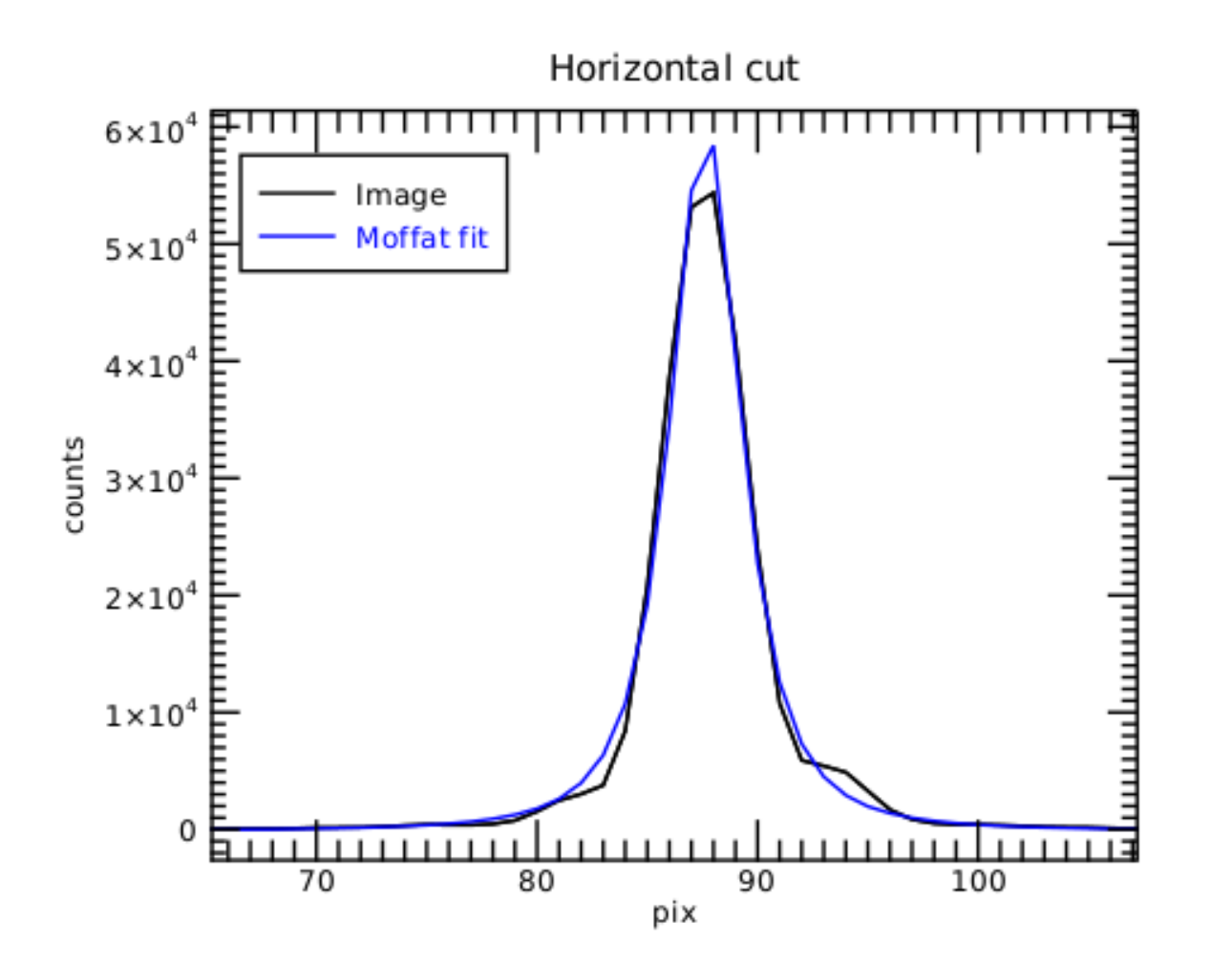}
      \caption{Profile of the PSF of the AO guide star along the right ascension axis (black line) in the SOUL H2 image overlaid with the best-fitting Moffat distribution (blue line). The plate scale is $0\farcs 015$ pix$^{-1}$.
              }
         \label{fit:psf}
   \end{figure}

The seeing correction achieved with SOUL was assessed by
fitting two Moffat distributions (along two perpendicular axes) to stars 2, 4, 5, 6, and 8
(see Fig.~\ref{nome:stelle}) in the final images. This allowed us to
determine the variation of the PSF width as a function of distance from the AO guide star (star 8). The same was done on the CIAO and FLAO(PISCES) images to evaluate the performance improvement. An example of Moffat fit is shown in Fig.~\ref{fit:psf}.

\begin{figure}
   \centering
   \includegraphics[width=\hsize]{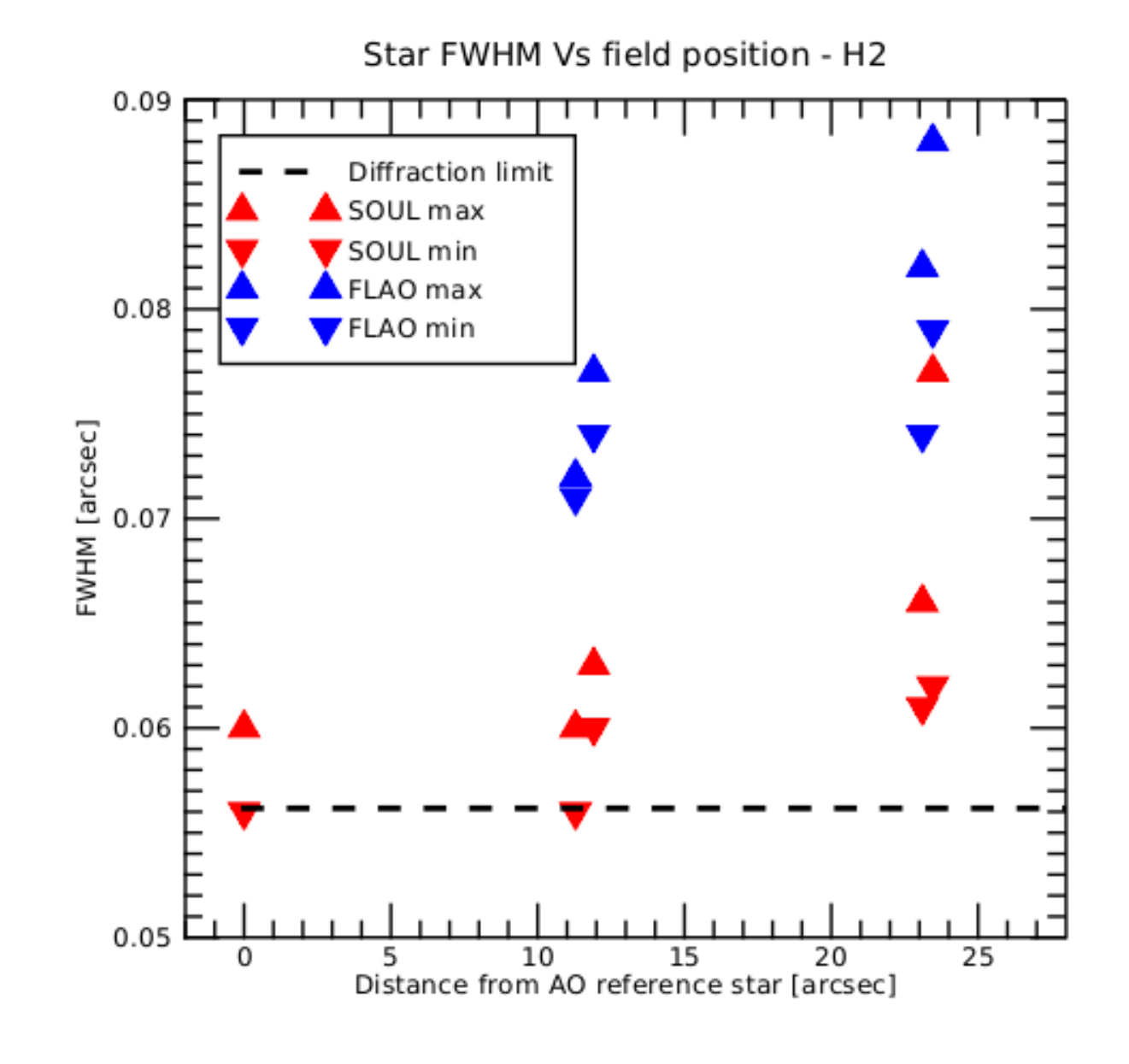}
      \caption{Maximum and minimum FWHMs obtained from the Moffat fits to the PSFs as a function of distance from the AO guide star in
      the SOUL (red triangles) and FLAO (blue triangles) H2 images.
              }
         \label{fwhm:dist}
   \end{figure}

The maximum and minimum FWHMs obtained from the five stars in the H2 images from SOUL and FLAO are shown in Fig.~\ref{fwhm:dist} 
as a function of distance from the AO guide star. The improvement in the correction degree achieved with SOUL compared to 
FLAO(PISCES) is evident. As expected, the FWHM increases with distance.

The overall results for the three sets of images (CIAO, FLAO, and SOUL) are listed in Table~\ref{AO:comp}. 
As for the H2 and Br$\gamma$ filters, the correction has clearly improved from 2003 to 2013, with SOUL achieving the best results (also considering that FLAO and SOUL operated roughly in the same seeing conditions). 
As for the K$_{\rm s}$ filter, our SOUL images were obtained in nights with seeing conditions not as good as those in the other 
available nights. The values listed were obtained on September 4, with the best seeing of the three nights occurring when K$_{s}$ was acquired.
%
%
\begin{table}
\begin{footnotesize}
\caption{Overall results from the Moffat fits to the PSFs in the CIAO, FLAO, and SOUL images.}             
\label{AO:comp}     
\centering                          
\begin{tabular}{l l l l}  
\hline\hline                 
  & CIAO & FLAO  & SOUL \\
  &      & PISCES & LUCI1 \\
     & (2003) & (2013)  & (2020) \\    
\hline                        
\multicolumn{4}{c}{H2 filter($^1$)}\\
Average FWHM of AO guide star & $0\farcs 123$ & -- & $0\farcs 058$ \\
Average FWHM of most distant star & $0\farcs195$  & $0\farcs 078$ & $0\farcs 064$ \\
Strehl ratio at the AO guide star & -- & -- & 47 \% \\
Strehl ratio at the most distant star & -- & -- & 35 \% \\
\hline
\multicolumn{4}{c}{Br$\gamma$ filter($^1$)}\\
Average FWHM of AO guide star & $0\farcs 113$ & -- & $0\farcs 062$ \\
Average FWHM of most distant star & $0\farcs189$  & $0\farcs 077$ & $0\farcs 069$ \\
Strehl ratio at the AO guide star & -- & -- & 42 \% \\
Strehl ratio at the most distant star & -- & -- & 33 \% \\
\hline
\multicolumn{4}{c}{K$_{\rm s}$ filter($^2$)}\\
Average FWHM of AO guide star & -- & -- & $0\farcs 066$ \\
Average FWHM of most distant star & --  & $0\farcs 071$ & $0\farcs 070$ \\
Strehl ratio at the AO guide star & -- & -- & 33 \% \\
Strehl ratio at the most distant star & -- & -- & 23 \% \\
\hline                                   
\end{tabular}
\tablefoot{(1) the FLAO and SOUL data were obtained 
in comparable seeing ($\sim 1\arcsec$) conditions;
(2) the SOUL data were obtained on September 4, in
slightly worse seeing conditions than H2, Br$\gamma$, and FLAO
K$_{\rm s}$.}
\end{footnotesize}
\end{table}
%

\section{Astrometry and image mapping}
\label{astro:map}

The astrometric calibration of the SOUL H2 image was achieved by a two-step process, using the iraf tasks {\it ccxymatch} and 
{\it ccmap}. First, we calibrated the large field H2 image obtained with the TNG in 2001 by matching the stars in the frame and the 
{\it Gaia} EDR3 catalogue (\citealt{2016A&A...595A...2G}, \citeyear{2021A&A...649A...1G}). From this calibrated image we derived the 
astrometric information for all the stars in the small FoV of the SOUL images. To minimise any degradation of the astrometric accuracy 
due to the proper motion of the stars in the two fields, we first noted that {\it Gaia} EDR3 provides the motion with good accuracy for two 
stars in the SOUL field, namely number 4 and 8 of Fig.~\ref{nome:stelle}. These are
$-4.307 \pm 0.048$ mas/yr (RA) and $-4.503 \pm 0.051$ mas/yr (DEC) for star 4, and
$-4.150 \pm 0.015$ mas/yr (RA) and $-4.544 \pm 0.017$ mas/yr (DEC) for star 8.
The proper motion of IRAS\,20126+4104 can be obtained from water masers (L. Moscadelli, private communication) and is
$\sim -4.1$ mas/yr (RA) and $\sim -4.1$ mas/yr (DEC), which is similar to that of the two stars. Therefore, we derived the astrometric 
solution for the TNG image using only stars with proper motions between 0 and $-8$ mas/yr both in RA and DEC from the {\it Gaia} catalogue, 
so that the reciprocal positions of the star selected should not have changed by more than $0.1\arcsec$ from 2001 to 2020. 

We checked that none of the eight stars in the SOUL image (Fig.~\ref{nome:stelle}) had shifted too much from each other by constructing 
triangles from each triplet of stars and measuring the relative changes of their sides between the CIAO and the SOUL image. We found that 
the relative changes imply side length differences $\sol 0.05\arcsec$, confirming that the stellar reciprocal positions have changed 
by $0.05\arcsec$ at most, making them an ideal reference grid for the astrometric calibration of all the small field images.   
The absolute positions in the calibrated H2 SOUL image are then in the ICRS standard (epoch $2015.5$) and should be more accurate than 
$\sim 0.1\arcsec$. 

Finally, we  registered and re-gridded all the other images to the H2 SOUL one by computing the transformations with the iraf tasks 
{\it geomap} (using the eight stars as the reference grid), and applying them with {\it geotran} so that all registered images matched 
each other. In the transformations we only considered a translation, a scale factor, and a rotation to avoid any possible deformation 
due to the small differences in the proper motions of the reference stars (although, as discussed above, they are probably not 
significant). 

\section{Photometric stability}
\label{phot:stabi}

\begin{figure}
   \centering
   \includegraphics[width=\hsize]{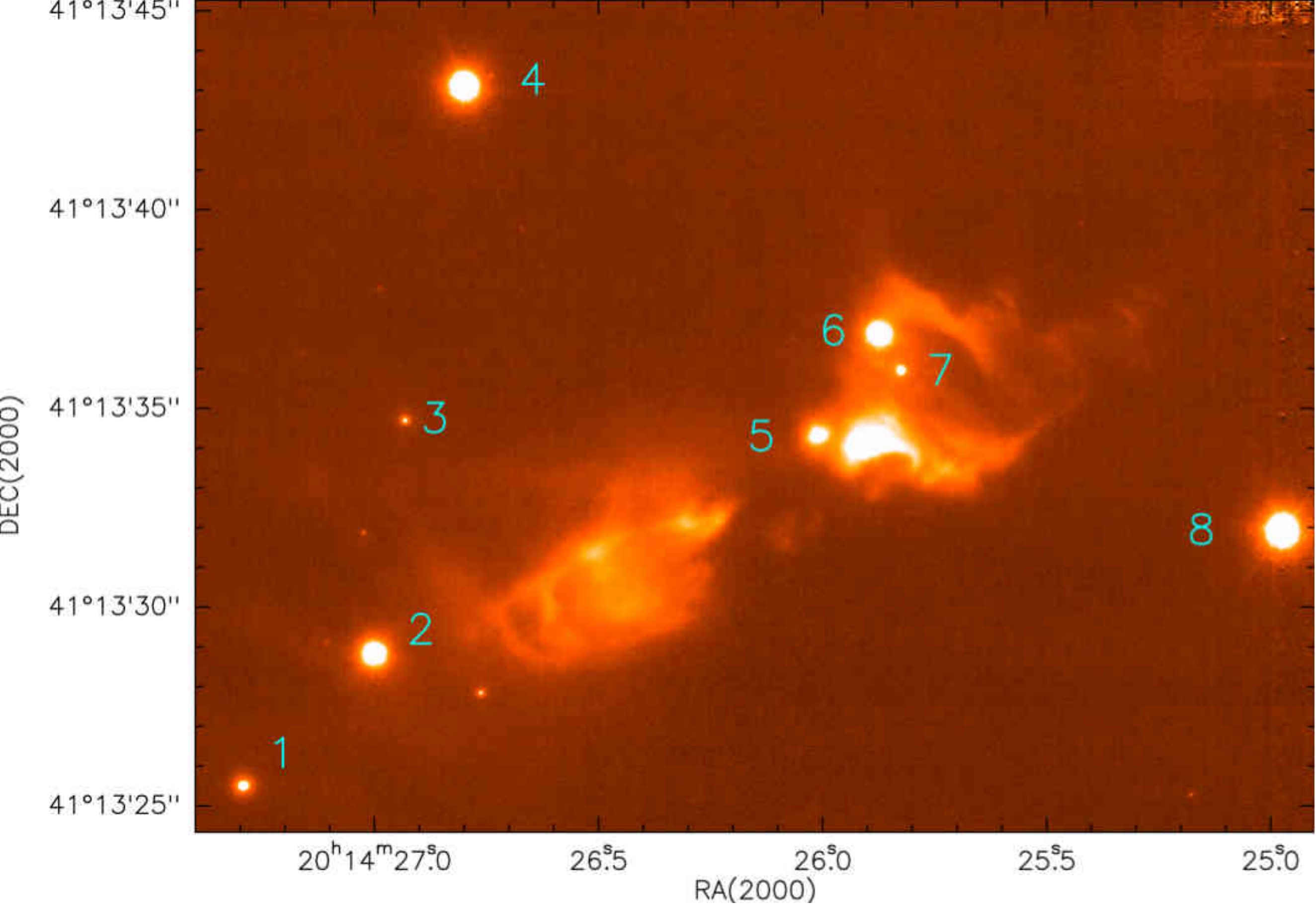}
      \caption{Br$\gamma$-filter image of IRAS\,20126+4104 obtained with LUCI1 and the AO system SOUL, with the identification number of the stars available as photometric references.
              }
         \label{nome:stelle}
   \end{figure}

As not all the images collected were obtained along with nearby standard stars' observations, a photometric monitoring of both line and 
continuum jet features must rely on stars common to all the frames, used as local standards. In turn, one has to check the photometric 
stability of these local stars to select the most stable ones. The candidates suitable to be used as secondary standards are labelled in 
Fig.~\ref{nome:stelle}. It is important to note that star 8 is the AO guide star in all AO-assisted observations examined in this work 
and is out of the PISCES frames. Our analysis, summed up in Appendix~\ref{app:ana}, shows that stars 2 and 4 are probably stable within 
$\sim 0.1$ mag, although they may have entered the non-linear regime in the CIAO frames. The jet photometry is therefore referenced to 
stars 2 and 4.

An effect to take into account in using only two local standards is the error in the obtained zero points caused by small differences in 
the PSF, the aperture, and the sky estimate from image to image. We have adopted apertures whose radii encompass the position where the 
signal is $\sol 5-10$ \%
of the peak, which means $\sog 1$ FWHM of the PSF for the lower-resolution images (UKIRT, TNG) and a few FWHMs of the PSF for the AO 
assisted images (to compensate for the varying Strehl ratios typical of these). To estimate the error due to any difference in the 
fractions of flux recovered in different images, we repeated the photometry doubling the aperture size, which should largely overestimate 
the effect we are concerned with. In fact this causes differences $\sol 0.2$ mag in all cases and bands, except for star 2 in the 
low-resolution images ($\sol 0.5$ mag), as in this case a larger aperture encompasses more diffuse emission from the south-eastern lobe 
of the jet resulting in a flux increase. We can therefore expect in this case as well that the fraction of stellar flux recovered by 
doubling the aperture would be $\sol 0.2$ mag if there was no diffuse emission.

\section{Band effects on the photometry}
\label{band:eff}

The photometric variability of the stars in the small FoV of the AO-assisted images (see Fig.~\ref{nome:stelle}) was measured in all
available bands ($K_{s}$, $K'$, H2, Br$\gamma$, and $K_{\rm cont}$; see Table~\ref{tab:bands}) by simply adopting the $K_{s}$ values 
provided by (or computed from) the 2Mass PSC, irrespectively of the band, to derive the zero point. A simple procedure can be used to 
estimate the effects of this simplified zero point estimate on a source magnitude in each band, at least at the first order.

Given a band centred at $\lambda_{c}$ with an FWHM of $\Delta \lambda$, it can be shown that a second-order approximation 
(in $\Delta \lambda/\lambda_{c}$)
is given by
\begin{equation}
{\rm mag_i} = C_i - 2.5 \times \log[F_{\lambda}(\lambda_{c})\Delta \lambda], 
\end{equation}
where $C_i$ depends on the band $i$ in the adopted standard system. This requires that the source has a continuum spectrum with no lines 
that is well approximated by $F_{\lambda}(\lambda)= A \times \lambda^{\gamma}$ inside the band being considered. 

In the NIR, $\Delta \lambda/\lambda_{c} < 1$ so the second-order approximation is valid.
The terms $C_{i}$ can be obtained if the zero magnitude flux $F_{\lambda,0}(\lambda_{c})$ is known, as in this case
\begin{equation}
C_i = 2.5 \times \log[F_{\lambda,0}(\lambda_{c})\Delta \lambda]. 
\end{equation}
The zero magnitude fluxes from 2Mass can be used to define a source which has ${\rm mag_i} = 0$ in every NIR band linked to this 
standard system. By simply fitting
\begin{equation}
\label{fitty:fit}
F_{\lambda,0}(\lambda_{c}) = A_0 \times \lambda_{c}^{\gamma_0} 
\end{equation}
to the 2Mass $J, H, K_{s}$ zero magnitude fluxes, one obtains
$\log(A_0) = -12.16$ and $\gamma_0 = -3.59$, with $F_{\lambda}$ in
W cm$^{-2}$ $\mu$m$^{-1}$, which approximates the zero magnitude fluxes by better than 5 \%.

This allows one to address the first issue, that is to say how the magnitude of a standard star should be changed to account for a 
different passband overlapping the original one. Given a filter $j$ with central wavelength $\lambda_j$ inside the $K$ band and 
${\rm FWHM} =\Delta \lambda_j$, its zero point in this generalised 2Mass standard system is
\begin{equation}
C_j = 2.5 \times \log[F_{\lambda,0}(\lambda_{j})\Delta \lambda_j] 
\end{equation}
and the magnitude of a continuum source with flux density $F_{\lambda}(\lambda_j)$ measured through a slightly different bandwidth 
$\Delta \lambda$ (standard star and target observed through slightly different passbands with the same central wavelength but 
a different width) is given by
\begin{equation}
\label{mag2mag}
{\rm mag'_j} = 2.5 \times \log[F_{\lambda,0}(\lambda_{j})/F_{\lambda}(\lambda_{j})] +
2.5\log[\Delta \lambda_j/\Delta \lambda] 
\end{equation}
\begin{equation}
{\rm mag'_j} - {\rm mag_j} = 2.5\log[\Delta \lambda_j/\Delta \lambda], 
\end{equation}
where ${\rm mag_j}$ is the magnitude that would be measured in the exact photometric system of the standard star:
\begin{equation}
\label{mag2mag2}
{\rm mag_j} = 2.5 \times \log[F_{\lambda,0}(\lambda_{j})/F_{\lambda}(\lambda_{j})]. 
\end{equation}
If the standard star and target are measured through the same filter, then 
Eq~(\ref{mag2mag2}) can be used to recompute the standard star magnitude once a new $F_{\lambda,0}$ is determined at the
modified central wavelength from Eq.~\ref{fitty:fit}.
The effects of absorption or emission lines in the spectra can
be neglected as long as the flux removed or added by the lines
is negligible compared to the total flux in the passband.

The standard star imaged in the SOUL run, FS149, is an A2 star and it can be seen that the slightly different $K_{s}$ band of the LUCI1 
filter and even the H2 LUCI1 band do not introduce a significant difference compared to its 2Mass $K_{s}$ value
(less than $0.01$ mag). However, since the jet photometry has been done using stars 2 and 4 as local standards
by assuming their $K_{s}$ values in all bands, the band effects must also be estimated for these stars. By fitting Eq~(\ref{fitty:fit}) 
to the 2Mass $J,H,K_{s}$ fluxes of star 4 we obtained  $\log(A_0) = -17.62$ and $\gamma = -1.47$ (with $F_{\lambda}$ in
W cm$^{-2}$ $\mu$m$^{-1}$). Unfortunately, this procedure does not yield a good approximation for star 2 (meaning that its spectral 
energy distribution is more complex), so, from $H$ and $K_{s}$ only, we obtained $\log(A_0) = -20.19$ and $\gamma_0 = 5.31$.

By using Eq.~(\ref{mag2mag2}), it can  easily be shown that the corrected magnitudes are within $\sim 0.05$ mag of the $K_{s}$ value for 
star 4 (except for $K_{\rm cont}$ which is $\sim 0.1$ mag brighter) and within $\sim 0.2$ mag for star 2 (except for $K_{\rm cont}$ 
which is $\sim 0.5$ mag brighter). The corrected magnitudes and the measured instrumental magnitudes of the two stars were then used 
to compute the zero points for each image. We note that the zero points obtained from each of the two stars
in the TNG and SOUL images, assuming the 2Mass $K_{s}$ values
in all bands,  always differ by $\sol 0.05$ mag (except for $K_{\rm cont}$ for which the difference is $\sim 0.1$ mag), indicating that 
the spectral energy distribution (SED) of star 2 is not as steep as in the above approximation when limited inside the $K$ band. 
Conversely, the same zero points estimated from the CIAO and PISCES images (i.e. derived separately from star 2 and 4) differ by 
$\sim 0.1-0.2$ mag in the H2, Br$\gamma$, and $K_{s}$ bands. As for CIAO, the PSF of the two stars suggests they are in a slightly 
non-linear regime. In addition, star 4 lies very close to the frame edge in the PISCES images.

Line emission is measured with photometry on narrow-band images encompassing the line after subtraction of the continuum contribution 
estimated from a nearby, ideally line-free, band. As continuum emission (mostly due to dust scattered emission) changes with wavelength, 
one expects a systematic error in the continuum estimate depending on the central wavelength of the band used.
From Eq.~(\ref{mag2mag}), one can derive a correction to apply to the estimated continuum magnitudes to obtain the right values in the 
2Mass $K_{s}$ band.
By assuming a continuum flux increasing as $\lambda^{4}$, this correction would be $\sim -0.04$ mag for the Br$\gamma$ filters, 
$\sim  -0.14$ mag for the H2 filters, $\sim -0.15$ mag for the $K'$ and $K98$ filters, and $\sim 0.43$ mag for the $K_{\rm cont}$ 
filter.

In conclusion, at least for filters $K_{s}$, $K98$, $K'$, H2, and Br$\gamma$ the continuum photometry can be carried out simply by 
assuming the 2Mass $K_{s}$ values for the two local standard stars
with errors $< 0.2$ mag. We also note that, provided the two stars do not vary significantly, the choice of the standard magnitudes 
assigned to the reference star for the zero point computation does not affect the {\rm relative} photometry in the field for a given band.

Nevertheless, it can be shown that to derive the line emission inside a passband one needs to account for
the band width. If $L = \int F_\lambda {\rm d} \lambda$ is the line flux, Eq.~\ref{mag2mag2} can be generalised as
\begin{equation}
{\rm mag_p} = -2.5 \times \log[L/(F_{\lambda,0}\Delta\lambda)]. 
\end{equation}
Since the width of the H2 passband of the LUCI filter is significantly different from that of
the TNG and CIAO filters (see Table~\ref{tab:bands}), this means that
\begin{equation}
{\rm mag_{\rm LBT}} = {\rm mag_{\rm TNG, CIAO}} + 2.5 \log(\Delta\lambda_{\rm LBT}/\Delta\lambda_{\rm TNG,CIAO})
\end{equation}
\begin{equation}
\label{biro:biro}
{\rm mag_{\rm LBT}} = {\rm mag_{\rm TNG, CIAO}} - 0.36
\end{equation}
\begin{equation}
\label{biro:biro2}
{\rm mag_{\rm LBT}} = {\rm mag_{\rm PISCES}} + 0.18.
\end{equation}
It is therefore important that this correction is applied to the H2 line magnitudes before they can be compared.
%
%
\begin{table}
\caption{Bandwidth and central wavelengths of the filters used in the
observations discussed in this work.}             
\label{tab:bands}     
\centering                          
\begin{tabular}{l l l l}  
\hline\hline                 
Band & Instrument & $\lambda_{c}$ & $\Delta \lambda$ \\
     &            &  ($\mu$m)  & ($\mu$m) \\    
\hline                        
   $K_{s}$ & 2Mass & $2.159$ & $0.262$ \\      
   $K_{s}$ & LUCI1(LBT) & $2.163$ & $0.270$ \\
   $K_{s}$ & PISCES(LBT) & $2.166$ & $0.328$ \\
  $K_{s}$ & CIAO(SUBARU) & $2.15$ & $0.35$ \\ 
   $K'$ & NICS(TNG) & $2.12$ & $0.35$ \\
   $K_{\rm cont}$ & NICS(TNG) & $2.275$ & $0.039$ \\
   $K98$ & UFTI(UKIRT) & $\sim 2.2$ & $0.34$ \\
   H2 & LUCI1(LBT) & $2.124$ & $0.023$ \\
   H2($^1$) & PISCES(LBT) & $2.118$ & $0.0195$ \\
   H2 & CIAO(SUBARU) & $2.122$ & $0.032$ \\
   H2 & NICS(TNG)  & $2.122$ & $0.032$ \\
   Br$\gamma$ & LUCI1(LBT) & $2.170$ & $0.024$ \\
   Br$\gamma$($^1$) & PISCES(LBT) & $2.169$ & $0.02-0.04$ \\
   Br$\gamma$ & CIAO(SUBARU) & $2.166$ & $0.032$ \\
   Br$\gamma$ & NICS(TNG)  & $2.169$ & $0.035$ \\
\hline                                   
\end{tabular}
\tablefoot{(1) D. McCarthy, private communication.}
\end{table}
%

\section{Photometric variability of local standard stars}
\label{app:ana}

As not all the observations were performed providing a suitable nearby standard star, one needs to resort to local stars that are common 
to all frames to do photometry. Thus, only seven stars are available (see Fig.~\ref{nome:stelle}, as star 8 was outside the PISCES frame).
To be used as secondary standards, their photometric variability must be checked. In fact, stars 2, 4, 6, and 8 were detected in X-rays 
with {\it Chandra} (sources 50, 49, 41, and 37, respectively, in the catalogue of \citealt{2015ApJS..219...41M}; and sources 192, 187, 175, 
and 166, respectively, in the catalogue of \citealt{2018ApJS..235...43T}). This is consistent with these stars being Weak-line T-Tauri 
Stars (WTTSs) associated with the protostar cluster. Typically, WTTSs exhibit very stable long-time variability at optical wavelengths 
rarely exceeding a few tenths of magnitude \citep{2008A&A...479..827G}. On the other hand, a small fraction of classical T-Tauri stars (CTTSs) can exhibit larger long-time photometric variations \citep{2007A&A...461..183G}. All four stars did not display X-ray variations, 
but star 8 is listed as 'possibly variable' \citep{2018ApJS..235...43T}. The lack of X-ray flares during the {\it Chandra} exposure may 
lead one to discard the identification as T-Tauri stars or magnetically active dwarfs; nevertheless, one has to keep in mind that this 
holds only for the $\sim 39$ Ks of integration. In any case, as photometric variations of a few tenths of magnitude in the secondary 
standard stars can significantly bias the light curves of the jet features, the brightness stability of the seven available stars must be checked. 

\begin{figure}
   \centering
   \includegraphics[width=\hsize]{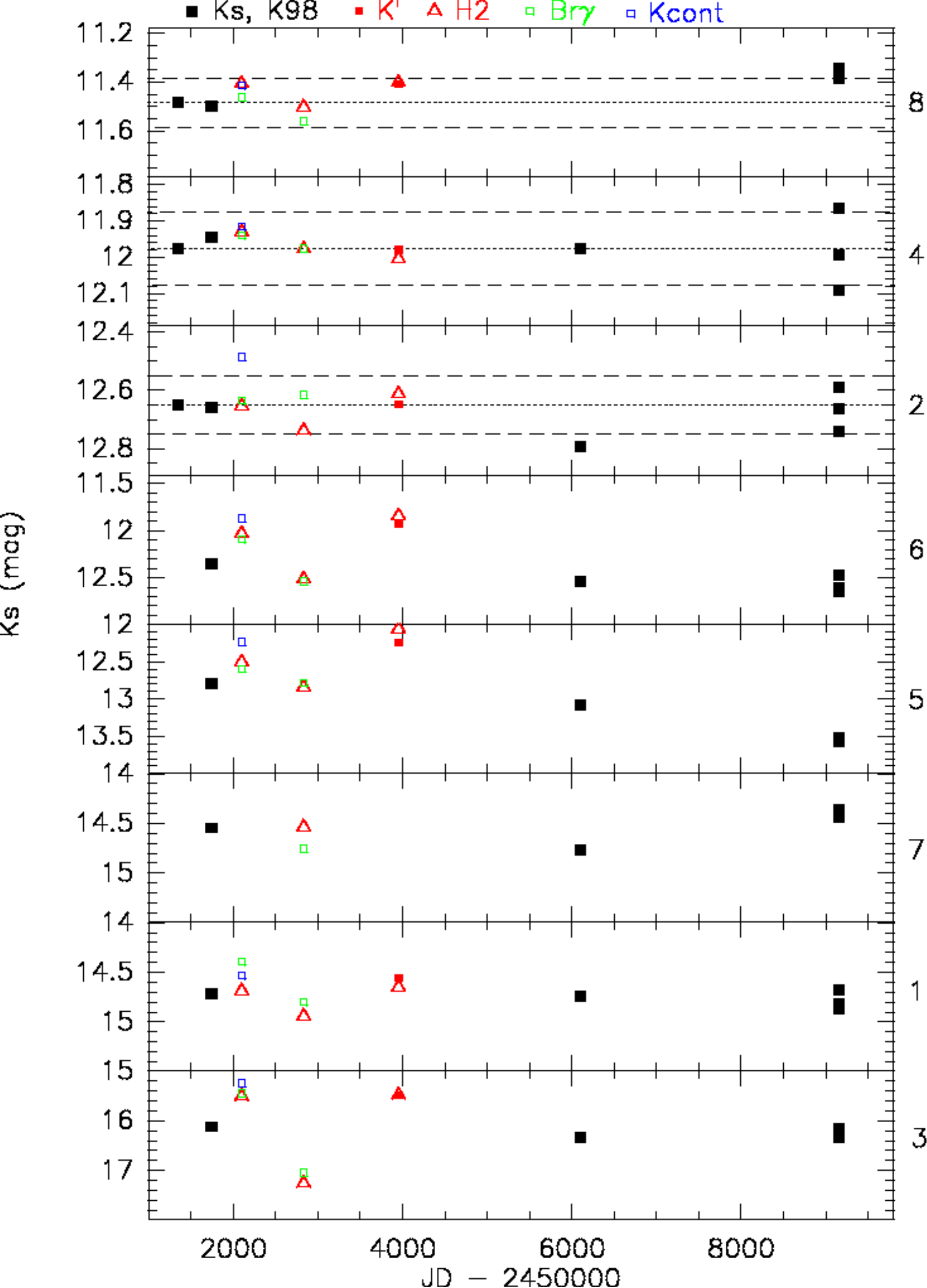}
      \caption{Photometric variability of the stars in the SOUL FoV. The star ID (with reference to Fig.~\protect\ref{nome:stelle}) 
is indicated on the right side of the panel. The filters are indicated with symbols as explained at the top of the panel. The data 
taken on JD $2452830.5$ (CIAO) and JD $2456099.8$ (PISCES) have not been calibrated and the magnitudes were computed by setting 
star 4 to $11.977$ mag in all bands. 
              }
         \label{FigStarVa}
   \end{figure}

We show our photometry of the eight stars in Fig.~\ref{FigStarVa}, as summarised in Table~\ref{table:jd}. 
The zero points to calibrate them were derived in different ways. The SOUL $K_{s}$ image was calibrated using standard star observations.
The TNG and UKIRT images were calibrated by cross-correlating the relatively large field photometry and the 2Mass PSC. We note
that for all
bands we adopted the 2Mass $K_{s}$ magnitudes (for an estimate of the error implied, see Appendix~\ref{band:eff}).
Finally, as the small field PISCES and CIAO images do not have standard star observations associated with them, we used star 4 as a
reference assuming its brightness is $K_{s} = 11.977$, the 2Mass value.

%
\begin{table}
\caption{Observation date of the data points used in Fig~\protect\ref{FigStarVa}.}            
\label{table:jd}      
\centering                          
\begin{tabular}{c c c c}        
\hline\hline                
JD & Data source & Bands & AO-assisted \\     
\hline                        
$2451352.8$ & 2Mass PSC & $K_{s}$ & no \\
$2451741.9$ & UFTI(UKIRT) & K98 & no \\
$2452098.7$ & NICS(TNG) & H2,$K_{\rm cont}$ & no \\
$2452099.7$ & NICS(TNG) &  Br$\gamma$ & no \\
$2452830.5$ & CIAO(SUBARU) & H2,Br$\gamma$ & yes \\
$2453953.5$ & NICS(TNG) & H2, $K'$ & no \\
$2456099.8$ & PISCES(LBT) & $K_{s}$ & yes \\
$2459155.6$ & LUCI1(LBT) & $K_{s}$ & yes \\
$2459157.6$ & LUCI1(LBT) & $K_{s}$  & yes \\
$2459159.6$ & LUCI1(LBT) & $K_{s}$  & yes \\
\hline                                   
\end{tabular}
\end{table}
%

In the last run (SOUL), the target and the standard star FS 149 were imaged in $K_{s}$ during three different nights (see 
Tables~\ref{obs:log} and \ref{table:jd}), so the different values shown in Fig.~\ref{FigStarVa} are essentially due to errors 
on the zero point caused by non-perfectly photometric nights.  On JD
$2459155.6$ and $2459157.6$ the standard star was imaged in five different areas of the detector and its instrumental
magnitudes remained within $\sim 0.03$ and $\sim 0.2$ mag, respectively. On JD $2459159.6$, the standard star was imaged in only 
two positions on the detector and its photometry differs by $\sim 0.02$ mag.

Figure~\ref{FigStarVa} indicates that stars 2 and 4 should be stable within $\sim 0.1$ mag and these were therefore used as the 
local standards. Nevertheless, we caution that stars 2, 4, and 8 may be in a non-linear regime in the CIAO images, leading to an 
overestimate of the fluxes measured for the other stars on that date.

%

\section{Photometry of the continuum-emitting jet regions with polygons}
\label{poly:des}

The photometric variability of the jet in the NIR continuum was analysed by performing aperture photometry. The used images were 
first registered to and projected onto the SOUL H2 frame grid. The jet was decomposed into different parts and, due to the image 
transformation, this only required the definition of a polygon for each part in the reference grid of the SOUL H2 image. The adopted 
polygons and their names are shown in Fig.~\ref{nome:poligoni:lowline}. 

The selected polygons are large enough to be insensitive to changes in the morphology of the emitting regions due to their proper 
motions. The continuum flux inside each polygon was measured using the task {\it polyphot} in IRAF. This was done on the Br$\gamma$ 
images; when unavailable, we used the $K$ images (UKIRT 2000 and TNG 2006) after subtracting the derived H2 line emission (we used 
TNG 2001 to estimate the correction for UKIRT 2000). For the epochs when only lower spatial resolution images are available,
we also subtracted the stars from the frames by using {\em daophot}. We note that using the transformed images also allows 
one to select the same
sky regions to estimate the background to subtract from the raw flux inside the polygons.

We carried out a few tests to assess the errors affecting our photometry. First, we checked on the UKIRT K98 image that performing 
the photometry on the transformed image would be equivalent to transforming the
polygons to fit the frame grid of the original image and perform the photometry on this image (i.e. the task {\it geomap}
conserves the flux), with only the photometric errors slightly underestimated as a result.
Then, we measured the differences between fluxes in the low spatial resolution images with and without star subtraction,
which are $\sol 0.05$ mag,
except for CN2 in the TNG 2001 image where the difference is $\sim 0.1$ mag. We also estimated the effects of the different 
spatial resolutions by convolving the SOUL Br$\gamma$ image with a Gaussian
distribution to obtain a spatial resolution similar to that of the TNG images and repeating the photometry. The flux differences
between the two images are $\sol 0.05$ mag, except for the ring-like feature whose flux is overestimated by $0.2$ mag in the
resolution-degraded image. Finally, we measured the flux differences for TNG 2003 between the Br$\gamma$ and the $K_{\rm cont}$ images,
which are $0.3 \pm 0.1$ mag, hence they are consistent
with the estimate given in Sect.~\ref{band:eff} (continuum flux increasing with wavelength).

As for the K98 (UKIRT 2000) and $K'$ (TNG 2006) images used to measure the continuum flux due
to the unavailability of Br$\gamma$ images,
it is important to determine how accurate the subtraction of the H$_2$ line emission falling in the
filter band is. Firstly, the flux differences
before and after H$_2$ line subtraction are always $\sol 0.1$ mag, except for N3 ($\sim 0.3-0.4$ mag). Then, the flux
differences between the Br$\gamma$ and the $K_{\rm s}$ images are $0.24 \pm 0.11$
(PISCES 2012) and $-0.07 \pm 0.13$ (SOUL 2020), respectively. The differences are
dominated by S1a and N3, which enclose an area of fainter continuum superimposed by strong H$_2$ line emission (compare Fig.~\ref{nome:poligoni:lowline} and Fig.~\ref{3col:fig}).

This analysis suggests that the derived continuum fluxes may be affected by a systematic epoch-to-epoch error (i.e. constant in 
each epoch) of up to $0.1$-$0.2$ mag related to the zero point error. In addition, the fluxes from CIAO 2003 are possibly 
being overestimated due to 
the slightly non-linear regime of the secondary standard stars (see Sect.~\ref{app:ana}). Finally, a photometric error 
(i.e. depending 
on the polygon inside each epoch) of up to $\sim 0.1$ mag, which is slightly larger for the data points obtained from K-corrected images 
(UKIRT 2000 and TNG 2006), has to be taken into account as well.

\section{Line photometry of the jet with polygons: The low- plus high-spatial-resolution case}
\label{poly:line:low}

We also investigated the flux variability in the H$_{2}$ line by adopting the same technique as used for the continuum emission. 
First, photometry was carried out with {\it polyphot} in IRAF by defining a set of larger polygons to allow data points from the 
low spatial resolution images to be included. The adopted polygons and their labelling are shown in Fig.~\ref{nome:poligoni:lowline}.

As for the photometric errors, one has to take into account an additional source of uncertainty compared to what is discussed in 
Appendix~\ref{poly:des}, namely how good the subtraction is of the continuum contamination due to dust scattered emission. In fact, 
two approaches have been followed to estimate the continuum contamination. The first consists in using a narrow-band filter near 
the H2 passband which does not encompass the H$_{2}$ line, namely Br$\gamma$ and K$_{\rm cont}$. The narrow-band images taken into 
the same runs as H2 exhibit roughly the same spatial resolution as the H2 image and were just scaled to the H2 image using the stars 
in the field (assuming their spectrum is essentially a continuum), then they were registered and subtracted from the H2 image. 
This appears as 
the more accurate correction, particularly when using the Br$\gamma$ filter, which is very close to the H2 filter passband.  One can 
use the data taken with the TNG in 2001 to compare the correction obtained with Br$\gamma$ and that obtained with K$_{\rm cont}$. 
The difference between the corrected line fluxes from the various polygons is $0.07 \pm 0.09$ mag, with the correction with 
$K_{\rm cont}$ resulting in fainter magnitudes. This is consistent with the fact that the dust scattered emission is expected to be 
more intense in the K$_{\rm cont}$ than in the Br$\gamma$ band. Nevertheless, the two corrections appear to be equivalent.

\begin{figure}
   \centering
   \includegraphics[width=\hsize]{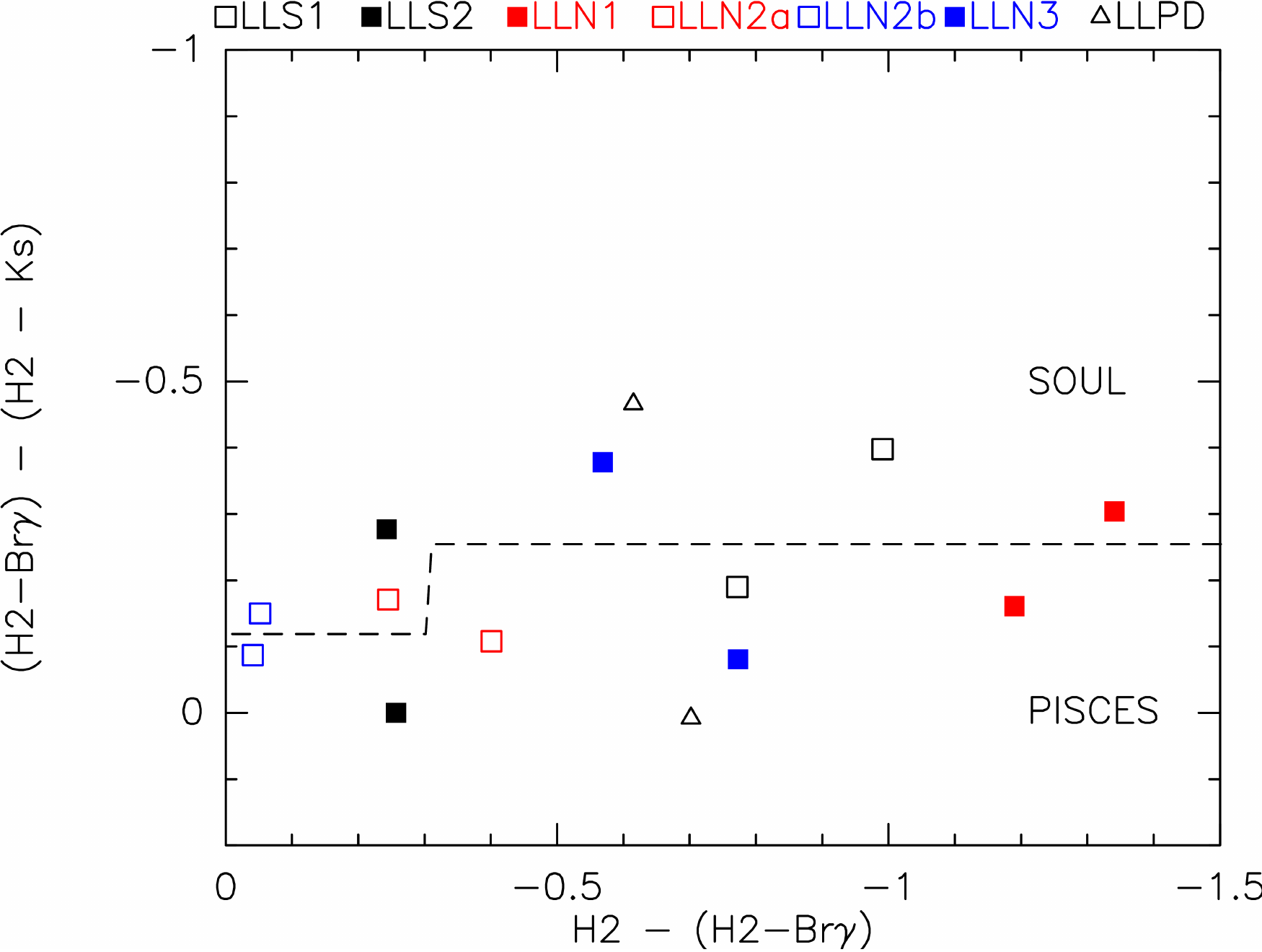}
      \caption{Difference (in magnitudes) between the pure line flux in the polygons after continuum correction from the 
      Br$\gamma$ image, (H2 - Br$\gamma$) and the K$_{s}$ image, 
      (H2 - K$_{s}$) vs the continuum contamination (in magnitudes) in the polygons estimated as the difference between the total 
flux measured in the H2 band and the line flux corrected using the Br$\gamma$ image (H2 - Br$\gamma$). The symbols refer to the 
polygons in Fig.~\protect\ref{nome:poligoni:lowline}, as explained at the top of the box. Two datasets have been used (PISCES 2012 
and SOUL 2020).}
         \label{err:line:fig}
   \end{figure}

The second approach consists in estimating the continuum contamination from the broad-band K$_{s}$ images. In this case the problem 
is that the band includes the H$_{2}$ $2.12$ $\mu$m line itself along with other H$_{2}$ lines. One has to assume that the continuum 
flux density is constant inside the band and that the band encompasses only the $2.12$ $\mu$m line. Then, from the H2 to K$_{s}$ flux 
ratios of the stars in the field, one can derive flux scaling factors for the H2 and K$_{s}$ images and subtract the latter from the 
former after flux scaling. We used the PISCES 2012 and SOUL 2020 data to compare the continuum corrections obtained from the 
Br$\gamma$ and K$_{s}$ band. In Fig.~\ref{err:line:fig} we plotted a raw estimate of the continuum contamination in the polygons 
(given by the difference of the total flux in magnitudes measured in the H2 band and the line flux corrected using the 
Br$\gamma$ image) versus
the difference (in magnitudes) of the pure line flux from the polygons corrected with the Br$\gamma$ image and the K$_{s}$ image.
It can be seen that the correction from the K$_{s}$ image tends to overestimate the continuum contribution compared to that from 
narrow band images and the difference (in magnitudes) is of the same order of the continuum contamination. So the uncertainty 
appears to be in the $0.2-0.5$ mag range, increasing with the continuum contamination level.

%

By using the spectra in \cite{2008A&A...485..137C}, we found that the inclusion of various H$_{2}$ lines in the $K_{s}$ band cannot 
contribute more than $0.1-0.2$ mag, even assuming that the continuum flux density increases as $\lambda^{4}$ would not affect 
the continuum correction significantly. This agrees with the results from the PISCES photometry, whereas
the SOUL photometry is affected by larger mismatches.
We have found that the source of the error lies in the accuracy of the ratio of the filter+atmospheric transmission (narrow band 
over broad band). An overestimate of this ratio by only 10--20 \% accounts for a correction error increasing with an increasing level 
of contamination as shown in Fig.~\ref{err:line:fig}. As we derived this ratio from the count ratios of the stars in the field, an 
error of 10--20 \% can be easily caused by their spectra containing absorption lines or other irregularities and/or significant 
differences in the PSFs of the two images (considering only a few stars are available in the AO fields). This highlights the 
need to check at least visually the level of continuum subtraction when using broad-band images. 

\section{Short movies}
\label{app:movies}

\begin{figure}
   \centering
   \includegraphics[width=8cm]{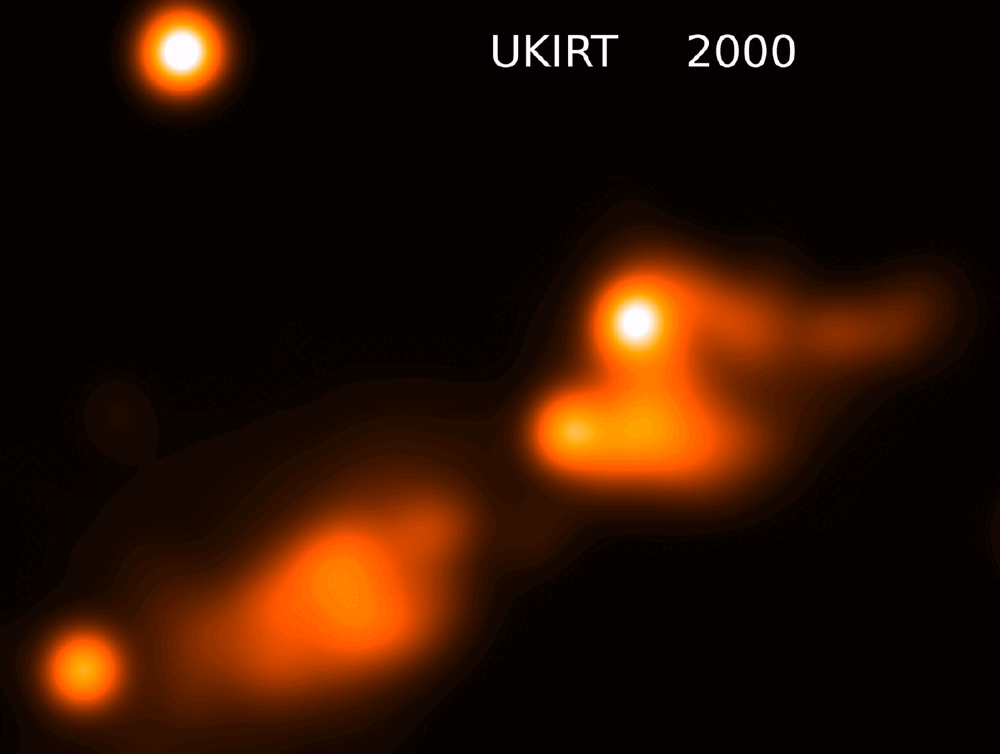}
   \caption{(Online movie) Continuum flux variability in
      the $K$ band.}
         \label{movie:cont}
   \end{figure}

To highlight the time variations in the field of IRAS20126+4104
through the various NIR images available from 2000 to 2020,
we have made two short movies by flux-scaling and aligning those
images. Figure~\ref{movie:cont} shows the flux variations in the continuum
$K$ band, obtained from the Br$\gamma$ filter frames (TNG 2001, CIAO 2003, PISCES 2012, and SOUL 2020) and the 
line-corrected $K$ frames (UKIRT 2000 and
TNG 2006). The resolution was degraded to the worst case and the photometric reference stars in the field were scaled 
to the same integrated fluxes. To slow down the time variations,
we added synthetic images in between each pair of consecutive epoch frames
obtained through an epoch-to-epoch linear interpolation of pixel values.
No assumptions were made as to possible periodical oscillations of the
light curves, so the flux changes were just linearly interpolated between 2006--2012 and 
2012--2020. The year is always indicated on the top of the frame as a rough clock marker.

\begin{figure}
   \centering
   \includegraphics[width=8cm]{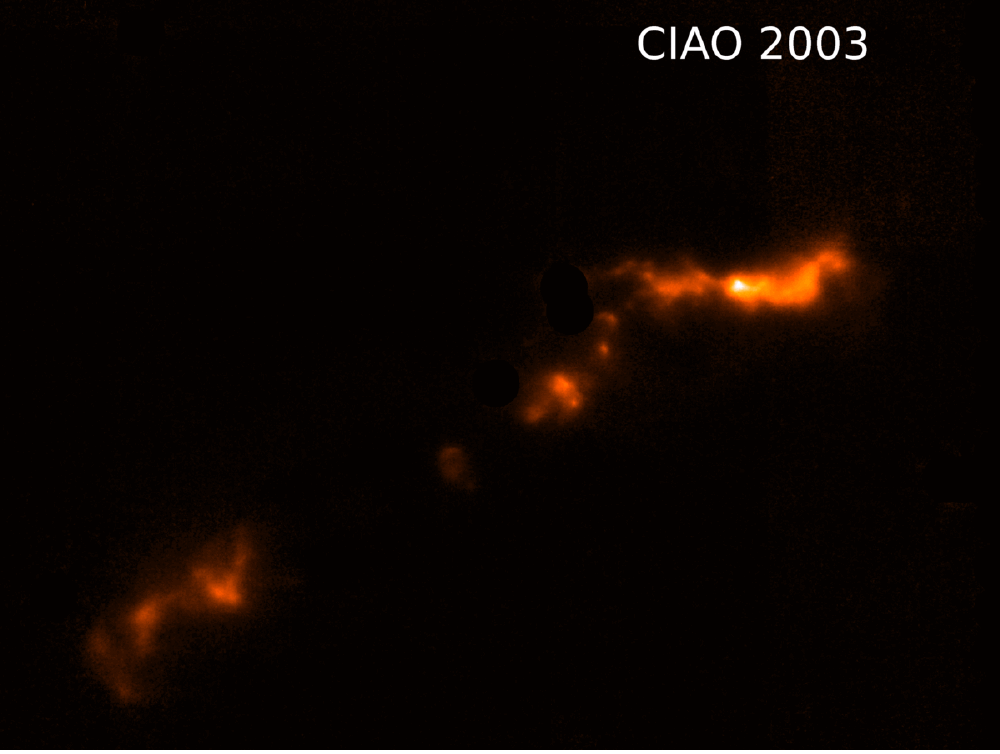}
      \caption{(Online movie) H$_{2}$ line structure variations
         between 2003 and 2020.}
         \label{movie:h2}
   \end{figure}

The movie in Fig.~\ref{movie:h2} only uses the high-resolution images
(CIAO 2003, PISCES 2012, and SOUL 2020) to show the variations
of the H$_{2}$ line-emitting structures from 2003 to 2020. The photometric reference stars were scaled to the same integrated 
fluxes, but the original resolution of the frames has not been modified. The proper motions of  knots B are evident, 
along with the development of bow-shock C1 
(HLN13, HLN14) and knot X2 (HLPD2).

\section{Kinematic measurements}
\label{appendix:morphology&PMs}

In Table~\ref{PMs:tab} we list the proper motions of the jet knots derived in this work. By combining these with the radial 
velocities measured by \citet{2008A&A...485..137C}, we obtained the 3D velocities listed in Table~\ref{table:3Dkinematics}.

\FloatBarrier
\longtab[1]{
\begin{longtable}{cccccccc}
\caption{Derived proper motions, kinematics, and dynamical ages of the IRAS\,20126+4104 knots. 
\label{PMs:tab}}\\
\hline\\[-5pt]
knot & Polygon & R.A.(J2000) &        Dec.(J2000) &  PM$\pm$dPM &  $\varv_{\rm tg} \pm {\rm d}\varv_{\rm tg}$ &  PA$\pm$dPA & $\tau \pm d\tau$ \\
 ID  &  & (hh:mm:ss) &     ($\degr$:$\arcmin$:$\arcsec$)   &  (mas\,yr$^{-1}$) & (km\,s$^{-1}$) & ($\degr$)  & (yr) \\
\hline\\[-5pt]
\endfirsthead
\caption{continued.}\\
\hline\\[-5pt]
knot & Polygon & R.A.(J2000) &        Dec.(J2000) &  PM$\pm$dPM &  $\varv_{\rm tg} \pm {\rm d}\varv_{\rm tg}$ &  PA$\pm$dPA & $\tau \pm d\tau$ \\
 ID  &  & (hh:mm:ss) &     ($\degr$:$\arcmin$:$\arcsec$)   &  (mas\,yr$^{-1}$) & (km\,s$^{-1}$) & ($\degr$)  & (yr) \\
\hline\\[-5pt]
\endhead
\hline
\endfoot
X1 & HLPD3  & 20:14:26.0760& 41:13:31.692 &    $\cdots$    &       $\cdots$  &    $\cdots$    &      $\cdots$  	  \\ 
X2 & HLPD2  & 20:14:26.0752& 41:13:31.999 &5.1$\pm$0.3    &       40$\pm$2 &   153$\pm$3     &      170$\pm$10  	   \\ 
X3c & HLPD1 & 20:14:26.126 & 41:13:32.550 &5.5$\pm$0.3    &       43$\pm$2 &    64$\pm$3     &      258$\pm$12  	   \\ 
X3r & HLPD1 & 20:14:26.115 & 41:13:32.609 &5.0$\pm$0.3    &       39$\pm$2 &   332$\pm$3     &      254$\pm$14  	   \\ 
X4  & & 20:14:26.120 & 41:13:31.982 &4.3$\pm$0.5    &       33$\pm$4 &   135$\pm$7     &      343$\pm$41  	   \\ 
A2a & HLS8 & 20:14:26.613 & 41:13:28.960 &8.1$\pm$0.3    &       63$\pm$2 &    71$\pm$1     &      1166$\pm$39  	   \\
A2b & HLS8 & 20:14:26.617 & 41:13:29.050 &13.0$\pm$1.0    &       101$\pm$8 &   142$\pm$4    &     728$\pm$57  	   \\ 
A2c & HLS10 & 20:14:26.597 & 41:13:29.680 &8.9$\pm$1.6    &       69$\pm$12&   140$\pm$8     &      1007$\pm$182	  \\ 
A2d & HLS11 & 20:14:26.576 & 41:13:29.940 &5.3$\pm$1.8    &       41$\pm$14&   144$\pm$14    &      1623$\pm$538	  \\ 
A2e & HLS11 & 20:14:26.583 & 41:13:30.090 &7.0$\pm$0.9    &       55$\pm$7 &   135$\pm$7     &      1229$\pm$153	  \\ 
A2f & HLS9 & 20:14:26.671 & 41:13:29.420 &8.5$\pm$0.3    &       66$\pm$3 &   126$\pm$2     &      1192$\pm$48	  \\ 
A2g & HLS7 & 20:14:26.649 & 41:13:29.040 &13.1$\pm$0.1    &       102$\pm$1 &   77$\pm$1     &     755 $\pm$3 		  \\ 
A1a & HLS1 & 20:14:26.844 & 41:13:26.430 &5.8$\pm$0.1    &       45$\pm$1 &   146$\pm$1     &      2367$\pm$37	  \\ 
A1b & HLS2 & 20:14:26.896 & 41:13:27.600 &3.6$\pm$0.2    &       28$\pm$2 &   130$\pm$3     &      3846$\pm$206	  \\ 
A1aa  & & 20:14:26.898 & 41:13:26.822 &9.7$\pm$0.9    &       75$\pm$7 &   133$\pm$5     &      1469$\pm$133	  \\ 
A1ab & HLS1 & 20:14:26.860 & 41:13:26.628 &9.0$\pm$1.0    &       70$\pm$8 &   148$\pm$6     &      1527$\pm$171	  \\ 
A1ac & HLS1 & 20:14:26.850 & 41:13:26.567 &5.2$\pm$1.0    &       40$\pm$8 &   128$\pm$11    &      2635$\pm$513	  \\ 
A1c & HLS5 & 20:14:26.808 & 41:13:28.320 &9.1$\pm$0.2    &       71$\pm$1 &   124$\pm$1     &      1363$\pm$25	  \\ 
A1d & HLS3 & 20:14:26.807 & 41:13:27.870 &6.9$\pm$0.3    &       54$\pm$2 &   141$\pm$2     &      1820$\pm$71	  \\ 
B1  & HLN1  & 20:14:25.919 & 41:13:33.520 &9.2$\pm$0.1    &       71$\pm$1 &   303$\pm$1     &      209$\pm$2  		\\ 
B1a & HLN1 & 20:14:25.932 & 41:13:33.400 &7.1$\pm$0.4    &       55$\pm$3 &   318$\pm$3     &      270$\pm$15 	\\ 
B1b & HLN1 & 20:14:25.934 & 41:13:33.708 &10.2$\pm$0.8    &       80$\pm$6 &   324$\pm$4     &     180$\pm$13 \\
B2a & HLN2 & 20:14:25.873 & 41:13:33.470 &4.8$\pm$0.6    &       37$\pm$5 &   344$\pm$7     &      526$\pm$68       \\ 
B2b & HLN2 & 20:14:25.879 & 41:13:33.635 &16.5$\pm$0.2    &       129$\pm$2 &   286$\pm$3    &     152$\pm$2  	    \\ 
B2c & HLN5 & 20:14:25.867 & 41:13:34.279 &14.5$\pm$0.7    &       113$\pm$5 &   316$\pm$3    &     206$\pm$10       \\ 
B2d & HLN5 & 20:14:25.853 & 41:13:34.360 &12.9$\pm$0.7    &       100$\pm$5 &   320$\pm$3    &     248$\pm$13       \\ 
B2da & HLN5 & 20:14:25.852 & 41:13:34.305 &10.8$\pm$1.0    &       84$\pm$8 &   320$\pm$5     &     295$\pm$28       \\ 
B2db & HLN5 & 20:14:25.836 & 41:13:34.296 &13.8$\pm$1.3    &       108$\pm$10&   304$\pm$5    &     245$\pm$24       \\ 
B2dc & HLN5 & 20:14:25.804 & 41:13:34.340 &15.3$\pm$1.2    &       119$\pm$10&   312$\pm$5    &     251$\pm$21       \\ 
B2e & HLN4 & 20:14:25.825 & 41:13:33.861 &14.7$\pm$0.5    &       114$\pm$4 &   292$\pm$4    &     228$\pm$7  	    \\ 
B2ea & HLN4 & 20:14:25.814 & 41:13:33.962 &14.4$\pm$1.2    &       112$\pm$10&   293$\pm$5    &     245$\pm$21       \\ 
B2f  & HLN3 & 20:14:25.850 & 41:13:33.948 &16.6$\pm$1.2    &       129$\pm$10&   314$\pm$5    &     183$\pm$14       \\ 
B2fa & HLN3 & 20:14:25.844 & 41:13:33.869 &17.7$\pm$1.2    &       138$\pm$10&   315$\pm$5    &     174$\pm$12       \\ 
B3a & HLN7 & 20:14:25.765 & 41:13:35.010 &16.9$\pm$0.9    &       132$\pm$7 &   310$\pm$3    &     276$\pm$15       \\ 
B3b & HLN8 & 20:14:25.764 & 41:13:35.307 &16.3$\pm$0.1    &       127$\pm$1 &   319$\pm$1    &     297$\pm$2  	    \\ 
B3ba & HLN8 & 20:14:25.769 & 41:13:35.488 &20.3$\pm$1.5    &       158$\pm$12&   323$\pm$4    &     241$\pm$18       \\ 
B3bb & HLN8 & 20:14:25.768 & 41:13:35.667 &13.5$\pm$1.5    &       105$\pm$12&   312$\pm$6    &     370$\pm$41       \\ 
B3c & HLN9 & 20:14:25.749 & 41:13:36.070 &15.1$\pm$0.1    &       117$\pm$1 &   321$\pm$1    &     363$\pm$2  	    \\ 
B3d & HLN10 & 20:14:25.718 & 41:13:36.250 &6.3$\pm$0.3    &       49$\pm$2 &   309$\pm$3     &      951$\pm$47       \\ 
C1a0  & & 20:14:25.940 & 41:13:35.350 &11.1$\pm$1.1    &       86$\pm$9 &   316$\pm$6     &     278$\pm$28       \\ 
C1b  & & 20:14:25.841 & 41:13:36.440 &14.6$\pm$2.5    &       113$\pm$20&   300$\pm$10   &     329$\pm$57       \\ 
C1c  & & 20:14:25.882 & 41:13:36.820 &4.3$\pm$0.2    &       33$\pm$2 &    5$\pm$3      &      1060$\pm$55       \\ 
C1d  & & 20:14:25.832 & 41:13:36.882 &6.4$\pm$1.2    &       50$\pm$10&   326$\pm$1     &      816$\pm$159      \\ 
C1e & HLN11 & 20:14:25.792 & 41:13:37.100 &6.7$\pm$1.0    &       52$\pm$8 &   326$\pm$9     &      690$\pm$104      \\ 
C1f & HLN11 & 20:14:25.768 & 41:13:37.060 &12.0$\pm$1.0    &       93$\pm$8 &   291$\pm$5     &     591$\pm$50       \\ 
C1g & HLN11 & 20:14:25.750 & 41:13:37.060 &4.1$\pm$0.5    &       32$\pm$4 &   342$\pm$6     &      1507$\pm$197      \\ 
C1h1 & HLN12& 20:14:25.724 & 41:13:37.250 &3.2$\pm$0.5    &       25$\pm$4 &   315$\pm$10    &      2031$\pm$337      \\ 
C1h2 & HLN12 & 20:14:25.720 & 41:13:37.254 &6.0$\pm$1.0    &       46$\pm$8 &   314$\pm$10    &      1107$\pm$188      \\ 
C1i  & HLN15b & 20:14:25.681 & 41:13:37.256 &6.4$\pm$0.4    &       50$\pm$3 &   313$\pm$4     &      1096$\pm$76       \\ 
C1hh & HLN14 & 20:14:25.652 & 41:13:36.561 &16.9$\pm$1.2    &       131$\pm$10&   293$\pm$4    &     440$\pm$33       \\ 
C1hi & HLN14 & 20:14:25.660 & 41:13:36.609 &1.7$\pm$1.2    &       13$\pm$10&   316$\pm$42    &      4319$\pm$3128     \\ 
C1hj & HLN15a & 20:14:25.683 & 41:13:36.940 &8.1$\pm$1.2    &       63$\pm$10&   322$\pm$9     &      812$\pm$125      \\ 
C1hk & HLN14 & 20:14:25.697 & 41:13:36.600 &9.1$\pm$1.2    &       71$\pm$10&   318$\pm$8     &      708$\pm$97       \\ 
C1k1 & HLN13 & 20:14:25.632 & 41:13:36.300 &7.7$\pm$1.2    &       60$\pm$10&   244$\pm$9     &      914$\pm$148      \\ 
C1k & HLN13 & 20:14:25.634 & 41:13:36.330 &5.4$\pm$1.2    &       42$\pm$10&   276$\pm$13    &      1314$\pm$307      \\ 
C1l & HLN14 & 20:14:25.627 & 41:13:36.620 &10.2$\pm$1.2    &       79$\pm$10&   288$\pm$7     &     713$\pm$87       \\ 
C1m  & HLN16a & 20:14:25.607 & 41:13:36.650 &10.1$\pm$1.2    &       79$\pm$10&   292$\pm$7     &     747$\pm$92       \\ 
C1n  & HLN16b & 20:14:25.594 & 41:13:36.866 &8.2$\pm$0.8    &       64$\pm$6 &   323$\pm$5     &      954$\pm$88       \\ 
C1o  & HLN17a & 20:14:25.563 & 41:13:36.710 &7.7$\pm$0.3    &       60$\pm$3 &   322$\pm$3     &      914$\pm$41       \\ 
C1p  & HLN17b & 20:14:25.548 & 41:13:36.860 &5.9$\pm$0.8    &       46$\pm$6 &   303$\pm$7     &      1417$\pm$181      \\ 
C1qu & HLN13 & 20:14:25.645 & 41:13:36.026 &10.1$\pm$0.9    &       78$\pm$7 &   325$\pm$5     &     651$\pm$57       \\ 
C1qd & HLN13 & 20:14:25.617 & 41:13:35.913 &16.0$\pm$1.2    &       124$\pm$10&   297$\pm$5    &     410$\pm$32       \\ 
C1r & HLN13 & 20:14:25.671 & 41:13:36.170 &6.3$\pm$1.2    &       49$\pm$10&   315$\pm$11    &      1028$\pm$204      \\ 
C2a & HLN18 & 20:14:25.514 & 41:13:36.752 &5.8$\pm$0.6    &       45$\pm$5 &   312$\pm$5     &      1585$\pm$157      \\ 
C2b & HLN19 & 20:14:25.464 & 41:13:36.771 &12.9$\pm$0.5    &       100$\pm$4 &   287$\pm$2    &     735$\pm$29       \\ 
C2ba & HLN19 & 20:14:25.475 & 41:13:36.838 &8.8$\pm$1.2    &       69$\pm$10&   300$\pm$8     &      1060$\pm$150      \\ 
C2bb & HLN19 & 20:14:25.500 & 41:13:36.471 &17.2$\pm$1.2    &       134$\pm$10&   315$\pm$4    &     513$\pm$37       \\ 
C2bc & HLN19 & 20:14:25.472 & 41:13:36.504 &8.3$\pm$1.2    &       65$\pm$10&   303$\pm$9     &      1113$\pm$167      \\ 
C2bd & HLN19 & 20:14:25.490 & 41:13:36.690 &11.5$\pm$1.2    &       90$\pm$10&   283$\pm$6     &     891$\pm$96       \\ 
C2c & HLN20 & 20:14:25.413 & 41:13:36.810 &11.3$\pm$0.6    &       88$\pm$5 &   288$\pm$3     &     905$\pm$50       \\ 
C2ca & HLN20 & 20:14:25.408 & 41:13:36.889 &13.5$\pm$1.2    &       105$\pm$10&   278$\pm$5    &     761$\pm$70       \\ 
C2d & HLN20 & 20:14:25.401 & 41:13:36.580 &12.3$\pm$0.4    &       95$\pm$3 &   292$\pm$3     &     836$\pm$30       \\ 
C2da & HLN20 & 20:14:25.395 & 41:13:36.420 &11.4$\pm$1.2    &       88$\pm$10&   286$\pm$6     &     904$\pm$99       \\ 
C2e & HLN21 & 20:14:25.362 & 41:13:36.570 &7.2$\pm$0.3    &       56$\pm$2 &   284$\pm$3     &      1497$\pm$62       \\ 
C2es & HLN21 & 20:14:25.356 & 41:13:36.438 &11.9$\pm$1.0    &       92$\pm$8 &   293$\pm$4     &     911$\pm$74       \\ 
C2f & HLN22 & 20:14:25.341 & 41:13:36.670 &9.6$\pm$0.3    &       75$\pm$2 &   280$\pm$4     &      1156$\pm$32       \\ 
C2fa & HLN22 & 20:14:25.335 & 41:13:36.420 &13.1$\pm$1.2    &       102$\pm$10&   264$\pm$5    &     847$\pm$81       \\ 
C2g & HLN22 & 20:14:25.323 & 41:13:36.852 &7.6$\pm$0.9    &       59$\pm$7 &   279$\pm$5     &      1512$\pm$175      \\ 
C2ga & HLN22 & 20:14:25.325 & 41:13:36.730 &6.9$\pm$0.9    &       53$\pm$7 &   303$\pm$7     &      1658$\pm$212      \\ 
C2h1 & HLN22 & 20:14:25.323 & 41:13:37.110 &11.6$\pm$0.8    &       90$\pm$6 &   280$\pm$4     &     996$\pm$65       \\ 
C2h2 & HLN22 & 20:14:25.314 & 41:13:37.207 &15.0$\pm$1.0    &       117$\pm$8 &   292$\pm$4    &     779$\pm$52       \\ 
C2i & HLN23a & 20:14:25.288 & 41:13:37.435 &11.1$\pm$1.1    &       86$\pm$9 &   296$\pm$11    &     1097$\pm$112      \\ 
C2j & HLN23b &  20:14:25.259 & 41:13:37.310 &10.1$\pm$0.4    &       78$\pm$3 &   293$\pm$2     &     1240$\pm$44       \\ 
C2k  & HLN23d & 20:14:25.230 & 41:13:37.348 &14.0$\pm$1.1    &       109$\pm$8 &   293$\pm$4    &     923$\pm$72       \\ 
C2ka & HLN23b & 20:14:25.245 & 41:13:37.250 &8.0$\pm$1.2    &       62$\pm$10&   283$\pm$9     &      1581$\pm$246      \\ 
C2m  & HLN25 & 20:14:25.118 & 41:13:38.690 &12.6$\pm$1.0    &       98$\pm$8 &   297$\pm$4     &     1189$\pm$93       \\
\hline\\[-5pt]
\hline
\end{longtable}
}

\begin{table}
\caption{Total velocity and inclination to the line of sight of the vectors' velocity of the knots in IRAS20126+4104 
(i.e. $i < 90$ indicates a blue-shifted knot and $i > 90$ indicates a red-shifted knot).}             
\label{table:3Dkinematics}     
\centering                     
\begin{tabular}{c c c c}       
\hline\hline                 
Structure & knot & $\varv_{\rm tot} \pm {\rm d}\varv_{\rm tot}$ & i$\pm$di \\  
          &  ID  & (km\,s$^{-1}$) & ($\degr$)  \\
\hline                       
   A1 & A1a & 45$\pm$3 & 93$\pm$3 \\
   A2 & A2a & 64$\pm$4    & 98$\pm$3 \\
   B & B2d & 101$\pm$6     & 82$\pm$2 \\
   C1 & C1hh & 134$\pm$10   & 82$\pm$3 \\
   C2 & C2b & 103$\pm$5    & 76$\pm$3 \\ 
\hline                                   
\end{tabular}
\end{table}
%
\end{appendix}
\end{document}